\documentclass[aps,pre,reprint,twocolumn,superscriptaddress,showpacs]{revtex4-2}
\usepackage{amssymb,amsmath}
\usepackage[table]{xcolor}
\usepackage{graphicx}
\usepackage{mathtools}
\usepackage[export]{adjustbox}
\usepackage{overpic}
\usepackage[colorlinks=true,
 linkcolor=black,
 urlcolor=blue,
 citecolor=blue]{hyperref}
\usepackage{nicefrac}
\usepackage{multirow}
\usepackage{lipsum} 
\usepackage[normalem]{ulem}
\usepackage{pbox}
\newcolumntype{C}[1]{>{\centering\let\newline\\\arraybackslash\hspace{0pt}}m{#1}}
\usepackage{verbatim}
\usepackage{enumitem}
\usepackage{caption}
\usepackage{subcaption}
\usepackage{float}

\usepackage{framed,color}
\definecolor{shadecolor}{rgb}{0.85,0.80,0.80}

\definecolor{myorange}{RGB}{253, 184, 99}
\definecolor{mypurple}{RGB}{178, 171, 210}

\newcommand{\comments}[1]{}
\usepackage{multirow}

\usepackage{float}

\newcommand{\matj}{\mathcal{J}}
\newcommand{\mati}{\mathcal{I}}
\newcommand{\mathinv}{\mathcal{H}_0^{-1}}
\newcommand{\mathh}{\mathcal{H}}
\newcommand{\id}{{\openone}}

\allowdisplaybreaks
\begin{document}
\title{A path integral approach to sparse non-Hermitian random matrices}
\author{Joseph W. Baron}
\email{joseph-william.baron@phys.ens.fr}
\affiliation{Laboratoire de Physique de l’Ecole Normale Sup\`{e}rieure, ENS, Universit\'{e} PSL, CNRS, Sorbonne Universit\'{e}, Universit\'{e} de Paris, F-75005 Paris, France}

\begin{abstract}
The theory of large random matrices has proved an invaluable tool for the study of systems with disordered interactions in many quite disparate research areas. Widely applicable results, such as the celebrated elliptic law for dense random matrices, allow one to deduce the statistical properties of the interactions in a complex dynamical system that permit stability. However, such simple and universal results have so far proved difficult to come by in the case of sparse random matrices. Here, we perform an expansion in the inverse connectivity, and thus derive general modified versions of the classic elliptic and semi-circle laws, taking into account the sparse correction. This is accomplished using a dynamical approach, which maps the hermitized resolvent of a random matrix onto the response functions of a linear dynamical system. The response functions are then evaluated using a path integral formalism, enabling one to construct Feynman diagrams, which facilitate the perturbative analysis. Additionally, in order to demonstrate the broad utility of the path integral framework, we derive a generic non-Hermitian generalization of the Marchenko-Pastur law, and we also show how one can handle non-negligible higher-order statistics (i.e. non-Gaussian statistics) in dense ensembles.
\end{abstract}

\maketitle

\section{Introduction}
The central observation of random matrix theory (RMT) is that the eigenvalues of a large matrix can often be determined by knowing only the \textit{statistical} properties of the matrix entries, rather than the specific entries themselves \cite{mehta, taobook}. This powerful insight is responsible for RMT's broad applicability. In many-component dynamical systems, for example, RMT allows one to draw qualitative conclusions about how the statistics of the interactions between components contribute to the (in)stability of the system. 

Some of the first forays made by physicists into RMT were the studies of Wigner \cite{wigner1955, wigner1958distribution, wigner1967random} (and also notably Dyson \cite{dyson1953dynamics, dyson1962brownian} and Mehta \cite{mehta}). Wigner proposed to model the interactions in large nuclei as having random i.i.d. values. In so doing, he uncovered his eponymous semi-circle law for the distribution of the eigenvalues a symmetric random matrix. Since then, the study of large random matrices has found a panoply of uses in physics \cite{RMPhysicsBook} (e.g. spin glasses \cite{mezard1987}) and in other disciplines (ecology \cite{may, allesinatang1, allesinatang2}, neural networks \cite{kuczala2016eigenvalue, aljadeff2015transition, Coolen_NN, rajan2006eigenvalue, ahmadian2015properties}, and finance \cite{laloux2000random}, for example). 

Driven by a rapidly growing range of applications, the variety of random matrix ensembles that have been studied over the years has increased greatly. In particular, Wigner's semi-circle law for symmetric matrices was generalized to the elliptic law for asymmetric matrices \cite{sommers, girko1986elliptic}. The solitary outlier eigenvalue that results from the inclusion of a non-zero mean of the matrix entries was also characterized with similarly compact formulae \cite{orourke, benaych2011eigenvalues, edwardsjones}. Recently, matrices with non-trivial network structure \cite{baron2022networks, metz2020spectral, rodgers2005eigenvalue}, block structure \cite{baron2020dispersal, grilli2017, allesina2015}, generalized correlations \cite{aceitunorogersschomerus, baron2022eigenvalues}, and index-dependent statistics \cite{aljadeff2015transition, kuczala2016eigenvalue} have also been studied.

All of the aforementioned results are applicable to dense matrices, for which the number of non-zero entries per row $p$ scales with the dimension of the matrix $N \gg1$. However, there are many systems \cite{metz2019spectral, rogers2009cavity, susca2021cavity} where $p$ is small compared with the system size, i.e. $p \ll N$. Such matrices are termed sparse. Applications of sparse random matrices include supercooled liquids \cite{cavagna1999analytic}, Anderson localization and diffusion on networks \cite{hatano1996localization, biroli1999single, baron2021persistent, chalker1990anderson, harris1981mean, kim1985density}, and particularly complex ecosystems \cite{marcus2022local, mambuca2022dynamical, valigi2023local, akjouj2022feasibility}. 

A general characterization of the eigenvalue spectra of sparse random matrices has proved a somewhat more difficult challenge than for dense matrices \cite{metz2019spectral}. This is largely due to the fact that there is not the same degree of so-called universality \cite{taovukrishnapur2010, bai2010spectral, nguyen2015elliptic} of results. That is, in the dense case, the eigenvalue density (characterised by the elliptic law in the asymmetric case \cite{sommers, girko1986elliptic}) is seen only to depend on the variance and covariance of the matrix elements. In contrast, the shape of the eigenvalue spectrum can vary greatly for sparse matrices \cite{rogers2009cavity, metz2019spectral}, depending more intricately on the details of the distribution from which the non-zero elements are drawn (i.e. on the higher-order moments). 

In this work, we aim to address this issue by extending the path-integral approach of Ref. \cite{baron2022eigenvalues} to calculate the eigenvalue spectra of non-Hermitian \cite{JANIK1997603,fyodorov2003random} random matrices. We use this method to perform a perturbative analysis, with the help of Feynman diagrams \cite{hertz2016path}, and thus provide simple closed-form expressions for the sparse corrections to the elliptic and semi-circle laws \cite{sommers, girko1986elliptic} as a series in $1/p$. These results only depend on a small number of statistics of the non-zero matrix elements, and therefore apply to many different ensembles of sparse matrix. 

In addition to presenting new results for sparse matrices, the aim of this work is also to exhibit the strengths of the path integral method for calculating the eigenvalue spectrum \cite{baron2022eigenvalues}. This method has several advantages over others:

(i) Most importantly, the path integral approach facilitates a controlled perturbative analysis, in contrast to the effective medium and single defect approximation schemes \cite{semerjian2002sparse, biroli1999single}, which are uncontrolled. The compact formulae that we obtain are simple and broadly applicable, whereas the more exact results obtained from the cavity method \cite{rogers2009cavity, metz2019spectral, mambuca2022dynamical, kuhn2008spectra, neri2012spectra}, for example, often have to be evaluated in certain special cases or using numerical methods.

(ii) The independence of the results for the eigenvalue density from the exact distribution from which the matrix elements are drawn (so-called universality \cite{taovukrishnapur2010, bai2010spectral, nguyen2015elliptic}) is clear. Other approaches, such as the direct expansion of the resolvent using a Dyson series \cite{braymoore, kuczala2016eigenvalue}, must use a Gaussian distribution to derive the results. The applicability of these results to other distributions is often presumed subsequently. 

(iii) Finally, there is no need for replicas \cite{dedominicis1978dynamics}, and hence no replica symmetry ansatz (unlike the approaches used in Refs. \cite{rodgers1988density, sommers, semerjian2002sparse}, for example), nor do we require the additional Grassmann variables that enter in the supersymmetric formalism \cite{akara2022random, efetov1999supersymmetry}.

We emphasize that the perturbative approach presented here can be used to take into account many other aspects of random matrix ensembles in addition to the sparse corrections. For example, the path integral method has already been used in Ref. \cite{baron2022eigenvalues} to take into account so-called generalized correlations. We demonstrate in the present work that the same `ribbon' Feynman diagrams that encode the sparse corrections can also be used to take into account non-negligible higher-order statistics in dense ensembles \cite{azaele2024generalized}. We also show how the path integral approach simplifies problems involving random matrix products (recovering a non-hermitian generalization of the Marchenko-Pastur law \cite{akemann2021non, kanzieper2010non}) and block-structured random matrices. 

The rest of this work is set out as follows. We first present the general method. In Section \ref{section:setup}, we introduce an auxiliary dynamical system whose response functions can be used to derive the properties of the eigenvalue spectrum. We show how these response functions can be represented by a path integral expression in Section \ref{section:msrjd}, and we introduce the associated `rainbow' Feynman diagram representation in \ref{section:ellipticlawdiagrams}. As a straightforward example of the general procedure for evaluating the eigenvalue spectrum, we recover the well-known elliptic law. 

We then move on to the case of sparse matrices. In Section \ref{section:sparsefirstorder}, we present the additional `ribbon' Feynman diagrams that encode the sparse corrections to the eigenvalue spectrum, and derive the first-order sparse corrections to the elliptic law. Our general formulae are then used to draw broad conclusions about the stability of sparsely interacting dynamical systems in Section \ref{section:stability}. In Section \ref{section:higherorder}, we show that one can continue the perturbative expansion to higher order in $1/p$ to obtain progressively more accurate results. 

In Section \ref{section:otherapplications}, we present some further applications of the path integral formalism. Namely, we show how one can handle non-vanishing higher order moments, matrices with block structure and products of random matrices. Finally, in Section \ref{section:conclusion}, we discuss the significance of the results, the drawbacks and shortcomings of this method, and we conclude.

\section{General set-up}\label{section:setup}
\subsection{Hermitization procedure}
We begin by briefly recapitulating the standard hermitization method, which allows one to compute the resolvent of a non-Hermitian random matrix \cite{JANIK1997603, feinberg1997non}. 

Let $\underline{\underline{a}}$ be a large $N \times N$  matrix whose elements $\{a_{ij}\}$ are drawn from some (possibly joint) distribution, and let $\{\lambda_\nu\}$ be its eigenvalues. We define the disorder-averaged eigenvalue density as
\begin{align}
	\rho(\omega) = \overline{\frac{1}{N}\sum_\nu \delta(x- \mathrm{Re}(\lambda_\nu))\delta(y- \mathrm{Im}(\lambda_\nu)) }, 
\end{align}
where $\omega = x + i y$, and the over-bar $\overline{\cdots}$ indicates an average over realizations of the random matrix elements. The disorder-averaged resolvent of the random matrix $\underline{\underline{a}}$ is defined as
\begin{align}
	C(\omega, \omega^\star) = \frac{1}{N}\mathrm{Tr}\overline{\left[\underline{\underline{\id}} \omega - \underline{\underline{a}}\right]^{-1} } = \overline{\frac{1}{N} \sum_\nu \frac{1}{\omega - \lambda_\nu}}. \label{resolventdefinition}
\end{align}
The eigenvalue density in the complex plane is then given by \cite{sommers, girko1986elliptic, JANIK1997603, feinberg1997non}
\begin{align}
	\rho(\omega) = \frac{1}{\pi} \frac{\partial C}{\partial \omega^\star} . \label{densityfromres}
\end{align}
In the case where $\underline{\underline{a}}$ is Hermitian, and all the eigenvalues are consequently real, we can obtain the eigenvalue density on the real axis via \cite{edwardsjones, braymoore}
\begin{align}
	\rho_x(x) = \frac{1}{\pi}\lim_{\epsilon \to 0} \mathrm{Im} \, C(x - i \epsilon) . \label{realeigenvaluedensity}
\end{align}
When the matrix $\underline{\underline{a}}$ is Hermitian, the disorder-averaged resolvent is an analytic function of $\omega$ (aside from the section of the real axis where the eigenvalues reside). Thus, Eq.~(\ref{resolventdefinition}) can be expanded as a series in powers of $\omega$, which is a helpful trick when performing the disorder average. When $\underline{\underline{a}}$ is non-Hermitian however, the resolvent is non-analytic for values of $\omega$ in areas of the complex plane where $\rho(\omega)$ is non-zero, as can be seen from Eq.~(\ref{densityfromres}). Thus, a series expansion in $\omega$ is not valid. Instead, it is necessary to construct a `hermitized' resolvent, from which $C$ can later be extracted. 

We now define the $2N \times 2N$ Hermitian matrix
\begin{align}
	\underline{\underline{H}} = \begin{bmatrix}
		0 &  \omega \underline{\underline{\id}}_{N} -\underline{\underline{a}} \\
		( \omega \underline{\underline{\id}}_{N}- \underline{\underline{a}} )^\dagger & 0
	\end{bmatrix} ,
\end{align}
and the hermitized resolvent matrix
\begin{align}
	\underline{\underline{\mathcal{H}}}(\eta, \omega, \omega^\star) = \overline{\left\langle \left[\eta \underline{\underline{\id}}_{2N} + \underline{\underline{H}}\right]^{-1} \right\rangle } . \label{hermres}
\end{align}

From these definitions we see that we can recover the resolvent we seek via
\begin{align}
	C(\omega, \omega^\star) = \frac{1}{N} \lim_{\eta\to 0} \mathrm{Tr} \left[  \underline{\underline{\mathcal{H}}}^{21} (\eta, \omega, \omega^\star)\right], \label{desiredresolvent}
\end{align}
where the upper indices of $\mathcal{H}$ refer to its blocks. Let us now label the blocks of the resolvent matrix $\underline{\underline{\mathcal{H}}}$ as
\begin{align}
	\underline{\underline{\mathcal{H}}} &= 
	\begin{bmatrix}
		\underline{\underline{A}} & \underline{\underline{B}} \\
		\underline{\underline{C}} & \underline{\underline{D}}
	\end{bmatrix},\label{hdef}
\end{align}
so that $C \equiv N^{-1}\mathrm{Tr}\underline{\underline{C}}$, and similar definitions apply for the other blocks. We also define the following matrices for later use 
\begin{align}
	\underline{\underline{\mathh}}_0^{-1} = \begin{bmatrix}
		0 & \omega \underline{\underline{\id}}_{N} \\
		\omega^\star \underline{\underline{\id}}_{N} & 0
	\end{bmatrix} , \,\,\,\,\, \underline{\underline{\mathcal{I}}} = \begin{bmatrix}
		0& \underline{\underline{a}} \\
		\underline{\underline{a}}^\dagger & 0
	\end{bmatrix},  \label{matrixdefs}
\end{align}
such that we have $\underline{\underline{H}} = \underline{\underline{\mathh}}_0^{-1} -\underline{\underline{\mathcal{I}}}$. For the sake of having a more compact notation later, we also introduce the following $2\times 2$ matrices, which we denote \textit{without} underscores or lower indices 
\begin{align}
	\mathcal{H} = 
	\begin{bmatrix}
		A & B\\
		C & D
	\end{bmatrix}, \,\,\,\,\, 	
	\mathcal{H}_0^{-1} = 
	\begin{bmatrix}
		0 & \omega \\
		\omega^\star & 0
	\end{bmatrix}. \label{def2by2}
\end{align}
As we shall see, it is the matrix $\mathcal{H}_0$ that will be used to perform a series expansion of the hermitized resolvent, just as the resolvent of an Hermitian matrix can be expanded in powers of $\omega^{-1}$ (see e.g. Ref. \cite{braymoore}).

\subsection{Why the MSRJD path integral approach?}\label{section:whymsrjd}
The Martin-Siggia-Rose-Janssen-de Dominicis (MSRJD) \cite{altlandsimons, msr, janssen1976lagrangean, dominicis1976techniques, dedominicis1978dynamics} framework is well-established as a tool for handling problems with quenched disorder \cite{altlandsimons, hertz2016path, dedominicis1978dynamics}. As we shall see below, using an auxiliary field (in this case, one that obeys a given dynamics) simplifies the process of taking the disorder average of the hermitized resolvent in Eq.~(\ref{hermres}), which is the object at the centre of our calculation. However, there are a number of formalisms that use different tricks to achieve the same simplification. We discuss briefly why the MSRJD framework is particularly well-suited to the task in the present work.

One notes that the related Keldysh formalism, which involves dynamic variables integrated over two time contours (as opposed to the single time of the MSRJD framework), has also been used to tackle many problems involving disorder \cite{kamenev1999electron}, such as the calculation of two-point eigenvalue correlations of dense symmetric random matrices \cite{altland2000wigner}, for example. However, we opt to use the MSRJD dynamic formalism to evaluate the eigenvalue density, due to its comparative simplicity and its immediate relation to simple dynamical systems, which can facilitate the generalization of the results to more complicated ensembles (see Section \ref{section:densegeneralexamples}).

Other auxiliary field formalisms that lend themselves to handling disorder are the replica and supersymmetric approaches. While the replica approach is certainly crucial for understanding phenomena such as glassy behaviour and rough energy landscapes via replica symmetry breaking \cite{mezard1987, altieri2021properties, ros2019complex}, and the supersymmetric method is well-suited for non-perturbative calculations (such as the microscopic correlations of eigenvalues \cite{verbaarschot1985critique, efetov1999supersymmetry, fyodorov1998universality} or in the weakly-non-hermitian regime \cite{fyodorov2003random, fyodorov1997almost}), these methods introduce additional variables that enter the calculation in non-trivial ways, but are arguably superfluous for a perturbative treatment of the eigenvalue spectral density. When using the replica approach to construct diagrammatic series, for example, one must take account of the order of diagrams not only in $N$ (the matrix dimension), but also in $n$ (the number of replicas) \cite{dhesi1990asymptotic}. 

The auxiliary time coordinate that enters in the MSRJD approach is relatively benign in that it does not enter substantially into the calculation. It serves only only to eliminate certain Feynman diagrams with time loops, which vanish due to causality \cite{hertz2016path}. For this reason, the MSRJD formalism is particularly well-suited to producing the diagrammatic series that form the backbone of the present work. Of course, this is somewhat a matter of opinion, and certainly the results of this work could be derived using other formalisms, although arguably in a more protracted way.

\subsection{Corresponding dynamical system}
We now show how the hermitized resolvent can be found from the response functions of an auxiliary dynamical system, in a similar fashion to Ref. \cite{baron2022eigenvalues}. Consider the following coupled set of differential equations
\begin{align}
	\dot x^1_i &= - \omega x_i^2 + \sum_{j = 1}^N a_{ij} x^2_j + h_i^1, \nonumber \\
	\dot x^2_i &= - \omega^\star x_i^1 + \sum_{j = 1}^N a^T_{ij} x^1_j + h_i^2 , \label{dynamicalsystem}
\end{align}
where we note that $x_i^a(t)$ are complex quantities. One should also note that the stability of this system about the fixed point $x_i^a = 0$ is not determined by the eigenvalues of the matrix $\underline{\underline{a}}$. The system in Eq.~(\ref{dynamicalsystem}) is introduced solely as a tool for computing the eigenvalue spectrum of $\underline{\underline{a}}$. We discuss the kinds of system for which stability is determined by the eigenvalues of $\underline{\underline{a}}$ later in Section \ref{section:stability}.

After functional differentiation with respect to the external source fields $h_i^a(t)$, one finds
\begin{align}
	\partial_t K_{ij}^{ab}(t, t') &= - \sum_{k,c} H^{ac}_{ik}  \,\, K_{kj}^{cb}(t,t')  + \delta(t- t') \delta_{ab}\delta_{ij} .
\end{align}
where $K_{ij}^{ab}(t,t') = \delta x_i^a(t)/\delta h_j^b(t')$ are the response functions. We note here that the upper indices take values $a,b \in \{1, 2\}$ and the lower indices take values $i, j~\in~\{ 1, \cdots, N\}$. Finally, assuming time-translational invariance $K_{ij}^{ab}(t, t') = K_{ij}^{ab}(t - t')$, we take the Laplace transform and the disorder average to find
\begin{align}
\overline{\hat  K_{ij}^{ab}(\eta) }\equiv	\mathcal{L}_t\left\{ \overline{K_{ij}^{ab}(t) }\right\}(\eta)   = \mathcal{H}_{ij}^{ab}, \label{correspondance}
\end{align}
where we denote the Laplace transform $\mathcal{L}_t\left\{ f(t)\right\}(\eta) = \int_0^\infty dt e^{-\eta t}f(t)$. So, we see that if we can find the disorder-averaged response functions of the system in Eq.~(\ref{dynamicalsystem}), the hermitized resolvent is immediately available, and consequently we can deduce the eigenvalue spectrum. 

Our strategy for finding these objects is as follows. We construct the MSRJD functional integral for the system in Eq.~(\ref{dynamicalsystem}). The disorder-averaged response functions can be extracted from the MSRJD functional integral using a series of Feynman diagrams. This series can then be resummed in the thermodynamic limit $N \to \infty$ to obtain $N^{-1}\sum_i\mathcal{H}_{ii}^{ab}$. We will thus have access to the spectral properties of $\underline{\underline{a}}$.

\section{Path integral formulation}\label{section:msrjd}
We now introduce the MSRJD path integral that will be the cornerstone of our subsequent analysis and will provide us with the disorder-averaged response functions of the dynamical system in Eq.~(\ref{dynamicalsystem}).

The MSRJD path integral that we consider is the generating functional (essentially a functional analogue of the Fourier transform) for the dynamical process in Eq.~(\ref{dynamicalsystem}) \cite{altlandsimons}. As such, the time-dependent correlators and response functions of the quantities $x_i^a(t)$ can be found by taking appropriate functional derivatives of this object. For a specific realization of the matrix $\underline{\underline{a}}$, the functional integral is written 
\begin{align}
	&Z[\psi, h] = \int D[x, \hat x] \exp\left[i \sum_{i,a}\int dt \psi^a_i x^{a\star}_i + \psi^{a\star}_i x^{a}_i\right]\nonumber \\
	&\times \exp \left[ i\sum_{i,a} \int dt \, \hat x^{a\star}_i \left(\dot x_i^a + \sum_{b, j}H^{ab}_{ij} x^b_j - h_j^a \right)\right] \nonumber \\
	&\times \exp \left[ i \sum_{i,a}\int dt \, \hat x^{a}_i \left(\dot x_i^{a\star} + \sum_{b, j}H^{ab}_{ij} x^{b\star}_j - h^{a\star}_j \right)\right], \label{genfunct}
\end{align}
where $D[x, \hat x]$ indicates integration with respect to all possible trajectories of the variables $\{ x_i^a(t)\}$ and their conjugate `momenta' $\{ \hat x_i^a(t)\}$. Constant factors that ensure the normalization $Z[0,h] = 1$ have been absorbed into the integral measure. Aside from the source terms containing the variables $\psi$, the integrand in Eq.~(\ref{genfunct}) is merely a complex exponential representation of Dirac delta functions, which constrain the system to follow trajectories satisfying Eqs.~(\ref{dynamicalsystem}). The reader is directed to Refs. \cite{altlandsimons, hertz2016path} for further details.

The response functions of the system can be found from this object by differentiating as follows to obtain
\begin{align}
	K_{ij}^{ab}(t, t') &= \frac{\delta \langle x_i^a(t)\rangle}{\delta h_j^{b}(t')} \bigg\vert_{\psi = h = 0} = -i\frac{\delta^2 Z}{\delta \psi^{a\star}_{i}(t) \delta h^{b}_j(t')} \bigg\vert_{\psi = h = 0} \nonumber \\
	&= - i  \langle x_i^a(t) \hat x^{\star b}_j(t') \rangle \big\vert_{\psi = h = 0} , \label{responsefunctions}
\end{align}
where here the angular brackets indicate an average with respect to the dynamical process, i.e.
\begin{align}
	&\langle \mathcal{O} \rangle \big\vert_{\psi = h =0}  = \int D[x, \hat x] \,\,\mathcal{O} \nonumber \\
	&\times \exp \left[ i\sum_{i,a} \int dt \, \hat x^{a\star}_i \left(\dot x_i^a + \sum_{b, j}H^{ab}_{ij} x^b_j \right)\right] \nonumber \\
	&\times \exp \left[ i \sum_{i,a}\int dt \, \hat x^{a}_i \left(\dot x_i^{a\star} + \sum_{b, j}H^{ab}_{ij} x^{b\star}_j  \right)\right]. \label{pathaverage}
\end{align}
From now on, it is to be understood that all averages $\langle \cdot\rangle$ are to be evaluated at $\psi = h = 0$. One notes that the quenched random variables $\{a_{ij}\}$ enter the expression in Eq.~(\ref{pathaverage}) linearly in the exponent. Thus, by making the link between the resolvent and the response functions in Eq.~(\ref{correspondance}), and by writing the response functions as in Eq.~(\ref{responsefunctions}), we drastically simplify the task of taking the average over realisations of the random matrix in comparison to the original expression for the hermitized resolvent in Eq.~(\ref{hermres}).

\section{The dense limit: Rainbow diagrams and recovering the elliptic law}\label{section:ellipticlawdiagrams}
Now that we have introduced the path integral framework, we show how a series for the disorder-averaged response functions can be constructed in terms of Feynman diagrams. This is done in the context of a simple example. Namely, we recover the elliptic law for dense matrices. A more detailed introduction to the diagrammatic formalism is given in the Supplemental Material (SM) Section S3.

Although such rainbow diagrams have been obtained previously using other methods  \cite{brezin1994correlation, kuczala2016eigenvalue}, one notes that the `ribbon' diagrams that appear later are unique to the sparse (or non-Gaussian) random matrix case. 

\subsection{Disorder-averaged generating functional and series expansion of the response functions}
Let us take the simple example where the matrix $\underline{\underline{a}}$ is fully populated (i.e. all elements are non-zero). We suppose its elements have statistics
\begin{align}
	\overline{ a_{ij} } = 0, & \,\,\,\,\, \overline{ a_{ij}^2} = \sigma^2/N, \nonumber \\
	\overline{ a_{ij}a_{ji}}&= \Gamma \sigma^2/N.  \label{ellipticstats}
\end{align}
If the higher-order moments decay more quickly than $1/N$, they do not contribute to the response functions in the thermodynamic limit $N\to \infty$, and consequently they do not affect the eigenvalue spectrum. This means that one could treat $a_{ij}$ as correlated Gaussian random variables without loss of generality. How one can see this so-called universality principle \cite{taovukrishnapur2010, bai2010spectral, nguyen2015elliptic} using the present approach is discussed further in SM Section S2. 

Taking the disorder average of the generating functional using the statistics in Eq.~(\ref{ellipticstats}), one arrives at 
\begin{align}
	\overline{Z[\psi, h]} &= \int D[x, \hat x] \exp\left[i \sum_{i,a}\int dt \psi^a_i x^{a\star}_i + \psi^{a\star}_i x^{a}_i\right]e^S, \label{disorderaveragedgenfunct}
\end{align}
where the action $S= S_0 + S_\mathrm{int}$ is the sum of two contributions: a `bare' action $S_0$ and an interaction term $S_\mathrm{int}$
\begin{align}
	S_0 &= i \left\{ \sum_{i,a} \int dt \, \hat x^{a\star}_i \left(\dot x_i^a + \sum_{b, j}(\mathinv)^{ab}_{ij} x^b_j  \right) + \mathrm{c.c.} \right\}, \nonumber \\
	S_\mathrm{int} &= \frac{(-i)^2}{2!\,2} \, {}^2C_1 \int dt dt' \overline{\hat x^{a\star}_i \matj^{ab} x^{b}_j  \hat x^{a'\star}_{j} (\matj^\dagger)^{a'b'} x^{b'}_{i}}  + \cdots , \label{ellipticaction}
\end{align}
where the sums over repeated indices in $S_\mathrm{int}$ are implied, and we have introduced the $2\times 2$ \textit{non}-Hermitian matrix 
\begin{align}
	\matj = 
	\begin{bmatrix}
		0 & a_{12} \\
		a_{21} & 0
	\end{bmatrix}, \label{jdef}
\end{align}
where $a_{12}$ and $a_{21}$ obey the statistics given in Eq.~(\ref{ellipticstats}). We note that we have omitted terms in $S_\mathrm{int}$ that do not contribute to the calculation of the response functions (see SM Section S3 for a justification of this).

We now define the following average with respect to both the dynamics and disorder
\begin{align}
	\langle  \cdots \rangle_S = \int D[x,\widehat x] [\cdots] e^S . \label{averagedef}
\end{align}
Because we can write the response functions as in Eq.~(\ref{responsefunctions}), we can calculate the disorder-averaged response functions using a series expansion. Such an expansion is arrived at by noticing (following Ref. \cite{hertz2016path})
\begin{align}
	\overline{K_{ij}^{ab}(t, t')} &=-i\langle x^a_i(t) \hat x^{b\star}_j(t')  \rangle_S \nonumber \\
	&= -i\langle x^a_i(t) \hat x^{b\star}_j(t')  e^{S_\mathrm{int}} \rangle_0 \nonumber \\
	&= -i\sum_r \frac{1}{r!} \langle x^a_i(t) \hat x^{b\star}_j(t')  (S_\mathrm{int})^r\rangle_0 , \label{interactionactionexpansion}
\end{align} 
where $\langle  \cdots \rangle_0$ indicates an average with respect to the bare action only. From here on, we refer to quantities averaged with respect to $S_0$ as `bare' and those that are averaged with respect to $S$ as `dressed' \cite{hertz2016path}.

Since the bare action is quadratic in the dynamic variables $\{x_i(t),\hat x_j(t)\}$, Wick's theorem holds for the averages $\langle \cdots \rangle_0$ \cite{hertz2016path}. We therefore obtain a series that we can evaluate entirely in terms of quantities that are calculable from the non-interacting theory. The various Wick pairings that arise in this series are then kept track of using Feynman diagrams, as we show below.

The bare response functions can be calculated easily from the dynamic system without interactions, which obeys
\begin{align}
	\dot x^a_i =  -\sum_{j,b}(\mathinv)^{ab}_{ij}x^{b}_j + h_j^a .
\end{align}
One thus obtains for the bare resolvent
\begin{align}
	\lim_{\eta \to 0} \overline{(\hat K_0)_{ij}^{ab}} &= \lim_{\eta\to 0}\mathcal{L}_t\left[ -i\langle x^a_i(t) \hat x^{b\star}_j(0)  \rangle_0 \right]=  (\mathcal{H}_0)_{ij}^{ab} .
\end{align}

\begin{figure}[H]
	\centering 
	\includegraphics[scale = 0.38]{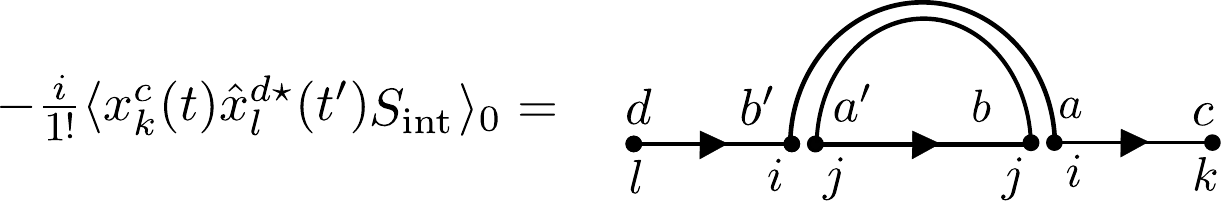}
	\captionsetup{justification=raggedright,singlelinecheck=false}
	\caption{Example rainbow diagram.}\label{fig:examplerainbow}
\end{figure}

\subsection{Constructing the Feynman diagrams}

We now discuss how the diagrammatic formalism can be constructed. Subsequently, we will see how this formalism can be used to efficiently identify and sum an infinite series of non-vanishing terms (for $N \to \infty$), and thus obtain an expression for the dressed response functions. Ultimately, we recover the well-known elliptic law for dense random matrices \cite{sommers, girko1986elliptic}.

Let us take for example the $r=1$ term in the expansion in Eq.~(\ref{interactionactionexpansion}). We have (neglecting vanishing Wick pairings)
\begin{widetext}

\begin{align}
	-\frac{i}{1! }\langle x^c_k(t) \hat x^{d\star}_l(t') S_{\mathrm{int}}\rangle_0
	=  \int_{T>T'} dT dT'(-i)^3 \left\langle  x^c_k(t)   \hat x^{a\star}_{i}(T) \right\rangle_0 \overline{ \matj^{ab} \left\langle x^{b}_{j}(T)    \, \hat x^{a'\star}_{j}(T') \right\rangle_0  (\matj^\dagger)^{a'b'}}\left\langle   x^{b'}_{i} (T') \hat x^{d\star}_l(t') \right\rangle_0  , \label{firstordersurviving}
\end{align}

\end{widetext}
where sums over repeated indices are implied. 

The surviving term in Eq.~(\ref{firstordersurviving}) can be represented diagrammatically as in Fig. \ref{fig:examplerainbow}, which should be interpreted as follows (see also Ref. \cite{baron2022eigenvalues}): A dot on the left-hand end of a directed edge represents an $\hat x$-type variable, and a dot on the right-hand end of a directed edge represents an $x$-type variable. Pairs of dots positioned together have the same time coordinate, and each pair of dots carries a matrix factor $\matj$ (on the right-hand side of an arc) or $\matj^\dagger$ (on the left-hand side). Double arcs connect pairs of $\matj$ and $\matj^\dagger$ matrices that are disorder-averaged together. The $x$- and $\hat x$-type variables connected by a single arc are also constrained to have the same lower indices. Points connected by horizontal edges are Wick-paired together (averaged with respect to the bare action), and thus evaluate to the bare response function. Because $(K_0)^{ab}_{ij}(t,t') = 0$ for $t<t'$, the time coordinate of an $x$-type variable must always be greater than that of the $\hat x$-type variable with which it is Wick-paired, hence the directionality of the edges. Finally, all internal (i.e. repeated) times and indices are integrated/summed over.

These diagrammatic representations are known as `rainbow' diagrams, and they have been obtained previously by other methods \cite{JANIK1997603,  feinberg1997non, brezin1994correlation}. The diagrammatic representation allows us to identify  easily the terms that survive in the limit $N\to \infty$. One finds that the only surviving rainbow diagrams are `planar' (i.e. with no crossing arcs) \cite{HOOFT1974461, brezin1978planar}. This is a consequence of the fact that the bare resolvent matrix $(\mathcal{H}_0)_{ij}^{ab}$ is diagonal in the indices $i$ and $j$. The reader is directed to SM Section S3 for further details of the diagrammatic representation, more examples of planar rainbow diagrams, and a more detailed explanation as to why planar diagrams are the terms that survive.

\begin{figure}[t]
	\centering 
	\includegraphics[scale = 0.20]{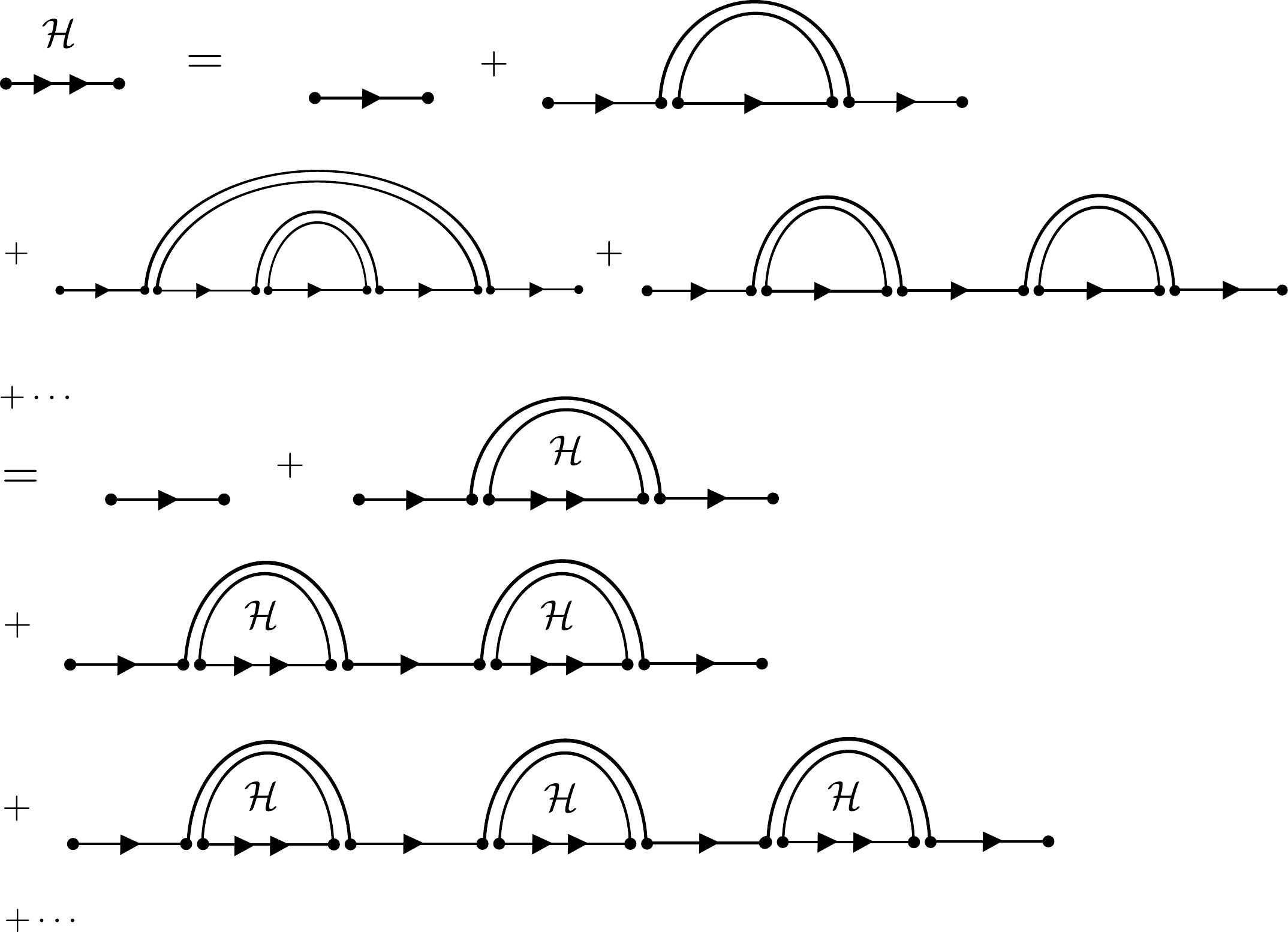}
	\captionsetup{justification=raggedright,singlelinecheck=false}
	\caption{Summing the full series of rainbow diagrams.} \label{fig:fullrainbowseries}
\end{figure}

The full series in Eq.~(\ref{interactionactionexpansion}) can thus be evaluated in the thermodynamic limit by evaluating the sum of all possible planar rainbow diagrams, which is illustrated in Fig. \ref{fig:fullrainbowseries}. We follow an approach similar to that discussed in Refs. \cite{kuczala2019dynamics, brezin1994correlation} in order to sum this diagrammatic series. Ultimately, one finds that the two series in Fig. \ref{fig:fullrainbowseries} are equivalent. One notes that we have employed an additional diagrammatic convention to encode the distributivity of an arc over a sum of diagrams (see SM Section S4), and that the dressed response function is denoted diagrammatically by an edge with a double arrow. The argument as to why the full series of planar diagrams can be resummed in this way is summarized in SM Section S4.

We recognize the second series of diagrams in Fig. \ref{fig:fullrainbowseries} as a geometric series. This series can be summed to yield the following compact expression for the hermitized resolvent 
\begin{align}
	\mathcal{H}
	&= \left[\mathcal{H}_0^{-1} - N\overline{\matj \mathcal{H} \matj^\dagger } \right]^{-1} , \label{selfconsistentseries}
\end{align}
where we have used the $2 \times 2$ matrices defined in Eqs.~(\ref{def2by2}) and (\ref{jdef}). We have thus succeeded in finding a self-consistent expression for the hermitized resolvent, which can be solved to yield $C$, and thus the eigenvalue density.

\subsection{Recovering the elliptic law}\label{section:ellipticcalculation} 
Let us now solve Eq.~(\ref{selfconsistentseries}) for the quantity $\mathcal{H}^{21} \equiv C$. After performing the disorder average of the object $N\overline{\matj \mathcal{H} \matj^\dagger }$ using the statistics in Eq.~(\ref{ellipticstats}), one finds that there are two possible solutions of Eq.~(\ref{selfconsistentseries}). First, we have $A = D= 0$, for which the corresponding solution for $C$ is an analytic function of $\omega$,
\begin{align}
	C = \frac{1}{\omega - \Gamma \sigma^2 C} . \label{analytic}
\end{align}
The other solution is $AD - BC = -1/\sigma^2$, for which the corresponding expression for $C$ is
\begin{align}
	C &= \frac{\omega^\star - \Gamma \omega}{(1-\Gamma^2)\sigma^2} . \label{nonanalytic}
\end{align}

The latter, non-analytic solution corresponds to the bulk region where there is a non-zero eigenvalue density, given by [using Eq.~(\ref{densityfromres})]
\begin{align}
	\rho(\omega) = \frac{1}{\pi\sigma^2(1-\Gamma^2)} .\label{ellipsedensity}
\end{align}
We note that in the region of the complex plane where Eq.~(\ref{analytic}) is the correct solution for the resolvent, the eigenvalue density must be nil. The boundary of the bulk region to which the eigenvalues are confined is thus given by the set of points $\omega$ that simultaneously satisfy Eqs.~(\ref{analytic}) and (\ref{nonanalytic}). By equating these two expressions for $C$, we recover the elliptic law (setting $\omega~=~x + i y$)
\begin{align}
	\left( \frac{x}{1+\Gamma}\right)^2 + \left( \frac{y}{1-\Gamma}\right)^2 &= \sigma^2 . \label{ellipse}
\end{align}
Finally, integrating Eq.~(\ref{ellipsedensity}) with respect to $y$ between the limits of the ellipse defined by Eq.~(\ref{ellipse}), one recovers the Wigner semi-circle law \cite{Wigner1958} for $\rho_x(x) = \int dy \rho(x,y)$ (generalized for asymmetric matrices) 
\begin{align}
	\rho_x(x) = \frac{2}{\pi \sigma^2 (1+\Gamma)^2}\sqrt{\sigma^2 (1+\Gamma)^2 - x^2}. \label{wignersemicircle}
\end{align}
Although what we have done may seem like a somewhat convoluted route to recovering these well-known results, the advantage of this method lies in its generalizability, as will be demonstrated in the rest of this text.

\section{The sparse correction to the elliptic and Wigner semi-circle laws} \label{section:sparsefirstorder}
We now turn our attention to sparse random matrices, where the mean number of non-zero elements per row, denoted by $p$ here, does not scale with $N$ (i.e. remains finite) as $N \to \infty$ \cite{metz2019spectral}. 

We consider the regime where $p$ is large enough that we can perform a systematic expansion in the inverse connectivity $1/p$. An expansion of this type was explored by Rodgers and Bray \cite{rodgers1988density} using the replica formalism, and has also been explored using the supersymmetric formalism in Ref. \cite{akara2022random}. However, both of these works considered only symmetric matrices, and the results obtained therein have yet to be extended to the asymmetric case. Such an extension is facilitated by the method presented in Sections \ref{section:msrjd} and \ref{section:ellipticlawdiagrams}, which is particularly amenable to perturbative treatments and can also handle non-Hermitian ensembles. 

Ultimately, we obtain simple closed-form expressions for the boundary of the support of the eigenvalue spectrum. We also derive similarly compact expressions for the eigenvalue density inside the bulk region and the locations of any outlier eigenvalues that arise due to a non-zero mean \cite{orourke, benaych2011eigenvalues, edwardsjones}. The expressions for the eigenvalue density that we obtain are universal in the following sense. Matrices drawn from an ensemble with a given set of moments [see Eq.~(\ref{loworderstats}) below] will have an expected eigenvalue density that agrees with the results that we derive, up to corrections of order $O(1/p^2)$, in the limit $N \to \infty$. We are thus able to see directly how the standard elliptic law is modified by virtue of the matrix being sparse, how the higher-order statistics of the non-zero elements affect this correction, and what this means for the stability of a sparse dynamical system.

\subsection{Random matrix ensemble and corresponding action}
For the purposes of this work, we consider sparse random matrices $\underline{\underline{a}}$ whose non-zero elements represent a weighted Erd\H{o}s-R\'enyi graph (see Ref. \cite{baron2022networks} for a perturbative treatment of more complex network structure). In other words, a link between two nodes $i$ and $j$ exists with probability $p/N$. If a link between two nodes exists, we draw $a_{ij}$ and $a_{ji}$ from a joint distribution $\pi(a_{ij}, a_{ji})$. All other entries of $\underline{\underline{a}}$ are set to zero. The joint distribution of the matrix elements $a_{ij}$ and $a_{ji}$ is therefore given by 
\begin{align}
	P(a_{ij}, a_{ji}) = \left(1 - \frac{p}{N}\right) \delta(a_{ij})\delta(a_{ji}) + \frac{p}{N} \pi(a_{ij}, a_{ji}) \label{sparsedef}
\end{align}
We see readily that $p$ is the mean number of connections per node on the network (i.e. the average number of non-zero random matrix elements per row/column). We denote the lower-order statistics of the distribution~$\pi(a_{ij}, a_{ji})$~by
\begin{align}
	\left\langle a_{ij} \right\rangle_\pi &= \frac{\mu}{p}, \nonumber \\
	\left\langle (a_{ij}- \mu/p)^2\right\rangle_\pi &= \frac{\sigma^2}{p}, \nonumber \\
	\left\langle (a_{ij}-\mu/p)(a_{ji}- \mu/p)\right\rangle_\pi &= \frac{\Gamma \sigma^2}{p},\nonumber \\
	\langle (a_{ij}-\mu/p)^4 \rangle_\pi &= \frac{\Gamma^{(1)}_4 \sigma^4}{p^2}, \nonumber \\
	\langle (a_{ij}-\mu/p)^3( a_{ji}-\mu/p) \rangle_\pi &= \frac{\Gamma^{(2)}_4 \sigma^4}{p^2}, \nonumber \\
	\langle (a_{ij}-\mu/p)^2 (a_{ji}-\mu/p)^2 \rangle_\pi &= \frac{\Gamma^{(3)}_4 \sigma^4}{p^2},  \label{loworderstats}
\end{align}
where $\langle \cdot \rangle_\pi$ indicates an average over the distribution $\pi$ (to be contrasted with $\overline{\cdots}$, which denotes an average over realizations of the network \textit{and} the weights of links). Scaling the variance of the interaction coefficients with $p$ as in Eq.~(\ref{loworderstats}) permits one to take the dense limit $p\to N$ in a sensible fashion. It also allows one to perform the perturbative expansion transparently. We note that one could also do the expansion without this scaling (see Ref. \cite{bray1982eigenvalue}). The scaling can easily be undone by substituting $\sigma^2 \to p \sigma^2$ and $\mu \to p \mu$. We also assume that higher order statistics scale with higher powers of $1/p$ such that $\langle a_{ij}^6 \rangle_\pi \sim 1/p^3$, and so on.

For the present, we consider the case $\mu = 0$, but we will generalize to the $\mu \neq 0$ case in Section \ref{section:nonzeromu}. In Section \ref{section:nongaussian}, we also discuss how one can include the possibility of the null entries of the matrix having fluctuations of order $\sim 1/\sqrt{N}$ about zero, as they would in a dense matrix.

We now evaluate the disorder averaged generating functional in Eq.~(\ref{disorderaveragedgenfunct}) and obtain the following contributions to the action $S = \sum_{r=0}^\infty S_r$, assuming $p/N\ll 1$ [c.f. Eq.~(\ref{ellipticaction})] 
\begin{align}
	S_0 &= i \left\{ \sum_{i,a} \int dt \, \hat x^{a\star}_i \left(\dot x_i^a + \sum_{b, j}(\mathinv)^{ab}_{ij} x^b_j  \right)  \right\}, \nonumber \\
	S_1 &= \frac{p}{2N} \frac{(-i)^2}{2!}\, {}^2C_1  \int dt_1 dt_2 \left\langle\hat x^{a\star}_i \matj^{ab} x^{b}_j  \hat x^{a'\star}_{j} (\matj^\dagger)^{a'b'} x^{b'}_{i} \right\rangle_\pi \nonumber \\
	S_2 &=\frac{p}{2N} \frac{(-i)^4}{4!} \, {}^4C_2 \int dt_1 \cdots dt_4 \Big\langle\hat x^{a_1\star}_{i} \matj^{a_1b_1} x^{b_1}_{j}  \nonumber \\
	& \times\hat x^{a_2\star}_{j} (\matj^\dagger)^{a_2b_2} x^{b_2}_{i} \hat x^{a_3\star}_{i} \matj^{a_3b_3} x^{b_3}_{j}\hat x^{a_4\star}_{j} (\matj^\dagger)^{a_4b_4} x^{b_4}_{i}\Big\rangle_\pi , \nonumber \\
	S_3 &= \cdots , \label{actionseriesterms}
\end{align}
where sums over the repeated indices in $S_1$ and $S_2$ are implied. Here, the matrix $\matj$ [defined in Eq.~(\ref{jdef})] now has elements $a_{12}$ and $a_{21}$ that are sampled from $\pi(a_{12}, a_{21})$. The first two terms in the action in Eq.~(\ref{actionseriesterms}) correspond to the bare action and the elliptic interaction term in Eq.~(\ref{ellipticaction}). That is, the higher order terms $S_2$ and $S_3$ constitute small corrections to the elliptic law when $p$ is large. If we ignored these terms, we would recover the result of the dense case. 

The response functions of the system are now found by evaluating the following series diagrammatically~[c.f.~Eq.~(\ref{interactionactionexpansion})]
\begin{align}
	-i\langle x^a_i(t) \hat x^{b\star}_j(t')  \rangle_S& = \sum_{r_1, r_2, \dots} \bigg[ \frac{-i}{r_1!r_2! \cdots} \cdot \nonumber \\
	&\times\left\langle x^a_i(t) \hat x^{b\star}_j(t')  [S_1^{r_1}S_2^{r_2} \cdots]\right\rangle_0 \bigg]. \label{interactionactionexpansion2}
\end{align} 
Our approach will be to truncate this series and find the response functions to the desired order in $1/p$. From these response functions, we will then obtain expressions for the eigenvalue spectrum that are accurate to the same order.

\subsection{Expansion in the inverse connectivity: Ribbon diagrams}\label{section:firstorderdiagrams}
We now proceed as in Section \ref{section:ellipticlawdiagrams} by constructing a series of non-vanishing diagrams that we can resum in the thermodynamic limit. As before, the diagrams allow us to keep track of the Wick pairings that arise from the averages in Eq.~(\ref{interactionactionexpansion2}). The diagrams also permit one to spot the self-similarity of the series and deduce a self-consistent expression for the response functions.

Because of the higher-order terms in the action in Eq.~(\ref{actionseriesterms}), we have to take new types of diagram into account, other than just the usual `rainbow' diagrams \cite{JANIK1997603}. Due to their shape (see below), we refer to these generalizations as `ribbon' diagrams.

The strategy is as follows. We separate the series into diagrams that contribute to different orders in $1/p$. We then sum each of these sub-series separately. Afterwards, we combine these sub-series to deduce a self-consistent expression for the hermitized resolvent to a particular accuracy in $1/p$.

\begin{figure}[t]
	\centering 
	\includegraphics[scale = 0.165]{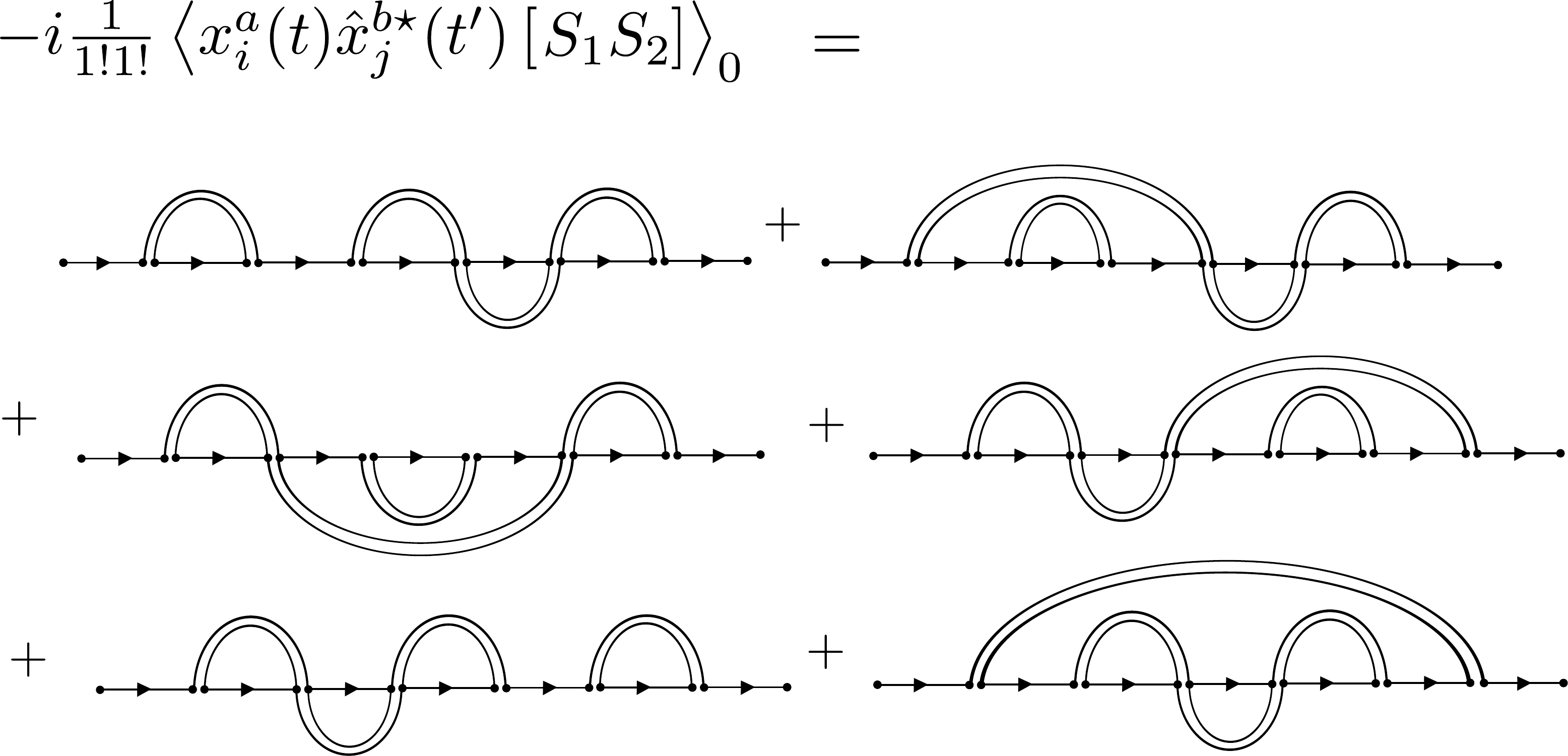}
	\captionsetup{justification=raggedright,singlelinecheck=false}
	\caption{Example ribbon diagrams.}\label{fig:examplefirstorder}
\end{figure}

Suppose that we deconstruct the hermitized resolvent as follows
\begin{align}
	\underline{\underline{\mathh}} = \underline{\underline{\mathh}}_1 + \frac{1}{p}\underline{\underline{\mathh}}_2 +  \frac{1}{p^2}\underline{\underline{\mathh}}_3 + \cdots . \label{hseriesdef}
\end{align}
One can use Eq.~(\ref{interactionactionexpansion2}) to find diagrammatic series expansions for each of $\mathh_1$, $\mathh_2$, ..., in a similar way to Section \ref{section:ellipticlawdiagrams}. For example, we have the contribution to $\mathcal{H}_2/p$ that is depicted in Fig. \ref{fig:examplefirstorder}. We have not included diagrams that vanish in the thermodynamic limit in this figure; once again, only planar diagrams survive. 

We see that the $1/p$ sparse correction to the action $S_2$ [see Eq.~(\ref{actionseriesterms})] has given rise to a new kind of diagram. We recall that arcs are used to connect matrix factors that are disorder-averaged together, and they also connect $x$- and $\hat x$-type variables with the same lower index. The action $S_2$ [see Eq.~(\ref{actionseriesterms})] contains two matrix factors of $\matj$ and two of $\matj^\dagger$ that are averaged together. This manifests diagrammatically as a `ribbon' of three concatenated arcs. More specifically, a ribbon with 3 concatenated arcs carries 4 simultaneously averaged $\matj$-matrix factors, together with a factor of $p/N$, and is consequently proportional to $1/p$. A ribbon with 5 concatenated arcs would carry 6 simultaneously averaged $\matj$-matrix factors, also with a factor of $p/N$, and would therefore be proportional to $1/p^2$, etc.

Let us now obtain the first-order correction in $1/p$ to the elliptic law. To zeroth order in $1/p$, we obtain the same self-similar series of rainbow diagrams as in Section \ref{section:ellipticlawdiagrams} and we have [similarly to Eq.~(\ref{selfconsistentseries})] 
\begin{align}
	\mathcal{H}_{1}&= \left[\mathcal{H}_0^{-1} - p \langle \matj \mathcal{H}_1 \matj^\dagger \rangle_\pi \right]^{-1},  \label{zerothorder}
\end{align}
where once again we define $2\times2$ matrices analogous to Eqs.~(\ref{jdef}) and (\ref{def2by2}), where now the entries $a_{12}$ and $a_{21}$ of $\matj$ are drawn from $\pi(a_{12}, a_{21})$.

\begin{figure}[t]
	\centering 
	\includegraphics[scale = 0.22]{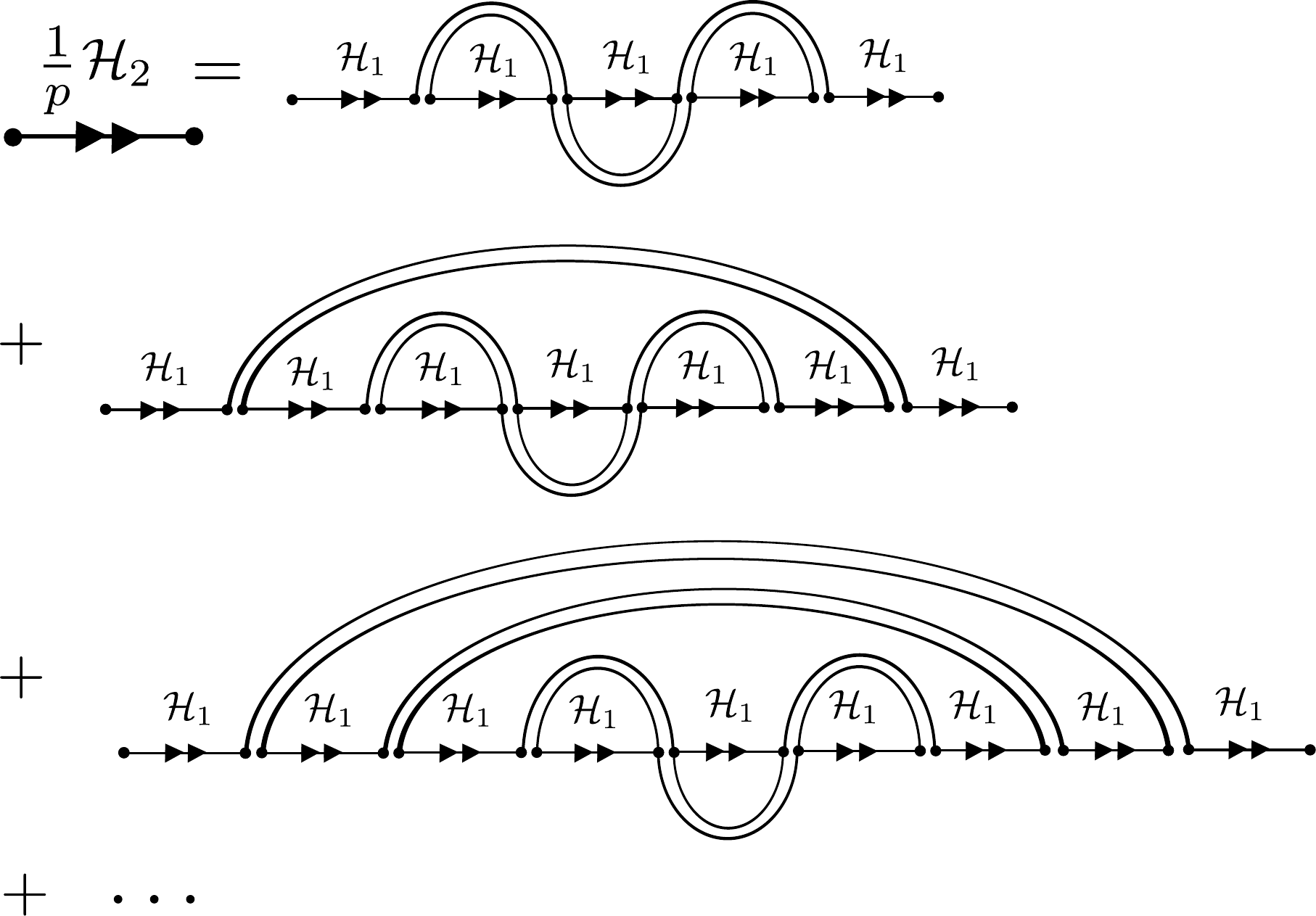}
	\captionsetup{justification=raggedright,singlelinecheck=false}
	\caption{The complete series of $O(1/p)$ ribbon diagrams.}\label{fig:fullfirstorderseries}
\end{figure}
One also obtains a series of ribbon diagrams for the sparse correction, which is given in Fig. \ref{fig:fullfirstorderseries}. In this series, we note that we have summed several sub-series to obtain factors of the zeroth-order hermitized resolvent $\mathcal{H}_1$ (denoted diagrammatically with a double arrowed horizontal line). This is similar to the way that factors of $\mathh$ appeared in the elliptic law calculation in Section \ref{section:ellipticlawdiagrams}. 

\begin{figure*}[t]
	\centering 
	\includegraphics[scale = 0.37]{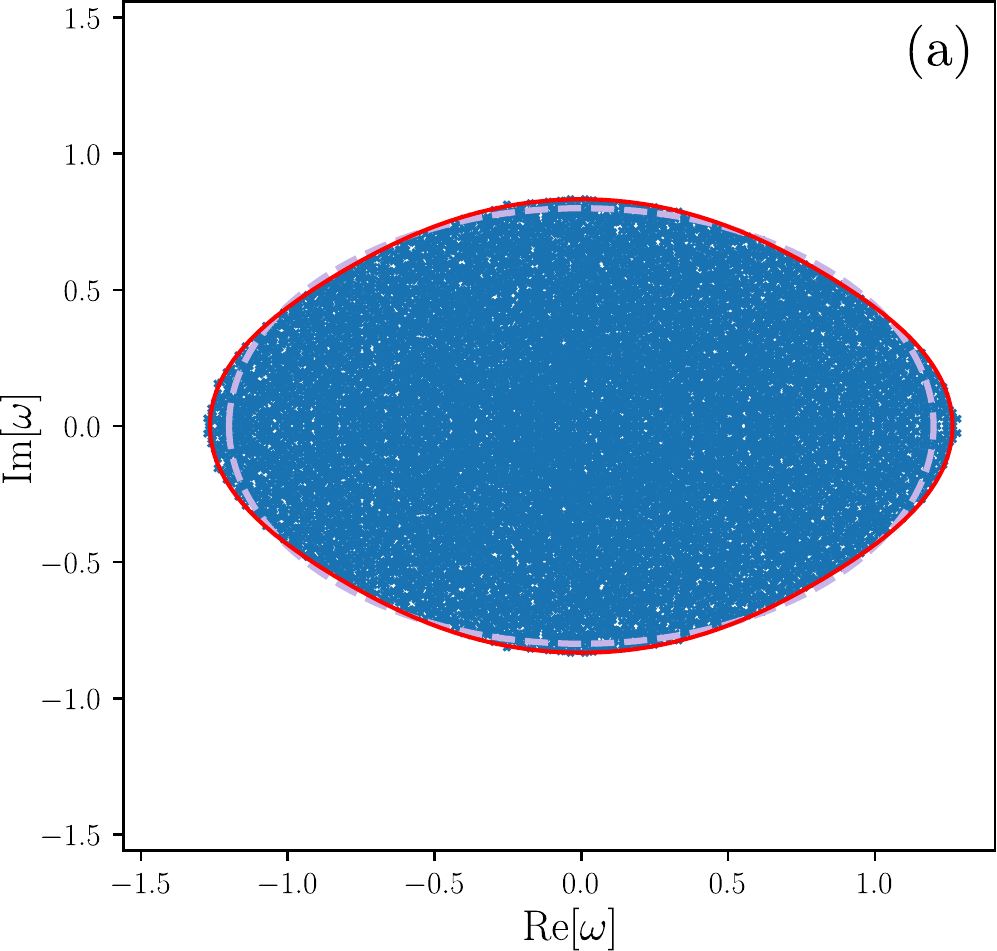} \hspace{5mm}
	\includegraphics[scale = 0.45]{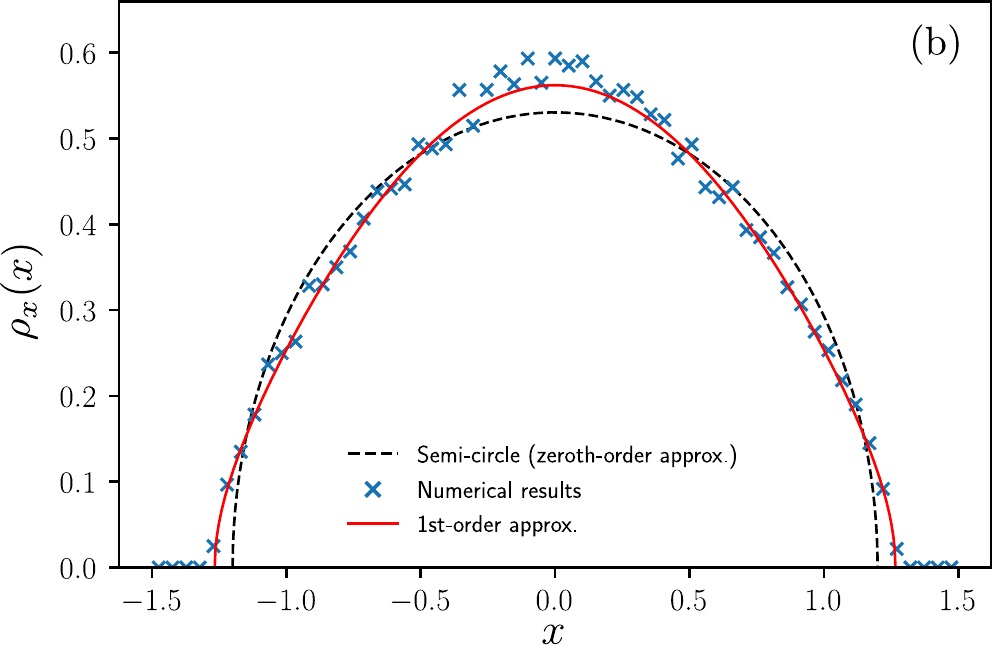}
	\captionsetup{justification=raggedright,singlelinecheck=false}
	\caption{Modified elliptic and semi-circular laws. Parameters are $p =30$, $N = 12000$, $\sigma =1$, $\Gamma = 0.2$, $\mu =0$ and a Gaussian distribution is used for the non-zero elements [see Eq.~(\ref{gaussstats})]. Panel (a): Boundary of the eigenvalue spectrum. Blue crosses are the results of numerical diagonalization and the red line is given by Eqs.~(\ref{ellipsemodcart}) using the coefficients given in Eq.~(\ref{xcmuzero}). The dashed line is the na\"ive elliptic law obtained in the limit $p \to \infty$. Panel (b): Integrated eigenvalue density as a function of the real part. The red line is given by Eq.~(\ref{generalizedwigner}). }\label{fig:modifiedellipse}
\end{figure*}

Evaluating the series of diagrams in Fig. \ref{fig:fullfirstorderseries}, we obtain
\begin{align}
	\frac{1}{p}\mathcal{H}_{2}= \langle \mathcal{H}_1 \matj \mathcal{H}_{2} \matj^\dagger \mathcal{H}_1 \rangle_\pi + p \langle \mathcal{H}_1 \matj \mathcal{H}_1 \matj^\dagger \mathcal{H}_1 \matj\mathcal{H}_1 \matj^\dagger \mathcal{H}_1 \rangle_\pi . \label{firstorder}
\end{align}

Combining Eq.~(\ref{zerothorder}) and (\ref{firstorder}) using Eq.~(\ref{hseriesdef}), we thus deduce the following self-consistent expression for $\mathh$, which is accurate up to first order in $1/p$
\begin{align}
	\mathh &\approx \mathh_1 + \frac{1}{p}\mathh_2 \nonumber \\
	&\approx \left[\mathh_0^{-1} - p\langle \matj \mathh_1 \matj^\dagger \rangle_\pi - \langle \matj \mathh_2 \matj^\dagger \rangle_\pi  \right]^{-1} \nonumber \\
	&+ p \langle \mathcal{H}_1 \matj \mathcal{H}_1 \matj^\dagger \mathcal{H}_1 \matj \mathcal{H}_1 \matj^\dagger \mathcal{H}_1 \rangle_\pi \nonumber \\
	&\approx   \left[\mathh_0^{-1} - p\langle \matj \mathh \matj^\dagger \rangle_\pi - p \langle \matj \mathcal{H} \matj^\dagger \mathcal{H} \matj \mathcal{H} \matj^\dagger \rangle_\pi\right]^{-1} .\label{generalfirstorder}
\end{align}
This is the self-consistent expression for the hermitized resolvent that we desired, analogous to Eq.~(\ref{selfconsistentseries}) for the dense case.

\begin{figure*}[t]
	\centering 
	\includegraphics[scale = 0.49]{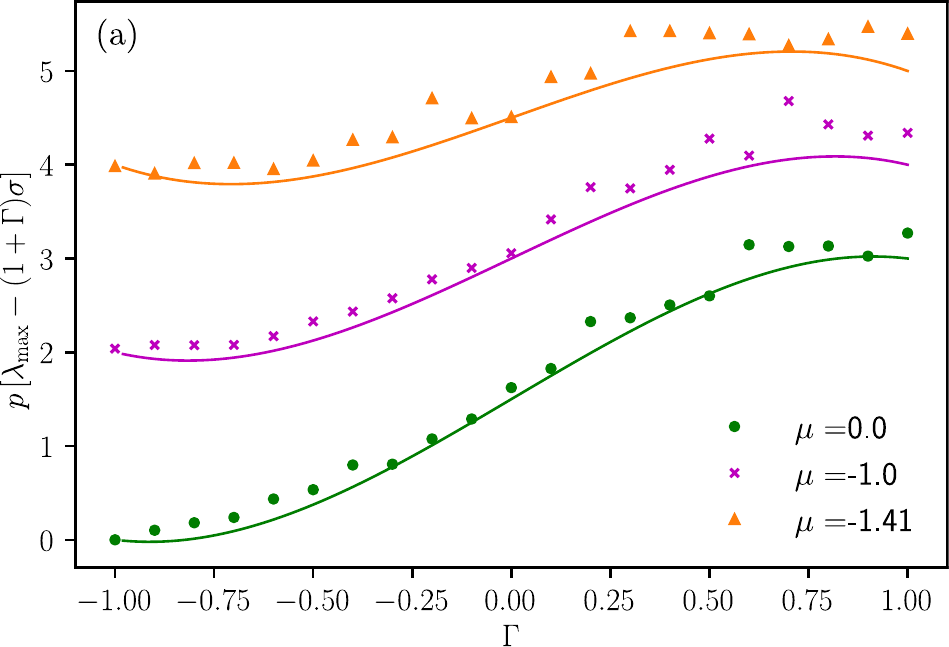}\hspace{3mm}
	\includegraphics[scale = 0.49]{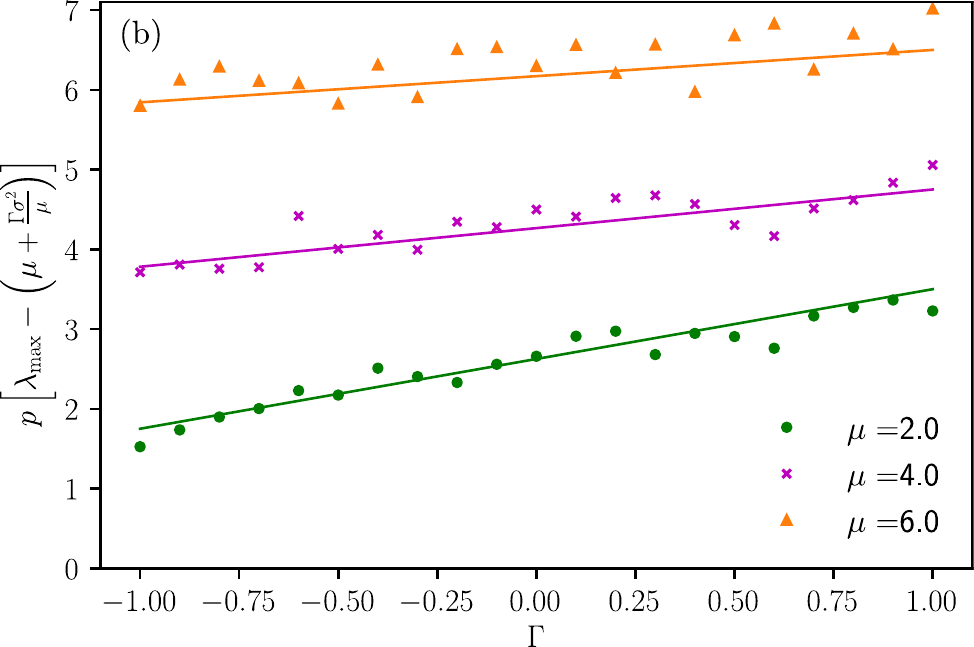}
	\captionsetup{justification=raggedright,singlelinecheck=false}
	\caption{Panel (a): Sparse correction to the rightmost edge of the bulk region of the spectrum. The solid lines are given by Eq.~(\ref{bulkcorrectiongaussian}). Panel (b): Sparse correction to the outlier eigenvalue. The solid lines are given by Eq.~(\ref{outliercorrectiongaussian}). In both panels, the markers are the results of numerical diagonalization for matrices with $p = 50$ and $N = 10000$ averaged over 10 realizations with $\sigma = 1$. }\label{fig:leadingeigenvaluecorrection}
\end{figure*}

\subsection{Modified elliptic and semi-circle laws}
We are now in a position to solve Eq.~(\ref{generalfirstorder}) for the trace of the resolvent $C \equiv \mathcal{H}^{21}$, and thus deduce the properties of the eigenvalue spectrum in a similar way to Section \ref{section:ellipticcalculation}. 

As is demonstrated in SM Section S5 A, Eq.~(\ref{generalfirstorder}) can be solved to yield two solutions for $C$: one that is non-analytic in $\omega$ (valid for the region of the complex plane with non-zero eigenvalue density) and one that is analytic (corresponding to the region with no eigenvalues). The set of values of $\omega$ at which these two expressions coincide corresponds to the boundary of the support of the eigenvalue spectrum. 

We thus can show that [see SM Section S5 B] the boundary of the eigenvalue spectrum is given by the following modified ellipse 
\begin{align}
	\frac{x^2}{x_c^2}+ \frac{y^2}{y_c^2} = 1 - \frac{16}{p} \frac{(\Gamma_4^{(3)}- \Gamma \Gamma_4^{(2)})}{(1-\Gamma^2)} \frac{x^2}{x_c^2} \frac{y^2}{y_c^2}\label{ellipsemodcart}
\end{align}
where we have identified the rightmost and topmost eigenvalues of the modified ellipse (respectively)
\begin{align}
	x_c &= \sigma(1+\Gamma)  + \frac{\sigma}{2p} \left[ (3-\Gamma) \Gamma_4^{(3)} + 2 (1-\Gamma) \Gamma^{(2)}_4\right] \nonumber \\
	y_c &= \sigma(1-\Gamma) + \frac{\sigma}{2p} \left[(3 + \Gamma) \Gamma_4^{(3)} - 2 (1+\Gamma) \Gamma_4^{(2)} \right]. \label{xcmuzero}
\end{align}
The expressions in Eqs.~(\ref{ellipsemodcart}) and (\ref{xcmuzero}) are compared with numerical results in Figs. \ref{fig:modifiedellipse}a and \ref{fig:leadingeigenvaluecorrection}a respectively.

Using Eq.~(\ref{densityfromres}), we can also find the density of eigenvalues within the support [see SM Section S5 C]. In SM Eq.~(S41), we see that the eigenvalue density is no longer uniform within the bulk region. We do not reproduce the expression here, which is lengthy but elementary.

Finally, we can also generalize the Wigner semi-circle law. By integrating the eigenvalue density with respect to $y$ between the limits given by Eq.~(\ref{ellipsemodcart}), one finds that for $\rho_x(x) = \int dy \rho(x,y)$ we obtain the surprisingly succinct expression [see SM Section S5 D]
\begin{align}
	\rho_x(x) &=  \frac{2}{\pi x_c^2} \left\{ 1 + \frac{\beta}{p} \frac{\sigma}{x_c} \left[1 - 4\frac{x^2}{x_c^2} \right]\right\} \sqrt{x_c^2 - x^2},  \label{generalizedwigner}
\end{align}
for $\vert x\vert <x_c$, where we have
\begin{align}
	\beta &= \frac{1}{3}  \left[ (1-\Gamma) \Gamma_4^{(1)} + 6 (1-\Gamma) \Gamma_4^{(2)} + 2 (4-\Gamma) \Gamma_4^{(3)}\right].
\end{align}
This result is compared with numerics in Figs. \ref{fig:modifiedellipse}b and \ref{fig:modifiedellipsemu}b. In the limit $\Gamma\to1$, the matrix $\underline{\underline{a}}$ becomes symmetric and the eigenvalues concentrate along the real axis. We thus have $\Gamma_4^{(1)} = \Gamma_4^{(2)} = \Gamma_4^{(3)}$. In this case, as a check, we can perform an alternative derivation of the modified semi-circle law using Eq.~(\ref{realeigenvaluedensity}) which agrees with the expression in Eq.~(\ref{generalizedwigner}) [see SM Section S5 E].

We note that the results in Eqs.~(\ref{generalfirstorder}), (\ref{ellipsemodcart}), and (\ref{generalizedwigner}) all reduce to their dense counterparts in Eqs.~(\ref{selfconsistentseries}), (\ref{ellipse}), (\ref{ellipsedensity}) and (\ref{wignersemicircle}) in the limit $p \to \infty$ as required. One also notes that by substituting $\sigma = \sqrt{p}$ and $\Gamma = \Gamma_4^{(1)} = 1$ into Eq.~(\ref{generalizedwigner}), we obtain the result of Rodgers and Bray in Ref. \cite{rodgers1988density}, which was derived for a sparse symmetric matrix with $\pi(a_{ij}, a_{ji}) = \delta(a_{ij}-a_{ji})[\delta(a_{ij}-1) + \delta(a_{ij}+1)]/2$.

\subsection{Inclusion of a non-zero mean ($\mu \neq 0$)}\label{section:nonzeromu}
We have so far obtained results for the case where $\langle a_{ij} \rangle_\pi = \mu/p = 0$ [defined in Eqs.~(\ref{sparsedef}) and (\ref{loworderstats})]. We now generalize these results to allow for the possibility of the non-zero elements having a non-zero mean.

It has been shown previously that when $\mu \neq 0$ for dense matrices, the eigenvalue spectrum may gain an additional outlier eigenvalue that strays from the bulk region to which most of the eigenvalues are confined \cite{tao2013outliers, orourke, benaych2011eigenvalues, baron2022eigenvalues}. We show here that this is also true in the sparse case. In contrast to the dense case however, we also show that the bulk spectrum itself is affected by the introduction of a non-zero mean. 

\begin{figure*}[t]
	\centering 
	\includegraphics[scale = 0.37]{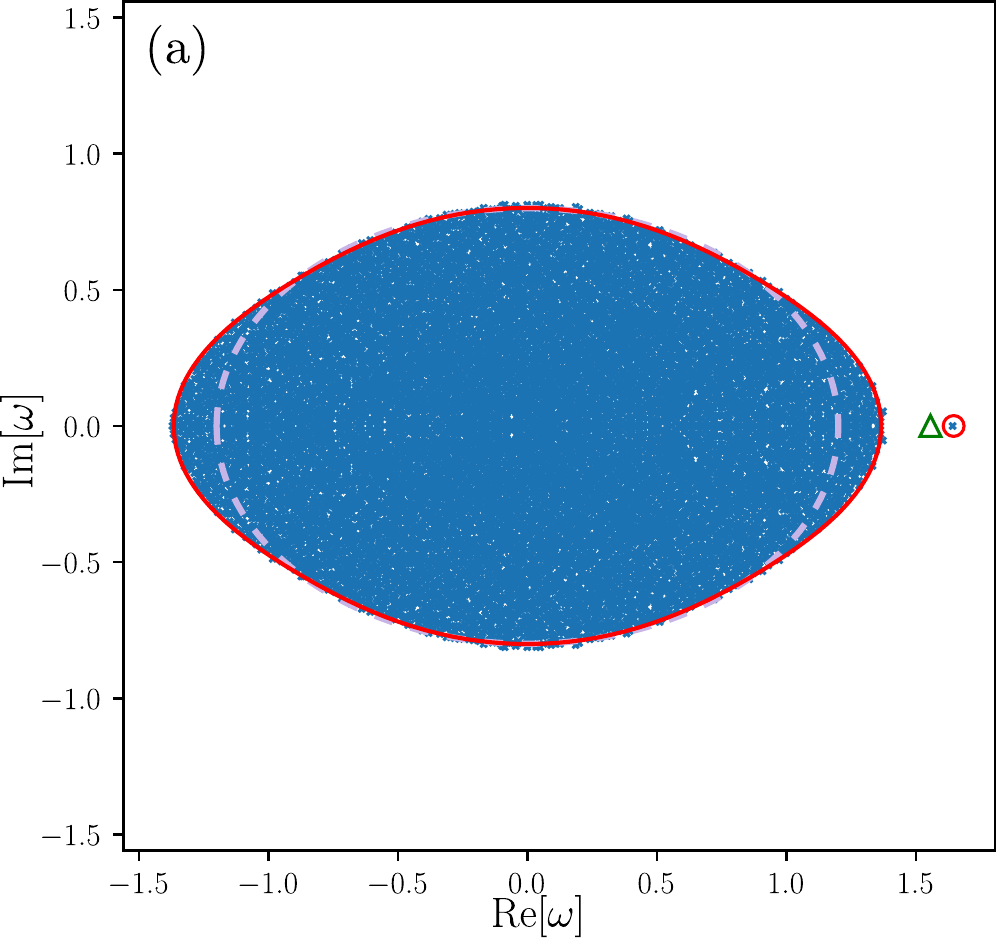} \hspace{5mm}
	\includegraphics[scale = 0.45]{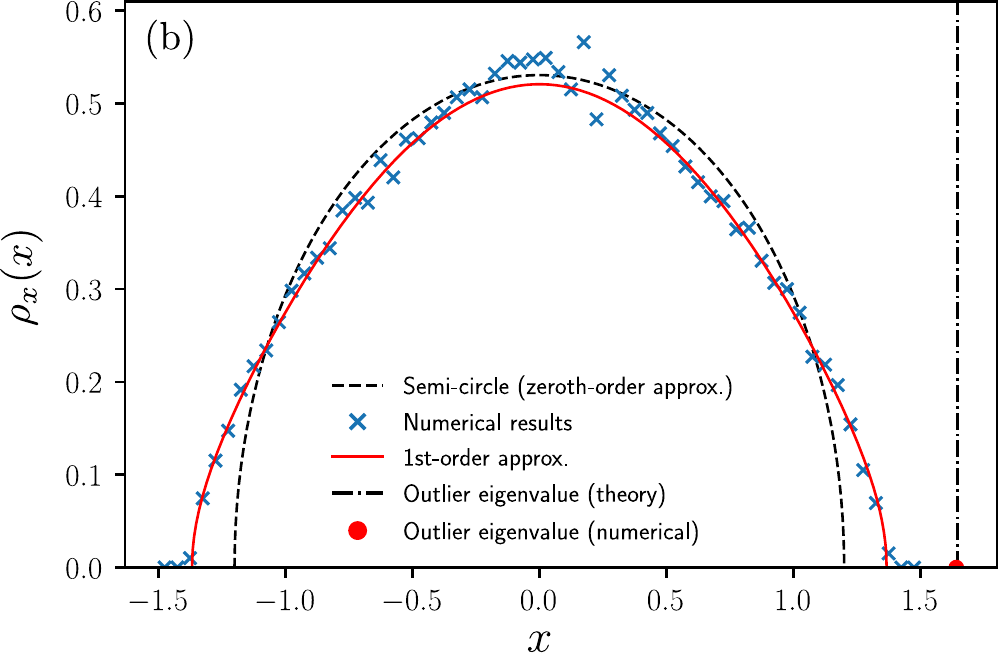}
	\captionsetup{justification=raggedright,singlelinecheck=false}
	\caption{Modified elliptic and semi-circular laws for $\mu \neq 0$. Parameters are the same as in Fig. \ref{fig:modifiedellipse}, but now with $\mu = \sqrt{2}$. Panel (a): Boundary of the eigenvalue spectrum. Blue crosses are the results of numerical diagonalization and the red line is given by Eqs.~(\ref{ellipsemodcart}) using the expressions for $x_c$ and $y_c$ given in Eq.~(\ref{xcmu}). The dashed line is the na\"ive elliptic law obtained in the limit $p \to \infty$. The red circle is the prediction for the outlier given in Eq.~(\ref{outliercorrectiongaussian}), whereas the green triangle is the na\"{i}ve prediction for $p\to \infty$. Panel (b): Integrated eigenvalue density as a function of the real part. The red line is given by Eq.~(\ref{generalizedwigner}) using the expression for $x_c$ given in Eq.~(\ref{xcmu}). }\label{fig:modifiedellipsemu}
\end{figure*}

In SM Section S6, we identify the new contributions to the action that give rise to the outlier eigenvalue and affect the bulk spectrum. For the bulk of the eigenvalue spectrum, we find that the expressions given in Eqs.~(\ref{ellipsemodcart}) and (\ref{generalizedwigner}) are unaltered explicitly by the introduction of a non-zero value of $\mu$. However, the rightmost and uppermost eigenvalues of the bulk, which enter into Eqs.~(\ref{ellipsemodcart}) and (\ref{generalizedwigner}), are now modified to be
\begin{align}
	x_c =& \sigma(1+\Gamma)  \nonumber \\
	&+ \frac{\sigma}{2p} \left[ (3-\Gamma) \left(\Gamma_4^{(3)} + \frac{\mu^2}{\sigma^2} \right) + 2 (1-\Gamma) \Gamma^{(2)}_4\right] , \nonumber \\
	y_c =& \sigma(1-\Gamma) \nonumber \\
	&+ \frac{\sigma}{2p} \left[(3 + \Gamma) \Gamma_4^{(3)} - 2 (1+\Gamma) \Gamma_4^{(2)} - (1- \Gamma)\frac{\mu^2}{\sigma^2} \right]. \label{xcmu}
\end{align}
The leading edge of the bulk region $x_c$, the boundary of the bulk and the generalized semi-circle law in the case of $\mu \neq 0$ are tested against numerics in Figs. \ref{fig:leadingeigenvaluecorrection}a, \ref{fig:modifiedellipsemu}a  and \ref{fig:modifiedellipsemu}b respectively. 

Let us now summarily address the additional outlier eigenvalue. For more information on the calculation of $\lambda_\mathrm{outlier}$, the reader is referred to SM Section S6. By inspecting the action for $\mu\neq 0$, we find that there is a term that would also have arisen if we had simply added $\mu/N$ to every element of the matrix. This effective rank-1 perturbation is responsible for the outlier eigenvalue \cite{orourke, benaych2011eigenvalues}. If we define $z_{ij} = a_{ij} - \mu/N$ and identify $\underline{\underline{C}} \equiv \overline{[\lambda_{\mathrm{outlier}}\id - \underline{\underline{z}}]^{-1}}$, one can show that the outlier eigenvalue satisfies \cite{orourke, benaych2011eigenvalues, baron2022eigenvalues}
\begin{align}
	\frac{1}{N}\sum_{ij}C_{ij}(\lambda_{\mathrm{outlier}}) = \frac{1}{\mu} . \label{sylvester}
\end{align}
Along with the effective rank-1 perturbation however, other terms in the action arise that encode new contributions to the off-diagonal elements of the resolvent. Crucially, because one sums over all elements of the resolvent to find the outlier in Eq.~(\ref{sylvester}), these off-diagonal contributions affect the outlier, but not the bulk spectrum. This results in the need for additional diagrams to compute the sum $N^{-1}\sum_{ij} C_{ij}$ \cite{baron2022eigenvalues}.

Ultimately, one solves Eq.~(\ref{sylvester}) to obtain
\begin{align}
	&\lambda_{\mathrm{outlier}}  = \mu+\frac{\Gamma \sigma^2}{ \mu} \nonumber \\
	&+\frac{1}{p}\left[ \mu+ \frac{(1+2\Gamma) \sigma^2}{\mu}+ (\Gamma^{(3)}_4- \Gamma- 2 \Gamma^2) \frac{\sigma^4}{\mu^3} \right] . \label{outliergeneral}
\end{align}
In order for this expression to be valid, one requires that
\begin{align}
	 \mu^2 \geq \sigma^2 + \frac{\sigma^2}{p} \left[ 1 - 2(1+2\Gamma)+ 2 \Gamma_4^{(2)} + \Gamma_4^{(3)}\right].
\end{align}
One can show that when this bound on $\mu$ is saturated, the outlier eigenvalue in Eq.~(\ref{outliergeneral}) and the edge of the bulk in Eq.~(\ref{xcmu}) coincide. We also note that the $p\to \infty$ limit of the expression in Eq.~(\ref{outliergeneral}) agrees with results found previously in Refs. \cite{orourke, benaych2011eigenvalues, edwardsjones} in the dense case. The expression in Eq.~(\ref{outliergeneral}) is verified in Figs. \ref{fig:leadingeigenvaluecorrection} and \ref{fig:modifiedellipsemu}. 

One notes further that, in principle, the position of the outlier eigenvalue can also be affected by the third moments of $a_{ij}$, but we do not consider this possibility here. That is, Eq.~(\ref{outliercorrectiongaussian}) assumes $\langle (a_{ij}-\mu/p)^3\rangle_\pi = \langle (a_{ij}-\mu/p)^2 (a_{ji}-\mu/p)\rangle_\pi = 0$. The bulk of the eigenvalue spectrum is not affected by these statistics.

\subsection{Tests against numerics}
To test the results for the sparse corrections that we have obtained so far, we examine the case where $\pi(a_{ij}, a_{ji})$ is a Gaussian distribution. We will study the alternative example of a dichotomous distribution in Section \ref{section:higherorder}. In the Gaussian case, we have
\begin{align}
	\Gamma_4^{(1)} = 3, \,\,\,\, \Gamma_4^{(2)} = 3 \Gamma, \,\,\,\, \Gamma_4^{(3)} = 1 + 2 \Gamma^2, \label{gaussstats}
\end{align}
meaning that the statistics of $\pi(a_{ij}, a_{ji})$ can be written entirely in terms of $\sigma^2$, $\Gamma$ and $\mu$. An example of a typical eigenvalue spectrum is presented in Fig. \ref{fig:modifiedellipsemu}. We see in panel (a) that the generalized elliptic law in Eq.~(\ref{ellipsemodcart}), with $x_c$ and $y_c$ given by Eq.~(\ref{xcmu}), is indeed a good approximation to the boundary of the eigenvalue spectrum. Panel (b) verifies the sparse correction to the Wigner semi-circle law given by Eq.~(\ref{generalizedwigner}) with the expression for $x_c$ given by Eq.~(\ref{xcmu}). 

We also test the prediction for the sparse correction to the leading edge of the bulk of the eigenvalue spectrum that is given in Eq.~(\ref{xcmu}) in Fig. \ref{fig:leadingeigenvaluecorrection}a for various $\Gamma$. From the general expression in Eq.~(\ref{xcmu}), we obtain for the leading-order sparse correction
\begin{align}
	p\left[\lambda_{\mathrm{max}} - \sigma(1+\Gamma)\right]= \frac{\sigma}{2}\left[3  + 5 \Gamma - 2 \Gamma^{ 3}   + \frac{\mu^2}{\sigma^2}\left(3 - \Gamma \right) \right]. \label{bulkcorrectiongaussian}
\end{align}
Similarly, the sparse correction to the outlier eigenvalue, which is tested in Fig. \ref{fig:leadingeigenvaluecorrection}b, can be derived from Eq.~(\ref{outliergeneral}) and is given by
\begin{align}
	p\left[\lambda_{\mathrm{max}} - \left(\mu + \frac{\Gamma\sigma^2}{\mu}\right)\right]= \mu + \frac{(1+2\Gamma)\sigma^2}{\mu} + \frac{(1-\Gamma)\sigma^4}{\mu^3} . \label{outliercorrectiongaussian}
\end{align}

\subsection{Implications of the sparse correction for the  stability of dynamical systems}\label{section:stability}
Let us now comment on the significance of our findings for the stability of complex dynamical systems. Let us suppose that the matrix $\underline{\underline{a}}$ encodes the off-diagonal elements of a Jacobian matrix of some system linearized about a fixed point. Like May's seminal work on complex ecosystems \cite{may}, let us suppose that the diagonal elements of the Jacobian are equal to a negative constant so that the system would be stable in the absence of interactions. That is, we imagine that the Jacobian is $\underline{\underline{J}} = -d\underline{\underline{\id}}_N + \underline{\underline{a}}$. 

One sees that if any of the eigenvalues of $\underline{\underline{a}}$ are greater than $d$, then the system is unstable. Therefore, if we alter the statistics of the matrix $\underline{\underline{a}}$ in such a way that the change increases the rightmost eigenvalue, then we say that this alteration is destabilizing. The simple formulae for the leading eigenvalue in Eqs.~(\ref{xcmu}) and (\ref{outliergeneral}) provide a transparent way for us to see how the sparse corrections affect stability.

For example, we see directly from the first of Eqs.~(\ref{xcmu}) that making $\mu$ large and negative can only serve to broaden the eigenvalue spectrum along the real axis and thus destabilize the system. This is in contrast to the dense case \cite{orourke, benaych2011eigenvalues, allesinatang2, allesinatang1} (and to studies of non-linear systems with dense interactions \cite{bunin, galla2018dynamically, garcialorenzana2022competitive}), where decreasing $\mu$, i.e. making interactions more `competitive', usually only serves to stabilize the system. This is a clear instance where the behaviour of a sparsely interacting system differs substantially from its densely interacting counterpart.

More generally, from Eq.~(\ref{xcmu}) and the definitions of $\Gamma_{4}^{(2)}$ and $\Gamma_{4}^{(3)}$ in Eq.~(\ref{loworderstats}), we see that the sparse correction to the rightmost edge of the bulk region is positive unless the matrix entries are very negatively correlated and $\mu$ is small in magnitude. That is, the term proportional to $\Gamma_4^{(2)}$ must be sufficiently negative to cancel the other terms, which are constrained to be positive, in order for the sparse correction to the bulk edge to be negative. In the case of Gaussian distributed elements [see Eq.~(\ref{bulkcorrectiongaussian}) and also Fig. \ref{fig:leadingeigenvaluecorrection}a], one requires $\Gamma<-(\sqrt{7}-1)/2 \approx -0.82$ when $\mu = 0$ in order for the sparse correction to be stabilizing. If on the other hand the outlier is the rightmost eigenvalue (which requires $\mu>\sigma$), then we see that the sparse correction in Eq.~(\ref{outliergeneral}) is in fact always positive. This can be seen from the fact that $\vert \Gamma\vert <1$ and $\Gamma_4^{(3)}>0$. 

Thus, broadly speaking, one tends to obtain a more conservative estimate of the interaction statistics that would permit stability by including the sparse correction.

\begin{figure}[H]
	\centering 
	\includegraphics[scale = 0.13]{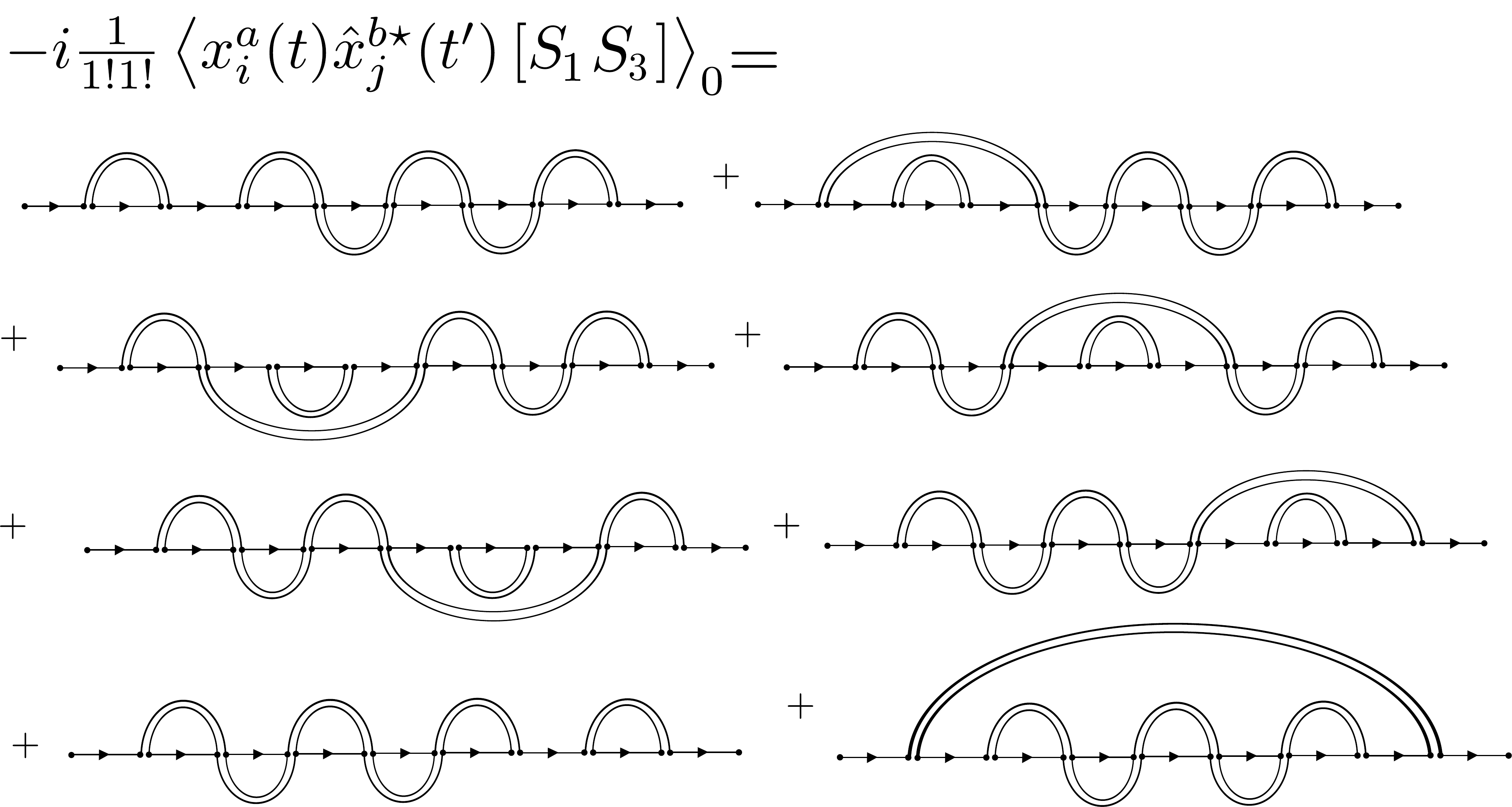}
	\captionsetup{justification=raggedright,singlelinecheck=false}
	\caption{Example $O(1/p^2)$ ribbon diagrams.}\label{fig:ribbonsecondorder1}
\end{figure}

\begin{figure}[t]
	\centering 
	\includegraphics[scale = 0.19]{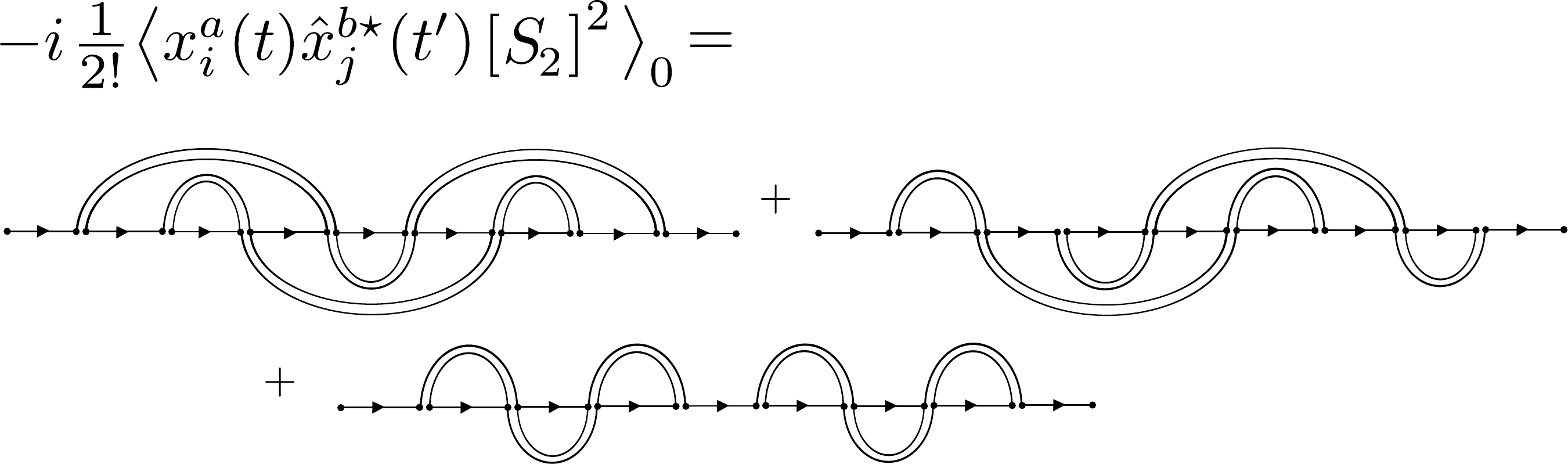}
	\captionsetup{justification=raggedright,singlelinecheck=false}
	\caption{Products of $O(1/p)$ diagrams that give a $O(1/p^2)$ contribution. }\label{fig:ribbonsecondorder2}
\end{figure}
\section{Higher-order sparse corrections}\label{section:higherorder}
In the previous section, we found the first-order sparse correction to the eigenvalue spectrum. In this section, we demonstrate how higher-order corrections can also be calculated. By evaluating these higher-order terms, we are able to obtain accurate results for values of the connectivity as low as $p \sim 10$ in our examples. 

The evaluation of higher order terms also allows us to compare with alternative approximation schemes. We show here that the `effective medium approximation' obtained in previous works \cite{semerjian2002sparse, slanina2011equivalence, slanina2012localization} is only accurate to first order in $1/p$ for the example used here.

\begin{figure*}[t]
	\centering 
	\includegraphics[scale = 0.34]{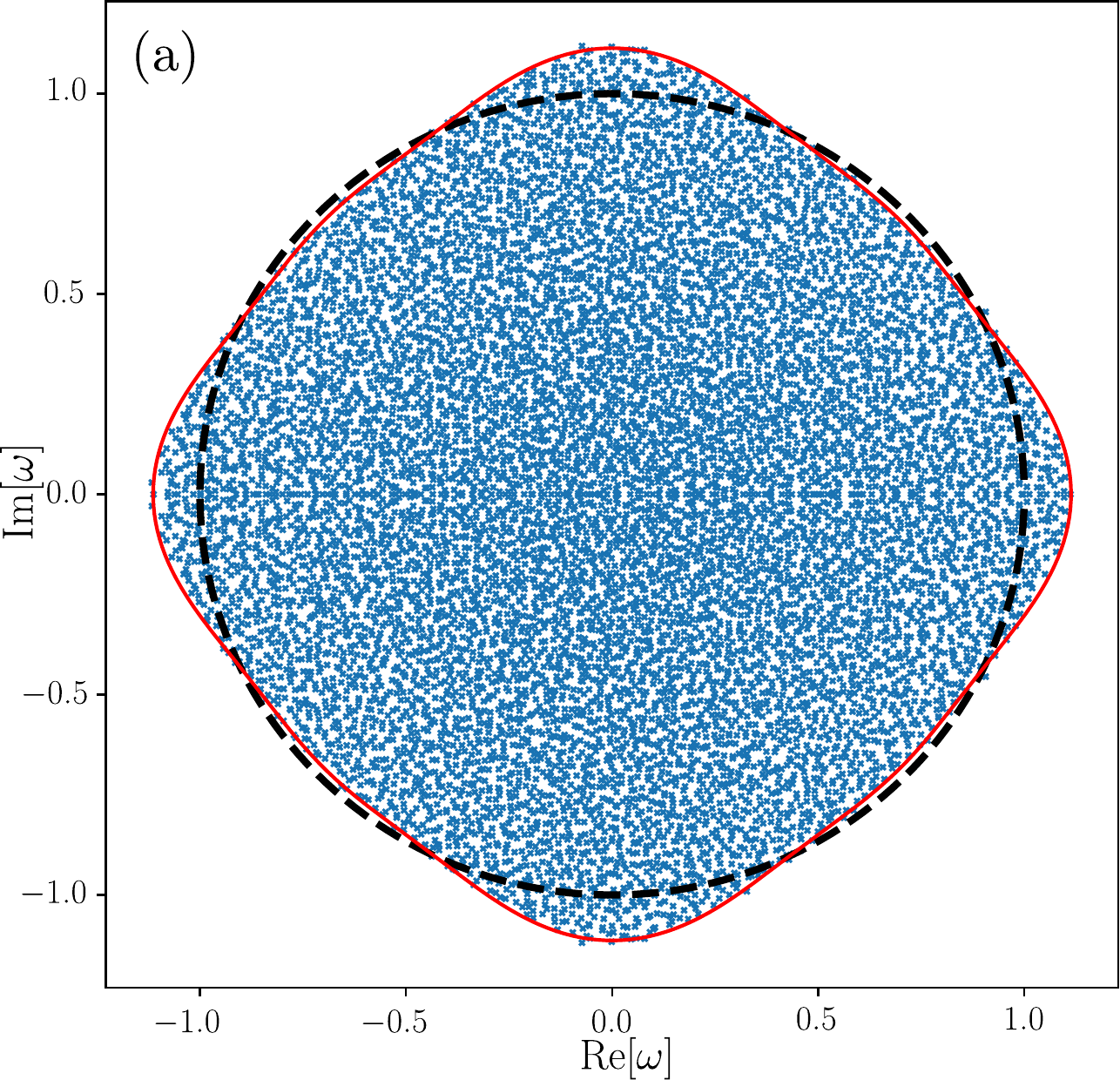} \hspace{3mm}
	\includegraphics[scale = 0.47]{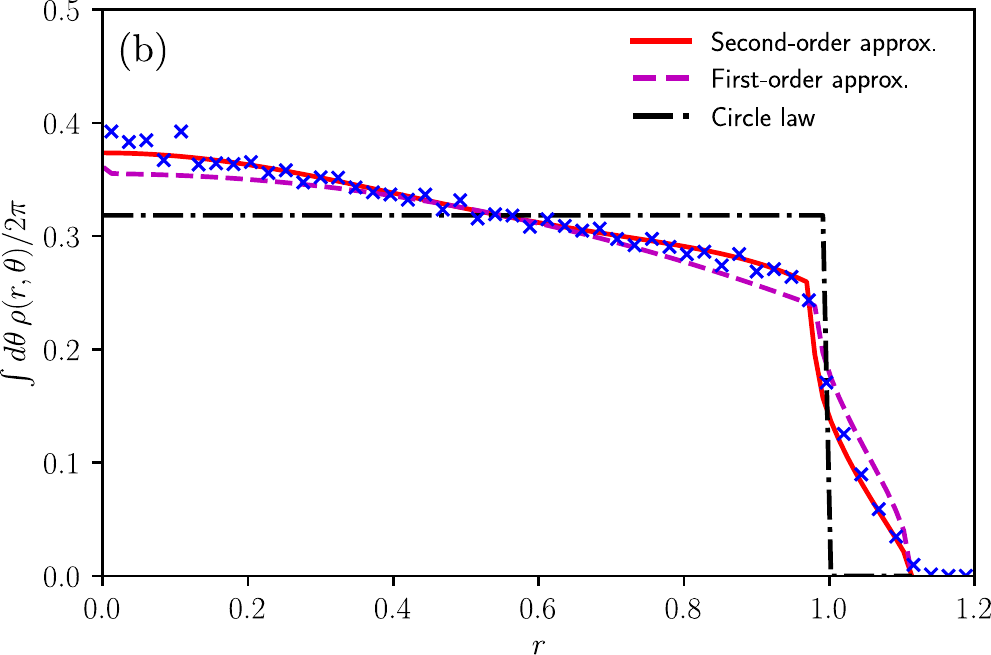}
	\captionsetup{justification=raggedright,singlelinecheck=false}
	\caption{Testing the second-order perturbative result in Eq.~(\ref{secondorderselfconsistent}). The parameters used here were $N = 10^4$, $p = 15$. Panel (a): The matrix $\underline{\underline{a}}$ was drawn from the ensemble defined in Eq.~(\ref{dichotomous}) with $\Gamma = 0$. The red line is given by Eq.~(\ref{secondorderboundary}). A dashed circle with radius 1 (the dense limit $p\to \infty$) is given for reference. Panel (b): The average eigenvalue density as a function of the distance from the origin. The red line is the result of integrating Eq.~(\ref{secondorderdensity}) over $\theta$, and the dashed purple line is the result in Eq.~(\ref{secondorderdensity}) ignoring the term proportional to $1/p^2$. The results were averaged over 10 trials. }\label{fig:secondorder}
\end{figure*}

\subsection{Second-order diagrams}
As we go to higher order in $1/p$, one has to be careful to take into account all the possible non-vanishing ribbon diagrams that contribute to the response functions, the variety of which increases at higher order. We now return to the general expansion of the response functions in Eq.~(\ref{interactionactionexpansion2}) and truncate the series at $O(1/p^2)$.

Let us take two examples of terms that are of order $1/p^2$. First, in Fig. \ref{fig:ribbonsecondorder1}, we present a term that comes about due to the second-order contribution to the action $S_3$. This term gives rise to ribbon diagrams that are very much analogous to those explored in Section \ref{section:firstorderdiagrams}, albeit with more concatenated arcs. However, the second-order terms that arise due to combinations of the first-order ribbon diagrams are more complicated. These are shown in Fig. \ref{fig:ribbonsecondorder2}. 

From the diagrams in Fig. \ref{fig:ribbonsecondorder2}, we see that there is no clear pattern to how the ribbon diagrams of lower order will combine to produce non-vanishing planar diagrams in thermodynamic limit as we go to higher order in $1/p$. Enumerating all the possible ways for the ribbons to `fit together' in a planar topology is non-trivial. So, while we can evaluate the diagrammatic series to arbitrarily high order in $1/p$, it does not seem that a full resummation of the diagrammatic series for the resolvent is a simple task. With that being said, by evaluating diagrams up to $O(1/p^2)$, we are still able to obtain remarkably accurate results.

In SM Section S7, we perform the summation of all the diagrams that do not vanish in thermodynamic limit to obtain the following expression for the resolvent that is accurate to second order in $1/p$ [c.f. Eq.~(\ref{generalfirstorder})]
\begin{align}
	\mathh &\approx  \Bigg\{\mathh_0^{-1} - p\langle \matj \mathh \matj^\dagger \rangle_\pi - p \langle \matj \mathcal{H} \matj^\dagger \mathcal{H} \matj\mathcal{H} \matj^\dagger  \rangle_\pi \nonumber \\
	&- p \langle \matj \mathcal{H} \matj^\dagger \mathcal{H} \matj\mathcal{H} \matj^\dagger\mathcal{H}  \matj \mathcal{H} \matj^\dagger \rangle_\pi \nonumber \\
	&- p^2  \left\langle  \matj_1 \mathcal{H} \matj_2 \mathcal{H} \matj_2^\dagger \mathcal{H} \matj_1^\dagger \mathcal{H}  \matj_1 \mathcal{H}  \matj_2 \mathcal{H} \matj_2^\dagger \mathcal{H} \matj_1^\dagger  \right\rangle_\pi\nonumber \\
	&- p^2 \left\langle \matj_1 \mathcal{H} \matj_1^\dagger \mathcal{H} \matj_2\mathcal{H} \matj_2^\dagger \mathcal{H} \matj_1 \mathcal{H}  \matj_1^\dagger \mathcal{H} \matj_2 \mathcal{H} \matj_2^\dagger  \right\rangle_\pi \Bigg\}^{-1}, \label{secondorderselfconsistent}
\end{align}
where here we have defined the $2 \times 2$ matrices $\matj_1$ and $\matj_2$, which each individually have the same statistics as $\matj$ [defined in Eq.~(\ref{jdef}), but with elements drawn from $\pi(a_{12},a_{21})$]. However, they are statistically independent of one another, so that $\langle \matj_1^{ab} \matj_2^{a'b'} \rangle_\pi = 0$ for all combinations of upper indices. All other $2 \times 2$ matrices here are as described in Eq.~(\ref{def2by2}). We now test this result with two examples.

\begin{figure*}[t]
	\centering 
	\includegraphics[scale = 0.33]{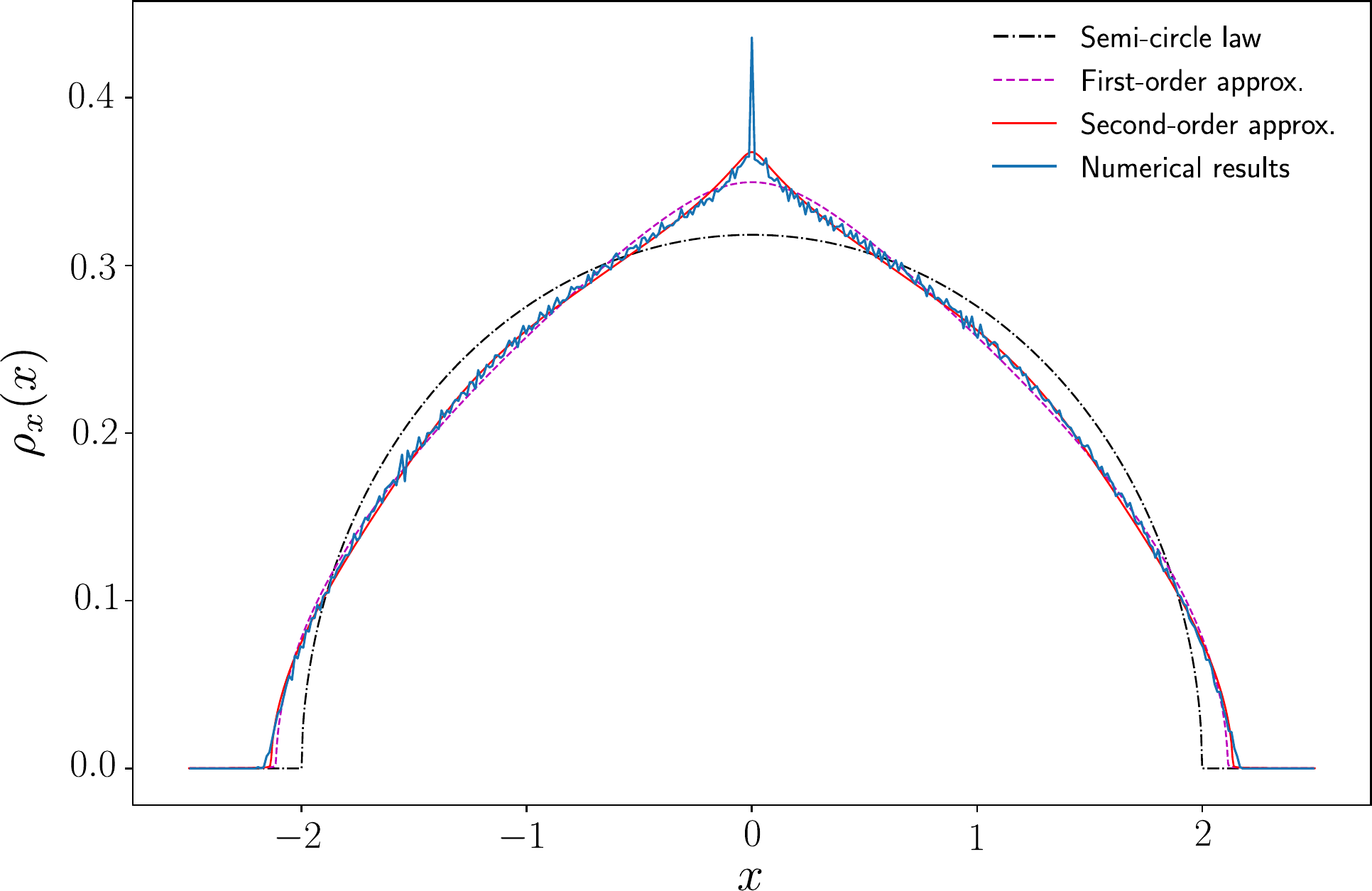}
	\captionsetup{justification=raggedright,singlelinecheck=false}
	\caption{Modified semi-circular law for the dichotomous distribution of non-zero elements defined in Eq.~(\ref{dichotomous}) with $\Gamma = 1$. Here, the connectivity is only $p = 7$, yet we see that the second-order perturbative approximation to the eigenvalue density [found using Eq.~(\ref{secondordersymmetricresolvent})] is still accurate. We plot the zeroth- and first-order approximations to the eigenvalue density to demonstrate the convergence of the result.}\label{fig:secondordersymmetric}
\end{figure*}

\subsection{Example: Asymmetric dichotomous distribution}\label{section:higherorderasymmetric}
Let us first consider the following asymmetric random matrix ensemble
\begin{align}
	P(a_{ij}, a_{ji}) &= \left(1 - \frac{p}{N}\right) \delta(a_{ij})\delta(a_{ji}) + \frac{p}{N} \pi(a_{ij}, a_{ji}), \nonumber \\
	\pi(a_{ij}, a_{ji}) &= \frac{(1+\Gamma)}{4} \left( \delta_{++} + \delta_{--}\right) \nonumber \\&+ \frac{(1-\Gamma)}{4} \left(\delta_{+-} + \delta_{-+} \right), \label{dichotomous}
\end{align}
where $\delta_{+-} = \delta(a_{ij} -1/\sqrt{p})\delta(a_{ji} +1/\sqrt{p})$, and so on. This example permits the relatively straightforward computation of the averages $\langle \cdot \rangle_\pi$ in Eq.~(\ref{secondorderselfconsistent}). In this subsection, we take the case $\Gamma = 0$ for simplicity.

After some algebra along the lines of what was done in SM Section S5, one obtains two expressions for the resolvent, one of which is analytic and the other of which is non-analytic. These can once again be solved simultaneously to yield the boundary of support. The density can also be obtained using Eq.~(\ref{densityfromres}). 

In this case, it is more convenient to represent the eigenvalue spectrum in polar coordinates (letting $\omega = re^{i\theta}$). This allows one to see more readily the deviation from the circular law, which would apply in the limit $p \to \infty$ \cite{girko1985circular}. One obtains for the boundary of the support
\begin{align}
	r =& 1 + \frac{1}{2p}\left[ 1 + 2 \cos\left( 4\theta\right)\right] \nonumber \\
	&+ \frac{1}{8 p^2} \left[-7 + 4\cos\left( 4\theta\right) + 30\cos\left( 8\theta\right) \right], \label{secondorderboundary}
\end{align}
and one obtains the following for the eigenvalue density
\begin{align}
	\rho(r, \theta) =& \frac{1}{\pi}\bigg\{1 + \frac{1}{p}(2 - 6 r^2) + \frac{1}{p^2} \big[9 - 104 r^2 + 258 r^4  \nonumber \\
	&\,\,\,\,\,\,\,\,\,\,- 152 r^6+ (70 r^4- 72 r^6) \cos(4 \theta)  \big]\bigg\}. \label{secondorderdensity}
\end{align}
Noting that in this case $\sigma^2 = 1$, $\Gamma = 0$, $\Gamma_4^{(1)} = \Gamma_4^{(3)} = 1$ and $\Gamma_4^{(2)} = 0$, one can see that Eqs.~(\ref{secondorderboundary}) and (\ref{secondorderdensity}) agree with Eqs.~(\ref{ellipsemodcart}) and (S41) of the SM respectively to first order in $1/p$. 

Both the boundary of the support and the eigenvalue density inside the support are tested against computer generated results in Fig. \ref{fig:secondorder}, where we see that the second-order result fits the data very closely for $p = 15$.

\subsection{Example: symmetric dichotomous distribution}\label{section:higherordersymmetric}
Now, partially for the sake of comparing to previous works where the eigenvalue spectrum was approximated for symmetric sparse matrices using the so-called Effective Medium Approximation (EMA) \cite{semerjian2002sparse, slanina2011equivalence, slanina2012localization}, we consider the ensemble in Eq.~(\ref{dichotomous}) in the case $\Gamma = 1$.

In this case, one finds from Eq.~(\ref{secondorderselfconsistent}) the following analytic expression for the resolvent
\begin{align}\label{secondordersymmetricresolvent}
	C &\approx \frac{1}{\omega- C} + \frac{1}{p} C^5 + \frac{1}{p^2} (C^7 + C^9) .
\end{align}
This can solved numerically for $C(\omega)$, which yields the eigenvalue density via Eq.~(\ref{realeigenvaluedensity}). The results of doing so are shown in Fig. \ref{fig:secondordersymmetric} and we see remarkable agreement with numerics even for as low a connectivity as $p = 7$.

One notes that for this ensemble, $\sigma^2 = \Gamma = \Gamma_4^{(1)} = 1$. Comparing the first-order term in Eq.~(\ref{secondordersymmetricresolvent}) with Eq.~(S23) in the SM, we see again that the expression obtained here is consistent with the universal result that we derived earlier up to first order in $1/p$.

It was shown in Ref. \cite{semerjian2002sparse} that the EMA, which is an uncontrolled approximation scheme that essentially assumes a Gaussian field theory \cite{biroli1999single}, was able to replicate the leading order $1/p$ perturbative correction to the semi-circle law that was found previously in Ref. \cite{rodgers1988density}. It was then speculated that perhaps the EMA might capture the perturbative series for the resolvent to all orders. 

However, we take the opportunity to note here that the effective medium approximation of the resolvent for the ensemble in Eq.~(\ref{dichotomous}) with $\Gamma =1$ is \cite{semerjian2002sparse, slanina2011equivalence, slanina2012localization} (after appropriate rescaling with $p$)
\begin{align}
	C_\mathrm{EMA} &= \frac{1}{\omega - C_\mathrm{EMA}/(1 - C_\mathrm{EMA}^2/p) } \nonumber \\
	&\approx \frac{1}{\omega - C_\mathrm{EMA}} + \frac{1}{p} C_\mathrm{EMA}^5 +\frac{1}{p^2} \left( C_\mathrm{EMA}^7 - C_\mathrm{EMA}^9 \right) . \label{cema}
\end{align}
Comparing this expression to Eq.~(\ref{secondordersymmetricresolvent}), we see that while the EMA captures the first-order correction, it does not accurately replicate higher order terms. This is unsurprising, given the uncontrolled nature of the EMA.

\section{Other applications of the path integral method}\label{section:otherapplications}
Having shown how the path integral method can be used to handle sparse random matrix ensembles, we now discuss how it can also be used to simplify the calculations for  a range of other ensembles. The results (formulae and figures) are mostly presented in SM Sections S8, S9 and S10, but we briefly summarize the additional findings here.

\subsection{General non-negligible higher order moments (non-Gaussian statistics)}\label{section:nongaussian}
In some sense, the ensemble of sparse random matrices defined in Eq.~(\ref{sparsedef}), which formed the basis for the study here, can be thought of as a special case of a broader class of random matrix. In SM Section S8, we extend the consideration to matrices whose elements mostly fluctuate within a small distance of $a_{ij} = 0$, but also have a small number of elements per row that are of order $N^0$. 

Specifically, we study dense matrices whose elements are all drawn from a distribution with moments $\overline{(a_{ij}-\mu/N)^r}$ that all scale as $1/N$ for arbitrarily high $r \in \mathbb{N}$. We note that the distribution in Eq.~(\ref{sparsedef}) indeed falls into this category. A similar observation to this was also made in Ref. \cite{azaele2024generalized} in the context of the random Lotka-Volterra equations. In particular, we take the example of elements sampled from a truncated Cauchy distribution, and we also consider a simple generalization of Eq.~(\ref{sparsedef}) where the null elements are allowed to fluctuate.

The perturbative method that we have developed can be applied to these ensembles without much additional effort. All that is required for the perturbative approach to be valid is for progressively higher-order moments to be decreasing in magnitude so that one can truncate the series expansion. With that being said, there are some non-trivial differences between the sparse matrices that we have discussed here and the dense non-Gaussian matrices that are highlighted in the SM. For example, when $\{a_{ij}\}$ are i.i.d. random variables drawn from a truncated Cauchy distribution, the boundary of the eigenvalue spectrum satisfies the na\"ive circle law (albeit with a modified density inside the circle), so the higher-order moments have no effect on stability in this case. However, when we constrain $a_{ij} = a_{ji}$, there is a distinct deviation from the semi-circle law, with a modified leading eigenvalue. 

\subsection{Generalized Marchenko-Pastur law and block-structured matrices}\label{section:densegeneralexamples}
We also highlight in SM Sections S9 and S10 that different ensembles of dense random matrices can be handled using the approach in Section \ref{section:ellipticlawdiagrams}, without the need for additional diagrams. The method used here permits one to see easily that the same formula in Eq.~(\ref{selfconsistentseries}) can be used for many matrix ensembles. To demonstrate this, we recover a generalization of the Marchenko-Pastur law for asymmetric products of random matrices \cite{akemann2021non, kanzieper2010non}. We also recover previous results for block-structured random matrices \cite{baron2020dispersal} in a similar way. We note that the eigenvalue density of products of random matrices has been derived previously using diagrammatic techniques \cite{burda2010eigenvalues, burda2010spectrum}, and is known to be universal \cite{o2011products}. 

The central idea is to introduce additional dynamical variables into the system in Eqs.~(\ref{dynamicalsystem}). These additional variables can be used to decouple products of random matrices (reminiscent of the Hubbard-Stratonovich transformation \cite{hubbard}), or they can be used to correspond to different blocks of the random matrix $\underline{\underline{a}}$. A similar trick to this was used in Ref. \cite{cui2020perturbative} in the case of symmetric matrices. 

By carefully identifying appropriate expressions for the matrices $\matj$, $\mathh$ and $\mathinv$ for the system at hand, one can then write the action in exactly the same form as Eq.~(\ref{ellipticaction}). One thus arrives immediately at the same equation for the hermitized resolvent as for the elliptic law in Eq.~(\ref{selfconsistentseries}). One notes however that the spectrum still varies from the elliptic law, due to the different forms of $\matj$, $\mathh$ and $\mathinv$. 

\section{Discussion}\label{section:conclusion}

To summarize, there are two main focuses of this work. First, we described how the MSRJD path integral approach can be used to find the hermitized resolvent of a non-Hermitian random matrix. This involved exploiting the correspondence between the hermitized resolvent and the response functions of a particular dynamical system, which could be expressed as a path integral. This approach has no need for replicas or Grassmann variables, the universality of the results is transparent, and we could utilize diagrammatic techniques to perform a perturbative analysis. We also demonstrated how the dynamic approach could be used to simplify calculations for ensembles that involved matrix products or block structure.

The second main contribution of this work was using the path integral approach to study non-Hermitian sparse random matrices. We saw that the sparse corrections could be considered as a perturbation to the dense case. These corrections could be accounted for (to arbitrary order in $1/p$) by considering `ribbon' Feynman diagrams in addition to the usual `rainbow' diagrams that arise in the dense case. Ultimately, we found concise universal expressions for the sparse corrections to the elliptic law. These allowed us to understand, in a transparent manner, how sparse interactions affect the stability of complex dynamical systems. For instance, we saw that `competitive' interactions can be destabilizing for sparse systems,  whereas they would not ordinarily be so in dense systems \cite{galla2018dynamically, bunin, garcialorenzana2022competitive, allesinatang2, allesinatang1}. We also demonstrated how the methods that we developed for sparse systems can be applied to dense systems with non-vanishing higher-order statistics. 

Because we used a perturbative approach in this work, it would be fairly straightforward to handle more intricacies alongside the sparse correction. For example, one could extend the work here to include more complex network structures \cite{metz2020spectral, silva2022analytic}, or more complicated correlations \cite{rogers2010universal}, both of which have also been handled in a perturbative fashion previously \cite{baron2022networks, baron2022eigenvalues}. The interplay of these factors could well lead to interesting effects. For example, just as we saw that the inclusion of a non-zero mean could broaden the bulk of the spectrum in the sparse case (whereas it does not in the dense case), one anticipates that more generalized correlations could well have a similar effect.

One drawback of the perturbative approach used here is that not all observables of interest can be treated perturbatively in all regimes of interest. In the weakly-non-hermitian regime where $\Gamma = 1- a/N$ with $a \sim O(1)$ \cite{fyodorov1997almost}, for example, the planar diagram approximation breaks down, and one finds that diagrams with crossing arcs begin to contribute in addition to the rainbow diagrams discussed in the main text. Summing the full series of relevant diagrams would present a likely-insurmountable task. However, methods such as the supersymmetric method or orthogonal polynomials have been shown to facilitate the necessary non-perturbative computation in this case \cite{fyodorov2003random, fyodorov1998universality}. Similarly, these same methods allow the computation of the microscopic correlations of eigenvalues, which also cannot be accessed by the perturbative approach \cite{mirlin1991universality, efetov1999supersymmetry, brezin1993universality}. With that being said, quantities such as the long-range eigenvalue correlations \cite{brezin1994correlation}, the eigenvector correlation functions \cite{nowak2018probing}, and of course the eigenvalue density in the strongly non-Hermitian regime considered here, can be treated perturbatively. 

It is hoped that the succinct results for the leading eigenvalues and the boundary of the spectrum presented here [see Eqs.~(\ref{ellipsemodcart}), (\ref{xcmu}) and (\ref{outliergeneral})] will be of immediate use in applications. For example, Ref. \cite{allesina2015predicting} fitted the elliptic law to the eigenvalue spectra corresponding to empirical food webs. Comparing whether or not the sparse correction fits better than the standard elliptic law could provide a measure of how effectively sparsely interacting various empirical food webs are.

Although we saw here that the eigenvalue spectra of the sparse matrices that we studied were bounded in the complex plane, it has been noted in recent works that (unless the matrix is locally sign-stable \cite{valigi2023local}) sparse matrices often have spectra that extend along the entire real axis \cite{mambuca2022dynamical}. The reason for this is that, as discussed for example in Ref. \cite{rodgers1988density}, there is also a non-perturbative contribution to the eigenvalue spectrum. This non-perturbative contribution takes the form of a Lifshitz tail, and is associated with large fluctuations in the network connectivity \cite{semerjian2002sparse}. The magnitude of this non-perturbative contribution scales roughly as $\sim e^{-p}$ \cite{semerjian2002sparse}, which is why it was negligible for the moderately high values of $p$ used in the present work, but was clearly visible for the relatively low values of $p$ used in Refs. \cite{mambuca2022dynamical, valigi2023local}. A more detailed study of the non-perturbative eigenvalue tails is the subject of ongoing collaborative work.

\acknowledgements
The author would like to thank Giulio Biroli, Chiara Cammarota, Tobias Galla, Giulia Garcia Lorenzana, Izaak Neri, Lyle Poley, Tim Rogers, and Pietro Valigi for insightful and helpful discussions. This work was supported by grants from the Simons Foundation (\#454935 Giulio Biroli). The author thanks the Leverhulme Trust for support through the Leverhulme Early Career Fellowship scheme.


\begin{thebibliography}{99}%
	\makeatletter
	\providecommand \@ifxundefined [1]{%
		\@ifx{#1\undefined}
	}%
	\providecommand \@ifnum [1]{%
		\ifnum #1\expandafter \@firstoftwo
		\else \expandafter \@secondoftwo
		\fi
	}%
	\providecommand \@ifx [1]{%
		\ifx #1\expandafter \@firstoftwo
		\else \expandafter \@secondoftwo
		\fi
	}%
	\providecommand \natexlab [1]{#1}%
	\providecommand \enquote  [1]{``#1''}%
	\providecommand \bibnamefont  [1]{#1}%
	\providecommand \bibfnamefont [1]{#1}%
	\providecommand \citenamefont [1]{#1}%
	\providecommand \href@noop [0]{\@secondoftwo}%
	\providecommand \href [0]{\begingroup \@sanitize@url \@href}%
	\providecommand \@href[1]{\@@startlink{#1}\@@href}%
	\providecommand \@@href[1]{\endgroup#1\@@endlink}%
	\providecommand \@sanitize@url [0]{\catcode `\\12\catcode `\$12\catcode
		`\&12\catcode `\#12\catcode `\^12\catcode `\_12\catcode `\%12\relax}%
	\providecommand \@@startlink[1]{}%
	\providecommand \@@endlink[0]{}%
	\providecommand \url  [0]{\begingroup\@sanitize@url \@url }%
	\providecommand \@url [1]{\endgroup\@href {#1}{\urlprefix }}%
	\providecommand \urlprefix  [0]{URL }%
	\providecommand \Eprint [0]{\href }%
	\providecommand \doibase [0]{https://doi.org/}%
	\providecommand \selectlanguage [0]{\@gobble}%
	\providecommand \bibinfo  [0]{\@secondoftwo}%
	\providecommand \bibfield  [0]{\@secondoftwo}%
	\providecommand \translation [1]{[#1]}%
	\providecommand \BibitemOpen [0]{}%
	\providecommand \bibitemStop [0]{}%
	\providecommand \bibitemNoStop [0]{.\EOS\space}%
	\providecommand \EOS [0]{\spacefactor3000\relax}%
	\providecommand \BibitemShut  [1]{\csname bibitem#1\endcsname}%
	\let\auto@bib@innerbib\@empty
	\bibitem [{\citenamefont {Mehta}(2004)}]{mehta}%
	\BibitemOpen
	\bibfield  {author} {\bibinfo {author} {\bibfnamefont {M.~L.}\ \bibnamefont
			{Mehta}},\ }\href@noop {} {\emph {\bibinfo {title} {Random matrices}}},\
	Vol.\ \bibinfo {volume} {142}\ (\bibinfo  {publisher} {Elsevier},\ \bibinfo
	{address} {London},\ \bibinfo {year} {2004})\BibitemShut {NoStop}%
	\bibitem [{\citenamefont {Tao}(2012)}]{taobook}%
	\BibitemOpen
	\bibfield  {author} {\bibinfo {author} {\bibfnamefont {T.}~\bibnamefont
			{Tao}},\ }\href@noop {} {\emph {\bibinfo {title} {Topics in Random Matrix
				Theory}}}\ (\bibinfo  {publisher} {American Mathematical Society},\ \bibinfo
	{address} {Providence, Rhode Island, US},\ \bibinfo {year}
	{2012})\BibitemShut {NoStop}%
	\bibitem [{\citenamefont {Wigner}(1955)}]{wigner1955}%
	\BibitemOpen
	\bibfield  {author} {\bibinfo {author} {\bibfnamefont {E.~P.}\ \bibnamefont
			{Wigner}},\ }\bibfield  {title} {\bibinfo {title} {Characteristic vectors of
			bordered matrices with infinite dimensions},\ }\href@noop {} {\bibfield
		{journal} {\bibinfo  {journal} {Annals of Mathematics}\ }\textbf {\bibinfo
			{volume} {62}},\ \bibinfo {pages} {548} (\bibinfo {year} {1955})}\BibitemShut
	{NoStop}%
	\bibitem [{\citenamefont
		{Wigner}(1958{\natexlab{a}})}]{wigner1958distribution}%
	\BibitemOpen
	\bibfield  {author} {\bibinfo {author} {\bibfnamefont {E.~P.}\ \bibnamefont
			{Wigner}},\ }\bibfield  {title} {\bibinfo {title} {On the distribution of the
			roots of certain symmetric matrices},\ }\href@noop {} {\bibfield  {journal}
		{\bibinfo  {journal} {Annals of Mathematics}\ ,\ \bibinfo {pages} {325}}
		(\bibinfo {year} {1958}{\natexlab{a}})}\BibitemShut {NoStop}%
	\bibitem [{\citenamefont {Wigner}(1967)}]{wigner1967random}%
	\BibitemOpen
	\bibfield  {author} {\bibinfo {author} {\bibfnamefont {E.~P.}\ \bibnamefont
			{Wigner}},\ }\bibfield  {title} {\bibinfo {title} {Random matrices in
			physics},\ }\href@noop {} {\bibfield  {journal} {\bibinfo  {journal} {SIAM
				review}\ }\textbf {\bibinfo {volume} {9}},\ \bibinfo {pages} {1} (\bibinfo
		{year} {1967})}\BibitemShut {NoStop}%
	\bibitem [{\citenamefont {Dyson}(1953)}]{dyson1953dynamics}%
	\BibitemOpen
	\bibfield  {author} {\bibinfo {author} {\bibfnamefont {F.~J.}\ \bibnamefont
			{Dyson}},\ }\bibfield  {title} {\bibinfo {title} {The dynamics of a
			disordered linear chain},\ }\href@noop {} {\bibfield  {journal} {\bibinfo
			{journal} {Physical Review}\ }\textbf {\bibinfo {volume} {92}},\ \bibinfo
		{pages} {1331} (\bibinfo {year} {1953})}\BibitemShut {NoStop}%
	\bibitem [{\citenamefont {Dyson}(1962)}]{dyson1962brownian}%
	\BibitemOpen
	\bibfield  {author} {\bibinfo {author} {\bibfnamefont {F.~J.}\ \bibnamefont
			{Dyson}},\ }\bibfield  {title} {\bibinfo {title} {A brownian-motion model for
			the eigenvalues of a random matrix},\ }\href@noop {} {\bibfield  {journal}
		{\bibinfo  {journal} {Journal of Mathematical Physics}\ }\textbf {\bibinfo
			{volume} {3}},\ \bibinfo {pages} {1191} (\bibinfo {year} {1962})}\BibitemShut
	{NoStop}%
	\bibitem [{\citenamefont {Edouard}\ \emph {et~al.}(2006)\citenamefont
		{Edouard}, \citenamefont {Kazakov}, \citenamefont {Serban}, \citenamefont
		{Wiegmann},\ and\ \citenamefont {Zabrin}}]{RMPhysicsBook}%
	\BibitemOpen
	\bibinfo {editor} {\bibfnamefont {B.}~\bibnamefont {Edouard}}, \bibinfo
	{editor} {\bibfnamefont {V.}~\bibnamefont {Kazakov}}, \bibinfo {editor}
	{\bibfnamefont {D.}~\bibnamefont {Serban}}, \bibinfo {editor} {\bibfnamefont
		{P.}~\bibnamefont {Wiegmann}},\ and\ \bibinfo {editor} {\bibfnamefont
		{A.}~\bibnamefont {Zabrin}},\ eds.,\ \href@noop {} {\emph {\bibinfo {title}
			{Applications of Random Matrices in Physics}}}\ (\bibinfo  {publisher}
	{Springer},\ \bibinfo {address} {Amsterdam, Netherlands},\ \bibinfo {year}
	{2006})\BibitemShut {NoStop}%
	\bibitem [{\citenamefont {M{\'e}zard}\ \emph {et~al.}(1987)\citenamefont
		{M{\'e}zard}, \citenamefont {Parisi},\ and\ \citenamefont
		{Virasoro}}]{mezard1987}%
	\BibitemOpen
	\bibfield  {author} {\bibinfo {author} {\bibfnamefont {M.}~\bibnamefont
			{M{\'e}zard}}, \bibinfo {author} {\bibfnamefont {G.}~\bibnamefont {Parisi}},\
		and\ \bibinfo {author} {\bibfnamefont {M.}~\bibnamefont {Virasoro}},\
	}\href@noop {} {\emph {\bibinfo {title} {Spin glass theory and beyond: An
				Introduction to the Replica Method and Its Applications}}},\ Vol.~\bibinfo
	{volume} {9}\ (\bibinfo  {publisher} {World Scientific Publishing Company},\
	\bibinfo {address} {London},\ \bibinfo {year} {1987})\BibitemShut {NoStop}%
	\bibitem [{\citenamefont {May}(1972)}]{may}%
	\BibitemOpen
	\bibfield  {author} {\bibinfo {author} {\bibfnamefont {R.~M.}\ \bibnamefont
			{May}},\ }\bibfield  {title} {\bibinfo {title} {Will a large complex system
			be stable?},\ }\href@noop {} {\bibfield  {journal} {\bibinfo  {journal}
			{Nature}\ }\textbf {\bibinfo {volume} {238}},\ \bibinfo {pages} {413}
		(\bibinfo {year} {1972})}\BibitemShut {NoStop}%
	\bibitem [{\citenamefont {Allesina}\ and\ \citenamefont
		{Tang}(2015)}]{allesinatang1}%
	\BibitemOpen
	\bibfield  {author} {\bibinfo {author} {\bibfnamefont {S.}~\bibnamefont
			{Allesina}}\ and\ \bibinfo {author} {\bibfnamefont {S.}~\bibnamefont
			{Tang}},\ }\bibfield  {title} {\bibinfo {title} {The stability--complexity
			relationship at age 40: a random matrix perspective},\ }\href@noop {}
	{\bibfield  {journal} {\bibinfo  {journal} {Population Ecology}\ }\textbf
		{\bibinfo {volume} {57}},\ \bibinfo {pages} {63} (\bibinfo {year}
		{2015})}\BibitemShut {NoStop}%
	\bibitem [{\citenamefont {Allesina}\ and\ \citenamefont
		{Tang}(2012)}]{allesinatang2}%
	\BibitemOpen
	\bibfield  {author} {\bibinfo {author} {\bibfnamefont {S.}~\bibnamefont
			{Allesina}}\ and\ \bibinfo {author} {\bibfnamefont {S.}~\bibnamefont
			{Tang}},\ }\bibfield  {title} {\bibinfo {title} {Stability criteria for
			complex ecosystems},\ }\href@noop {} {\bibfield  {journal} {\bibinfo
			{journal} {Nature}\ }\textbf {\bibinfo {volume} {483}},\ \bibinfo {pages}
		{205} (\bibinfo {year} {2012})}\BibitemShut {NoStop}%
	\bibitem [{\citenamefont {Kuczala}\ and\ \citenamefont
		{Sharpee}(2016)}]{kuczala2016eigenvalue}%
	\BibitemOpen
	\bibfield  {author} {\bibinfo {author} {\bibfnamefont {A.}~\bibnamefont
			{Kuczala}}\ and\ \bibinfo {author} {\bibfnamefont {T.~O.}\ \bibnamefont
			{Sharpee}},\ }\bibfield  {title} {\bibinfo {title} {Eigenvalue spectra of
			large correlated random matrices},\ }\href@noop {} {\bibfield  {journal}
		{\bibinfo  {journal} {Physical Review E}\ }\textbf {\bibinfo {volume} {94}},\
		\bibinfo {pages} {050101} (\bibinfo {year} {2016})}\BibitemShut {NoStop}%
	\bibitem [{\citenamefont {Aljadeff}\ \emph {et~al.}(2015)\citenamefont
		{Aljadeff}, \citenamefont {Stern},\ and\ \citenamefont
		{Sharpee}}]{aljadeff2015transition}%
	\BibitemOpen
	\bibfield  {author} {\bibinfo {author} {\bibfnamefont {J.}~\bibnamefont
			{Aljadeff}}, \bibinfo {author} {\bibfnamefont {M.}~\bibnamefont {Stern}},\
		and\ \bibinfo {author} {\bibfnamefont {T.}~\bibnamefont {Sharpee}},\
	}\bibfield  {title} {\bibinfo {title} {Transition to chaos in random networks
			with cell-type-specific connectivity},\ }\href@noop {} {\bibfield  {journal}
		{\bibinfo  {journal} {Physical review letters}\ }\textbf {\bibinfo {volume}
			{114}},\ \bibinfo {pages} {088101} (\bibinfo {year} {2015})}\BibitemShut
	{NoStop}%
	\bibitem [{\citenamefont {Coolen}\ \emph {et~al.}(2004)\citenamefont {Coolen},
		\citenamefont {K\"uhn},\ and\ \citenamefont {Sollich}}]{Coolen_NN}%
	\BibitemOpen
	\bibfield  {author} {\bibinfo {author} {\bibfnamefont {A.~C.~C.}\
			\bibnamefont {Coolen}}, \bibinfo {author} {\bibfnamefont {R.}~\bibnamefont
			{K\"uhn}},\ and\ \bibinfo {author} {\bibfnamefont {P.}~\bibnamefont
			{Sollich}},\ }\href@noop {} {\emph {\bibinfo {title} {Theory of Neural
				Information Processing Systems}}}\ (\bibinfo  {publisher} {Oxford University
		Press},\ \bibinfo {address} {Oxford, UK},\ \bibinfo {year}
	{2004})\BibitemShut {NoStop}%
	\bibitem [{\citenamefont {Rajan}\ and\ \citenamefont
		{Abbott}(2006)}]{rajan2006eigenvalue}%
	\BibitemOpen
	\bibfield  {author} {\bibinfo {author} {\bibfnamefont {K.}~\bibnamefont
			{Rajan}}\ and\ \bibinfo {author} {\bibfnamefont {L.~F.}\ \bibnamefont
			{Abbott}},\ }\bibfield  {title} {\bibinfo {title} {Eigenvalue spectra of
			random matrices for neural networks},\ }\href@noop {} {\bibfield  {journal}
		{\bibinfo  {journal} {Physical review letters}\ }\textbf {\bibinfo {volume}
			{97}},\ \bibinfo {pages} {188104} (\bibinfo {year} {2006})}\BibitemShut
	{NoStop}%
	\bibitem [{\citenamefont {Ahmadian}\ \emph {et~al.}(2015)\citenamefont
		{Ahmadian}, \citenamefont {Fumarola},\ and\ \citenamefont
		{Miller}}]{ahmadian2015properties}%
	\BibitemOpen
	\bibfield  {author} {\bibinfo {author} {\bibfnamefont {Y.}~\bibnamefont
			{Ahmadian}}, \bibinfo {author} {\bibfnamefont {F.}~\bibnamefont {Fumarola}},\
		and\ \bibinfo {author} {\bibfnamefont {K.~D.}\ \bibnamefont {Miller}},\
	}\bibfield  {title} {\bibinfo {title} {Properties of networks with partially
			structured and partially random connectivity},\ }\href@noop {} {\bibfield
		{journal} {\bibinfo  {journal} {Physical Review E}\ }\textbf {\bibinfo
			{volume} {91}},\ \bibinfo {pages} {012820} (\bibinfo {year}
		{2015})}\BibitemShut {NoStop}%
	\bibitem [{\citenamefont {Laloux}\ \emph {et~al.}(2000)\citenamefont {Laloux},
		\citenamefont {Cizeau}, \citenamefont {Potters},\ and\ \citenamefont
		{Bouchaud}}]{laloux2000random}%
	\BibitemOpen
	\bibfield  {author} {\bibinfo {author} {\bibfnamefont {L.}~\bibnamefont
			{Laloux}}, \bibinfo {author} {\bibfnamefont {P.}~\bibnamefont {Cizeau}},
		\bibinfo {author} {\bibfnamefont {M.}~\bibnamefont {Potters}},\ and\ \bibinfo
		{author} {\bibfnamefont {J.-P.}\ \bibnamefont {Bouchaud}},\ }\bibfield
	{title} {\bibinfo {title} {Random matrix theory and financial correlations},\
	}\href@noop {} {\bibfield  {journal} {\bibinfo  {journal} {International
				Journal of Theoretical and Applied Finance}\ }\textbf {\bibinfo {volume}
			{3}},\ \bibinfo {pages} {391} (\bibinfo {year} {2000})}\BibitemShut {NoStop}%
	\bibitem [{\citenamefont {Sommers}\ \emph {et~al.}(1988)\citenamefont
		{Sommers}, \citenamefont {Crisanti}, \citenamefont {Sompolinsky},\ and\
		\citenamefont {Stein}}]{sommers}%
	\BibitemOpen
	\bibfield  {author} {\bibinfo {author} {\bibfnamefont {H.-J.}\ \bibnamefont
			{Sommers}}, \bibinfo {author} {\bibfnamefont {A.}~\bibnamefont {Crisanti}},
		\bibinfo {author} {\bibfnamefont {H.}~\bibnamefont {Sompolinsky}},\ and\
		\bibinfo {author} {\bibfnamefont {Y.}~\bibnamefont {Stein}},\ }\bibfield
	{title} {\bibinfo {title} {Spectrum of large random asymmetric matrices},\
	}\href@noop {} {\bibfield  {journal} {\bibinfo  {journal} {Physical Review
				Letters}\ }\textbf {\bibinfo {volume} {60}},\ \bibinfo {pages} {1895}
		(\bibinfo {year} {1988})}\BibitemShut {NoStop}%
	\bibitem [{\citenamefont {Girko}(1986)}]{girko1986elliptic}%
	\BibitemOpen
	\bibfield  {author} {\bibinfo {author} {\bibfnamefont {V.}~\bibnamefont
			{Girko}},\ }\bibfield  {title} {\bibinfo {title} {Elliptic law},\ }\href@noop
	{} {\bibfield  {journal} {\bibinfo  {journal} {Theory of Probability \& Its
				Applications}\ }\textbf {\bibinfo {volume} {30}},\ \bibinfo {pages} {677}
		(\bibinfo {year} {1986})}\BibitemShut {NoStop}%
	\bibitem [{\citenamefont {O'Rourke}\ \emph {et~al.}(2014)\citenamefont
		{O'Rourke}, \citenamefont {Renfrew} \emph {et~al.}}]{orourke}%
	\BibitemOpen
	\bibfield  {author} {\bibinfo {author} {\bibfnamefont {S.}~\bibnamefont
			{O'Rourke}}, \bibinfo {author} {\bibfnamefont {D.}~\bibnamefont {Renfrew}},
		\emph {et~al.},\ }\bibfield  {title} {\bibinfo {title} {Low rank
			perturbations of large elliptic random matrices},\ }\href@noop {} {\bibfield
		{journal} {\bibinfo  {journal} {Electronic Journal of Probability}\ }\textbf
		{\bibinfo {volume} {19}} (\bibinfo {year} {2014})}\BibitemShut {NoStop}%
	\bibitem [{\citenamefont {Benaych-Georges}\ and\ \citenamefont
		{Nadakuditi}(2011)}]{benaych2011eigenvalues}%
	\BibitemOpen
	\bibfield  {author} {\bibinfo {author} {\bibfnamefont {F.}~\bibnamefont
			{Benaych-Georges}}\ and\ \bibinfo {author} {\bibfnamefont {R.~R.}\
			\bibnamefont {Nadakuditi}},\ }\bibfield  {title} {\bibinfo {title} {The
			eigenvalues and eigenvectors of finite, low rank perturbations of large
			random matrices},\ }\href@noop {} {\bibfield  {journal} {\bibinfo  {journal}
			{Advances in Mathematics}\ }\textbf {\bibinfo {volume} {227}},\ \bibinfo
		{pages} {494} (\bibinfo {year} {2011})}\BibitemShut {NoStop}%
	\bibitem [{\citenamefont {Edwards}\ and\ \citenamefont
		{Jones}(1976)}]{edwardsjones}%
	\BibitemOpen
	\bibfield  {author} {\bibinfo {author} {\bibfnamefont {S.~F.}\ \bibnamefont
			{Edwards}}\ and\ \bibinfo {author} {\bibfnamefont {R.~C.}\ \bibnamefont
			{Jones}},\ }\bibfield  {title} {\bibinfo {title} {The eigenvalue spectrum of
			a large symmetric random matrix},\ }\href@noop {} {\bibfield  {journal}
		{\bibinfo  {journal} {Journal of Physics A: Mathematical and General}\
		}\textbf {\bibinfo {volume} {9}},\ \bibinfo {pages} {1595} (\bibinfo {year}
		{1976})}\BibitemShut {NoStop}%
	\bibitem [{\citenamefont {Baron}(2022)}]{baron2022networks}%
	\BibitemOpen
	\bibfield  {author} {\bibinfo {author} {\bibfnamefont {J.~W.}\ \bibnamefont
			{Baron}},\ }\bibfield  {title} {\bibinfo {title} {Eigenvalue spectra and
			stability of directed complex networks},\ }\href@noop {} {\bibfield
		{journal} {\bibinfo  {journal} {Phys. Rev. E}\ }\textbf {\bibinfo {volume}
			{106}},\ \bibinfo {pages} {064302} (\bibinfo {year} {2022})}\BibitemShut
	{NoStop}%
	\bibitem [{\citenamefont {Metz}\ and\ \citenamefont
		{Silva}(2020)}]{metz2020spectral}%
	\BibitemOpen
	\bibfield  {author} {\bibinfo {author} {\bibfnamefont {F.~L.}\ \bibnamefont
			{Metz}}\ and\ \bibinfo {author} {\bibfnamefont {J.~D.}\ \bibnamefont
			{Silva}},\ }\bibfield  {title} {\bibinfo {title} {Spectral density of dense
			random networks and the breakdown of the wigner semicircle law},\ }\href@noop
	{} {\bibfield  {journal} {\bibinfo  {journal} {Phys. Rev. Research}\ }\textbf
		{\bibinfo {volume} {2}},\ \bibinfo {pages} {043116} (\bibinfo {year}
		{2020})}\BibitemShut {NoStop}%
	\bibitem [{\citenamefont {Rodgers}\ \emph {et~al.}(2005)\citenamefont
		{Rodgers}, \citenamefont {Austin}, \citenamefont {Kahng},\ and\ \citenamefont
		{Kim}}]{rodgers2005eigenvalue}%
	\BibitemOpen
	\bibfield  {author} {\bibinfo {author} {\bibfnamefont {G.~J.}\ \bibnamefont
			{Rodgers}}, \bibinfo {author} {\bibfnamefont {K.}~\bibnamefont {Austin}},
		\bibinfo {author} {\bibfnamefont {B.}~\bibnamefont {Kahng}},\ and\ \bibinfo
		{author} {\bibfnamefont {D.}~\bibnamefont {Kim}},\ }\bibfield  {title}
	{\bibinfo {title} {Eigenvalue spectra of complex networks},\ }\href@noop {}
	{\bibfield  {journal} {\bibinfo  {journal} {Journal of Physics A:
				Mathematical and General}\ }\textbf {\bibinfo {volume} {38}},\ \bibinfo
		{pages} {9431} (\bibinfo {year} {2005})}\BibitemShut {NoStop}%
	\bibitem [{\citenamefont {Baron}\ and\ \citenamefont
		{Galla}(2020)}]{baron2020dispersal}%
	\BibitemOpen
	\bibfield  {author} {\bibinfo {author} {\bibfnamefont {J.~W.}\ \bibnamefont
			{Baron}}\ and\ \bibinfo {author} {\bibfnamefont {T.}~\bibnamefont {Galla}},\
	}\bibfield  {title} {\bibinfo {title} {Dispersal-induced instability in
			complex ecosystems},\ }\href@noop {} {\bibfield  {journal} {\bibinfo
			{journal} {Nature communications}\ }\textbf {\bibinfo {volume} {11}},\
		\bibinfo {pages} {1} (\bibinfo {year} {2020})}\BibitemShut {NoStop}%
	\bibitem [{\citenamefont {Grilli}\ \emph {et~al.}(2017)\citenamefont {Grilli},
		\citenamefont {Adorisio}, \citenamefont {Suweis}, \citenamefont
		{Barab{\'a}s}, \citenamefont {Banavar}, \citenamefont {Allesina},\ and\
		\citenamefont {Maritan}}]{grilli2017}%
	\BibitemOpen
	\bibfield  {author} {\bibinfo {author} {\bibfnamefont {J.}~\bibnamefont
			{Grilli}}, \bibinfo {author} {\bibfnamefont {M.}~\bibnamefont {Adorisio}},
		\bibinfo {author} {\bibfnamefont {S.}~\bibnamefont {Suweis}}, \bibinfo
		{author} {\bibfnamefont {G.}~\bibnamefont {Barab{\'a}s}}, \bibinfo {author}
		{\bibfnamefont {J.~R.}\ \bibnamefont {Banavar}}, \bibinfo {author}
		{\bibfnamefont {S.}~\bibnamefont {Allesina}},\ and\ \bibinfo {author}
		{\bibfnamefont {A.}~\bibnamefont {Maritan}},\ }\bibfield  {title} {\bibinfo
		{title} {Feasibility and coexistence of large ecological communities},\
	}\href@noop {} {\bibfield  {journal} {\bibinfo  {journal} {Nature
				Communications}\ }\textbf {\bibinfo {volume} {8}},\ \bibinfo {pages} {14389}
		(\bibinfo {year} {2017})}\BibitemShut {NoStop}%
	\bibitem [{\citenamefont {Allesina}\ \emph
		{et~al.}(2015{\natexlab{a}})\citenamefont {Allesina}, \citenamefont {Grilli},
		\citenamefont {Barab{\'a}s}, \citenamefont {Tang}, \citenamefont {Aljadeff},\
		and\ \citenamefont {Maritan}}]{allesina2015}%
	\BibitemOpen
	\bibfield  {author} {\bibinfo {author} {\bibfnamefont {S.}~\bibnamefont
			{Allesina}}, \bibinfo {author} {\bibfnamefont {J.}~\bibnamefont {Grilli}},
		\bibinfo {author} {\bibfnamefont {G.}~\bibnamefont {Barab{\'a}s}}, \bibinfo
		{author} {\bibfnamefont {S.}~\bibnamefont {Tang}}, \bibinfo {author}
		{\bibfnamefont {J.}~\bibnamefont {Aljadeff}},\ and\ \bibinfo {author}
		{\bibfnamefont {A.}~\bibnamefont {Maritan}},\ }\bibfield  {title} {\bibinfo
		{title} {Predicting the stability of large structured food webs},\
	}\href@noop {} {\bibfield  {journal} {\bibinfo  {journal} {Nature
				Communications}\ }\textbf {\bibinfo {volume} {6}},\ \bibinfo {pages} {1}
		(\bibinfo {year} {2015}{\natexlab{a}})}\BibitemShut {NoStop}%
	\bibitem [{\citenamefont {Aceituno}\ \emph {et~al.}(2019)\citenamefont
		{Aceituno}, \citenamefont {Rogers},\ and\ \citenamefont
		{Schomerus}}]{aceitunorogersschomerus}%
	\BibitemOpen
	\bibfield  {author} {\bibinfo {author} {\bibfnamefont {P.~V.}\ \bibnamefont
			{Aceituno}}, \bibinfo {author} {\bibfnamefont {T.}~\bibnamefont {Rogers}},\
		and\ \bibinfo {author} {\bibfnamefont {H.}~\bibnamefont {Schomerus}},\
	}\bibfield  {title} {\bibinfo {title} {Universal hypotrochoidic law for
			random matrices with cyclic correlations},\ }\href@noop {} {\bibfield
		{journal} {\bibinfo  {journal} {Physical Review E}\ }\textbf {\bibinfo
			{volume} {100}},\ \bibinfo {pages} {010302} (\bibinfo {year}
		{2019})}\BibitemShut {NoStop}%
	\bibitem [{\citenamefont {Baron}\ \emph {et~al.}(2022)\citenamefont {Baron},
		\citenamefont {Jewell}, \citenamefont {Ryder},\ and\ \citenamefont
		{Galla}}]{baron2022eigenvalues}%
	\BibitemOpen
	\bibfield  {author} {\bibinfo {author} {\bibfnamefont {J.~W.}\ \bibnamefont
			{Baron}}, \bibinfo {author} {\bibfnamefont {T.~J.}\ \bibnamefont {Jewell}},
		\bibinfo {author} {\bibfnamefont {C.}~\bibnamefont {Ryder}},\ and\ \bibinfo
		{author} {\bibfnamefont {T.}~\bibnamefont {Galla}},\ }\bibfield  {title}
	{\bibinfo {title} {Eigenvalues of random matrices with generalized
			correlations: A path integral approach},\ }\href@noop {} {\bibfield
		{journal} {\bibinfo  {journal} {Phys. Rev. Lett.}\ }\textbf {\bibinfo
			{volume} {128}},\ \bibinfo {pages} {120601} (\bibinfo {year}
		{2022})}\BibitemShut {NoStop}%
	\bibitem [{\citenamefont {Metz}\ \emph {et~al.}(2019)\citenamefont {Metz},
		\citenamefont {Neri},\ and\ \citenamefont {Rogers}}]{metz2019spectral}%
	\BibitemOpen
	\bibfield  {author} {\bibinfo {author} {\bibfnamefont {F.~L.}\ \bibnamefont
			{Metz}}, \bibinfo {author} {\bibfnamefont {I.}~\bibnamefont {Neri}},\ and\
		\bibinfo {author} {\bibfnamefont {T.}~\bibnamefont {Rogers}},\ }\bibfield
	{title} {\bibinfo {title} {Spectral theory of sparse non-hermitian random
			matrices},\ }\href@noop {} {\bibfield  {journal} {\bibinfo  {journal}
			{Journal of Physics A: Mathematical and Theoretical}\ }\textbf {\bibinfo
			{volume} {52}},\ \bibinfo {pages} {434003} (\bibinfo {year}
		{2019})}\BibitemShut {NoStop}%
	\bibitem [{\citenamefont {Rogers}\ and\ \citenamefont
		{Castillo}(2009)}]{rogers2009cavity}%
	\BibitemOpen
	\bibfield  {author} {\bibinfo {author} {\bibfnamefont {T.}~\bibnamefont
			{Rogers}}\ and\ \bibinfo {author} {\bibfnamefont {I.~P.}\ \bibnamefont
			{Castillo}},\ }\bibfield  {title} {\bibinfo {title} {Cavity approach to the
			spectral density of non-hermitian sparse matrices},\ }\href@noop {}
	{\bibfield  {journal} {\bibinfo  {journal} {Physical Review E}\ }\textbf
		{\bibinfo {volume} {79}},\ \bibinfo {pages} {012101} (\bibinfo {year}
		{2009})}\BibitemShut {NoStop}%
	\bibitem [{\citenamefont {Susca}\ \emph {et~al.}(2021)\citenamefont {Susca},
		\citenamefont {Vivo},\ and\ \citenamefont {K{\"u}hn}}]{susca2021cavity}%
	\BibitemOpen
	\bibfield  {author} {\bibinfo {author} {\bibfnamefont {V.~A.}\ \bibnamefont
			{Susca}}, \bibinfo {author} {\bibfnamefont {P.}~\bibnamefont {Vivo}},\ and\
		\bibinfo {author} {\bibfnamefont {R.}~\bibnamefont {K{\"u}hn}},\ }\bibfield
	{title} {\bibinfo {title} {Cavity and replica methods for the spectral
			density of sparse symmetric random matrices},\ }\href@noop {} {\bibfield
		{journal} {\bibinfo  {journal} {SciPost Physics Lecture Notes}\ ,\ \bibinfo
			{pages} {033}} (\bibinfo {year} {2021})}\BibitemShut {NoStop}%
	\bibitem [{\citenamefont {Cavagna}\ \emph {et~al.}(1999)\citenamefont
		{Cavagna}, \citenamefont {Giardina},\ and\ \citenamefont
		{Parisi}}]{cavagna1999analytic}%
	\BibitemOpen
	\bibfield  {author} {\bibinfo {author} {\bibfnamefont {A.}~\bibnamefont
			{Cavagna}}, \bibinfo {author} {\bibfnamefont {I.}~\bibnamefont {Giardina}},\
		and\ \bibinfo {author} {\bibfnamefont {G.}~\bibnamefont {Parisi}},\
	}\bibfield  {title} {\bibinfo {title} {Analytic computation of the
			instantaneous normal modes spectrum in low-density liquids},\ }\href@noop {}
	{\bibfield  {journal} {\bibinfo  {journal} {Phys. Rev. Lett.}\ }\textbf
		{\bibinfo {volume} {83}},\ \bibinfo {pages} {108} (\bibinfo {year}
		{1999})}\BibitemShut {NoStop}%
	\bibitem [{\citenamefont {Hatano}\ and\ \citenamefont
		{Nelson}(1996)}]{hatano1996localization}%
	\BibitemOpen
	\bibfield  {author} {\bibinfo {author} {\bibfnamefont {N.}~\bibnamefont
			{Hatano}}\ and\ \bibinfo {author} {\bibfnamefont {D.~R.}\ \bibnamefont
			{Nelson}},\ }\bibfield  {title} {\bibinfo {title} {Localization transitions
			in non-hermitian quantum mechanics},\ }\href@noop {} {\bibfield  {journal}
		{\bibinfo  {journal} {Physical review letters}\ }\textbf {\bibinfo {volume}
			{77}},\ \bibinfo {pages} {570} (\bibinfo {year} {1996})}\BibitemShut
	{NoStop}%
	\bibitem [{\citenamefont {Biroli}\ and\ \citenamefont
		{Monasson}(1999)}]{biroli1999single}%
	\BibitemOpen
	\bibfield  {author} {\bibinfo {author} {\bibfnamefont {G.}~\bibnamefont
			{Biroli}}\ and\ \bibinfo {author} {\bibfnamefont {R.}~\bibnamefont
			{Monasson}},\ }\bibfield  {title} {\bibinfo {title} {A single defect
			approximation for localized states on random lattices},\ }\href@noop {}
	{\bibfield  {journal} {\bibinfo  {journal} {Journal of Physics A:
				Mathematical and General}\ }\textbf {\bibinfo {volume} {32}},\ \bibinfo
		{pages} {L255} (\bibinfo {year} {1999})}\BibitemShut {NoStop}%
	\bibitem [{\citenamefont {Baron}(2021)}]{baron2021persistent}%
	\BibitemOpen
	\bibfield  {author} {\bibinfo {author} {\bibfnamefont {J.~W.}\ \bibnamefont
			{Baron}},\ }\bibfield  {title} {\bibinfo {title} {Persistent individual bias
			in a voter model with quenched disorder},\ }\href@noop {} {\bibfield
		{journal} {\bibinfo  {journal} {Phys. Rev. E}\ }\textbf {\bibinfo {volume}
			{103}},\ \bibinfo {pages} {052309} (\bibinfo {year} {2021})}\BibitemShut
	{NoStop}%
	\bibitem [{\citenamefont {Chalker}\ and\ \citenamefont
		{Siak}(1990)}]{chalker1990anderson}%
	\BibitemOpen
	\bibfield  {author} {\bibinfo {author} {\bibfnamefont {J.}~\bibnamefont
			{Chalker}}\ and\ \bibinfo {author} {\bibfnamefont {S.}~\bibnamefont {Siak}},\
	}\bibfield  {title} {\bibinfo {title} {Anderson localisation on a cayley
			tree: a new model with a simple solution},\ }\href@noop {} {\bibfield
		{journal} {\bibinfo  {journal} {Journal of Physics: Condensed Matter}\
		}\textbf {\bibinfo {volume} {2}},\ \bibinfo {pages} {2671} (\bibinfo {year}
		{1990})}\BibitemShut {NoStop}%
	\bibitem [{\citenamefont {Harris}\ and\ \citenamefont
		{Lubensky}(1981)}]{harris1981mean}%
	\BibitemOpen
	\bibfield  {author} {\bibinfo {author} {\bibfnamefont {A.~B.}\ \bibnamefont
			{Harris}}\ and\ \bibinfo {author} {\bibfnamefont {T.~C.}\ \bibnamefont
			{Lubensky}},\ }\bibfield  {title} {\bibinfo {title} {Mean-field theory and
			$\ensuremath{\epsilon}$ expansion for anderson localization},\ }\href@noop {}
	{\bibfield  {journal} {\bibinfo  {journal} {Phys. Rev. B}\ }\textbf {\bibinfo
			{volume} {23}},\ \bibinfo {pages} {2640} (\bibinfo {year}
		{1981})}\BibitemShut {NoStop}%
	\bibitem [{\citenamefont {Kim}\ and\ \citenamefont
		{Harris}(1985)}]{kim1985density}%
	\BibitemOpen
	\bibfield  {author} {\bibinfo {author} {\bibfnamefont {Y.}~\bibnamefont
			{Kim}}\ and\ \bibinfo {author} {\bibfnamefont {A.~B.}\ \bibnamefont
			{Harris}},\ }\bibfield  {title} {\bibinfo {title} {Density of states of the
			random-hopping model on a cayley tree},\ }\href@noop {} {\bibfield  {journal}
		{\bibinfo  {journal} {Phys. Rev. B}\ }\textbf {\bibinfo {volume} {31}},\
		\bibinfo {pages} {7393} (\bibinfo {year} {1985})}\BibitemShut {NoStop}%
	\bibitem [{\citenamefont {Marcus}\ \emph {et~al.}(2022)\citenamefont {Marcus},
		\citenamefont {Turner},\ and\ \citenamefont {Bunin}}]{marcus2022local}%
	\BibitemOpen
	\bibfield  {author} {\bibinfo {author} {\bibfnamefont {S.}~\bibnamefont
			{Marcus}}, \bibinfo {author} {\bibfnamefont {A.~M.}\ \bibnamefont {Turner}},\
		and\ \bibinfo {author} {\bibfnamefont {G.}~\bibnamefont {Bunin}},\ }\bibfield
	{title} {\bibinfo {title} {Local and collective transitions in
			sparsely-interacting ecological communities},\ }\href@noop {} {\bibfield
		{journal} {\bibinfo  {journal} {PLoS computational biology}\ }\textbf
		{\bibinfo {volume} {18}},\ \bibinfo {pages} {e1010274} (\bibinfo {year}
		{2022})}\BibitemShut {NoStop}%
	\bibitem [{\citenamefont {Mambuca}\ \emph {et~al.}(2022)\citenamefont
		{Mambuca}, \citenamefont {Cammarota},\ and\ \citenamefont
		{Neri}}]{mambuca2022dynamical}%
	\BibitemOpen
	\bibfield  {author} {\bibinfo {author} {\bibfnamefont {A.~M.}\ \bibnamefont
			{Mambuca}}, \bibinfo {author} {\bibfnamefont {C.}~\bibnamefont {Cammarota}},\
		and\ \bibinfo {author} {\bibfnamefont {I.}~\bibnamefont {Neri}},\ }\bibfield
	{title} {\bibinfo {title} {Dynamical systems on large networks with
			predator-prey interactions are stable and exhibit oscillations},\ }\href@noop
	{} {\bibfield  {journal} {\bibinfo  {journal} {Phys. Rev. E}\ }\textbf
		{\bibinfo {volume} {105}},\ \bibinfo {pages} {014305} (\bibinfo {year}
		{2022})}\BibitemShut {NoStop}%
	\bibitem [{\citenamefont {Valigi}\ \emph {et~al.}(2023)\citenamefont {Valigi},
		\citenamefont {Neri},\ and\ \citenamefont {Cammarota}}]{valigi2023local}%
	\BibitemOpen
	\bibfield  {author} {\bibinfo {author} {\bibfnamefont {P.}~\bibnamefont
			{Valigi}}, \bibinfo {author} {\bibfnamefont {I.}~\bibnamefont {Neri}},\ and\
		\bibinfo {author} {\bibfnamefont {C.}~\bibnamefont {Cammarota}},\ }\bibfield
	{title} {\bibinfo {title} {Local sign stability and its implications for
			spectra of sparse random graphs and stability of ecosystems},\ }\href@noop {}
	{\bibfield  {journal} {\bibinfo  {journal} {arXiv preprint arXiv:2303.09897}\
		} (\bibinfo {year} {2023})}\BibitemShut {NoStop}%
	\bibitem [{\citenamefont {Akjouj}\ and\ \citenamefont
		{Najim}(2022)}]{akjouj2022feasibility}%
	\BibitemOpen
	\bibfield  {author} {\bibinfo {author} {\bibfnamefont {I.}~\bibnamefont
			{Akjouj}}\ and\ \bibinfo {author} {\bibfnamefont {J.}~\bibnamefont {Najim}},\
	}\bibfield  {title} {\bibinfo {title} {Feasibility of sparse large
			lotka-volterra ecosystems},\ }\href@noop {} {\bibfield  {journal} {\bibinfo
			{journal} {Journal of Mathematical Biology}\ }\textbf {\bibinfo {volume}
			{85}},\ \bibinfo {pages} {66} (\bibinfo {year} {2022})}\BibitemShut {NoStop}%
	\bibitem [{\citenamefont {Tao}\ \emph {et~al.}(2010)\citenamefont {Tao},
		\citenamefont {Vu}, \citenamefont {Krishnapur} \emph
		{et~al.}}]{taovukrishnapur2010}%
	\BibitemOpen
	\bibfield  {author} {\bibinfo {author} {\bibfnamefont {T.}~\bibnamefont
			{Tao}}, \bibinfo {author} {\bibfnamefont {V.}~\bibnamefont {Vu}}, \bibinfo
		{author} {\bibfnamefont {M.}~\bibnamefont {Krishnapur}}, \emph {et~al.},\
	}\bibfield  {title} {\bibinfo {title} {Random matrices: Universality of
			{ESD}s and the circular law},\ }\href@noop {} {\bibfield  {journal} {\bibinfo
			{journal} {The Annals of Probability}\ }\textbf {\bibinfo {volume} {38}},\
		\bibinfo {pages} {2023} (\bibinfo {year} {2010})}\BibitemShut {NoStop}%
	\bibitem [{\citenamefont {Bai}\ and\ \citenamefont
		{Silverstein}(2010)}]{bai2010spectral}%
	\BibitemOpen
	\bibfield  {author} {\bibinfo {author} {\bibfnamefont {Z.}~\bibnamefont
			{Bai}}\ and\ \bibinfo {author} {\bibfnamefont {J.~W.}\ \bibnamefont
			{Silverstein}},\ }\href@noop {} {\emph {\bibinfo {title} {Spectral analysis
				of large dimensional random matrices}}},\ Vol.~\bibinfo {volume} {20}\
	(\bibinfo  {publisher} {Springer},\ \bibinfo {year} {2010})\BibitemShut
	{NoStop}%
	\bibitem [{\citenamefont {Nguyen}\ and\ \citenamefont
		{O’Rourke}(2015)}]{nguyen2015elliptic}%
	\BibitemOpen
	\bibfield  {author} {\bibinfo {author} {\bibfnamefont {H.~H.}\ \bibnamefont
			{Nguyen}}\ and\ \bibinfo {author} {\bibfnamefont {S.}~\bibnamefont
			{O’Rourke}},\ }\bibfield  {title} {\bibinfo {title} {The elliptic law},\
	}\href@noop {} {\bibfield  {journal} {\bibinfo  {journal} {International
				Mathematics Research Notices}\ }\textbf {\bibinfo {volume} {2015}},\ \bibinfo
		{pages} {7620} (\bibinfo {year} {2015})}\BibitemShut {NoStop}%
	\bibitem [{\citenamefont {Janik}\ \emph {et~al.}(1997)\citenamefont {Janik},
		\citenamefont {Nowak}, \citenamefont {Papp},\ and\ \citenamefont
		{Zahed}}]{JANIK1997603}%
	\BibitemOpen
	\bibfield  {author} {\bibinfo {author} {\bibfnamefont {R.~A.}\ \bibnamefont
			{Janik}}, \bibinfo {author} {\bibfnamefont {M.~A.}\ \bibnamefont {Nowak}},
		\bibinfo {author} {\bibfnamefont {G.}~\bibnamefont {Papp}},\ and\ \bibinfo
		{author} {\bibfnamefont {I.}~\bibnamefont {Zahed}},\ }\bibfield  {title}
	{\bibinfo {title} {Non-hermitian random matrix models},\ }\href@noop {}
	{\bibfield  {journal} {\bibinfo  {journal} {Nuclear Physics B}\ }\textbf
		{\bibinfo {volume} {501}},\ \bibinfo {pages} {603} (\bibinfo {year}
		{1997})}\BibitemShut {NoStop}%
	\bibitem [{\citenamefont {Fyodorov}\ and\ \citenamefont
		{Sommers}(2003)}]{fyodorov2003random}%
	\BibitemOpen
	\bibfield  {author} {\bibinfo {author} {\bibfnamefont {Y.~V.}\ \bibnamefont
			{Fyodorov}}\ and\ \bibinfo {author} {\bibfnamefont {H.-J.}\ \bibnamefont
			{Sommers}},\ }\bibfield  {title} {\bibinfo {title} {Random matrices close to
			hermitian or unitary: overview of methods and results},\ }\href@noop {}
	{\bibfield  {journal} {\bibinfo  {journal} {Journal of Physics A:
				Mathematical and General}\ }\textbf {\bibinfo {volume} {36}},\ \bibinfo
		{pages} {3303} (\bibinfo {year} {2003})}\BibitemShut {NoStop}%
	\bibitem [{\citenamefont {Hertz}\ \emph {et~al.}(2016)\citenamefont {Hertz},
		\citenamefont {Roudi},\ and\ \citenamefont {Sollich}}]{hertz2016path}%
	\BibitemOpen
	\bibfield  {author} {\bibinfo {author} {\bibfnamefont {J.~A.}\ \bibnamefont
			{Hertz}}, \bibinfo {author} {\bibfnamefont {Y.}~\bibnamefont {Roudi}},\ and\
		\bibinfo {author} {\bibfnamefont {P.}~\bibnamefont {Sollich}},\ }\bibfield
	{title} {\bibinfo {title} {Path integral methods for the dynamics of
			stochastic and disordered systems},\ }\href@noop {} {\bibfield  {journal}
		{\bibinfo  {journal} {Journal of Physics A: Mathematical and Theoretical}\
		}\textbf {\bibinfo {volume} {50}},\ \bibinfo {pages} {033001} (\bibinfo
		{year} {2016})}\BibitemShut {NoStop}%
	\bibitem [{\citenamefont {Semerjian}\ and\ \citenamefont
		{Cugliandolo}(2002)}]{semerjian2002sparse}%
	\BibitemOpen
	\bibfield  {author} {\bibinfo {author} {\bibfnamefont {G.}~\bibnamefont
			{Semerjian}}\ and\ \bibinfo {author} {\bibfnamefont {L.~F.}\ \bibnamefont
			{Cugliandolo}},\ }\bibfield  {title} {\bibinfo {title} {Sparse random
			matrices: the eigenvalue spectrum revisited},\ }\href@noop {} {\bibfield
		{journal} {\bibinfo  {journal} {Journal of Physics A: Mathematical and
				General}\ }\textbf {\bibinfo {volume} {35}},\ \bibinfo {pages} {4837}
		(\bibinfo {year} {2002})}\BibitemShut {NoStop}%
	\bibitem [{\citenamefont {K{\"u}hn}(2008)}]{kuhn2008spectra}%
	\BibitemOpen
	\bibfield  {author} {\bibinfo {author} {\bibfnamefont {R.}~\bibnamefont
			{K{\"u}hn}},\ }\bibfield  {title} {\bibinfo {title} {Spectra of sparse random
			matrices},\ }\href@noop {} {\bibfield  {journal} {\bibinfo  {journal}
			{Journal of Physics A: Mathematical and Theoretical}\ }\textbf {\bibinfo
			{volume} {41}},\ \bibinfo {pages} {295002} (\bibinfo {year}
		{2008})}\BibitemShut {NoStop}%
	\bibitem [{\citenamefont {Neri}\ and\ \citenamefont
		{Metz}(2012)}]{neri2012spectra}%
	\BibitemOpen
	\bibfield  {author} {\bibinfo {author} {\bibfnamefont {I.}~\bibnamefont
			{Neri}}\ and\ \bibinfo {author} {\bibfnamefont {F.~L.}\ \bibnamefont
			{Metz}},\ }\bibfield  {title} {\bibinfo {title} {Spectra of sparse
			non-hermitian random matrices: An analytical solution},\ }\href@noop {}
	{\bibfield  {journal} {\bibinfo  {journal} {Phys. Rev. Lett.}\ }\textbf
		{\bibinfo {volume} {109}},\ \bibinfo {pages} {030602} (\bibinfo {year}
		{2012})}\BibitemShut {NoStop}%
	\bibitem [{\citenamefont {Bray}\ and\ \citenamefont {Moore}(1979)}]{braymoore}%
	\BibitemOpen
	\bibfield  {author} {\bibinfo {author} {\bibfnamefont {A.~J.}\ \bibnamefont
			{Bray}}\ and\ \bibinfo {author} {\bibfnamefont {M.~A.}\ \bibnamefont
			{Moore}},\ }\bibfield  {title} {\bibinfo {title} {Evidence for massless modes
			in the `solvable model' of a spin glass},\ }\href@noop {} {\bibfield
		{journal} {\bibinfo  {journal} {Journal of Physics C: Solid State Physics}\
		}\textbf {\bibinfo {volume} {12}},\ \bibinfo {pages} {L441} (\bibinfo {year}
		{1979})}\BibitemShut {NoStop}%
	\bibitem [{\citenamefont {De~Dominicis}(1978)}]{dedominicis1978dynamics}%
	\BibitemOpen
	\bibfield  {author} {\bibinfo {author} {\bibfnamefont {C.}~\bibnamefont
			{De~Dominicis}},\ }\bibfield  {title} {\bibinfo {title} {Dynamics as a
			substitute for replicas in systems with quenched random impurities},\
	}\href@noop {} {\bibfield  {journal} {\bibinfo  {journal} {Phys. Rev. B}\
		}\textbf {\bibinfo {volume} {18}},\ \bibinfo {pages} {4913} (\bibinfo {year}
		{1978})}\BibitemShut {NoStop}%
	\bibitem [{\citenamefont {Rodgers}\ and\ \citenamefont
		{Bray}(1988)}]{rodgers1988density}%
	\BibitemOpen
	\bibfield  {author} {\bibinfo {author} {\bibfnamefont {G.~J.}\ \bibnamefont
			{Rodgers}}\ and\ \bibinfo {author} {\bibfnamefont {A.~J.}\ \bibnamefont
			{Bray}},\ }\bibfield  {title} {\bibinfo {title} {Density of states of a
			sparse random matrix},\ }\href@noop {} {\bibfield  {journal} {\bibinfo
			{journal} {Physical Review B}\ }\textbf {\bibinfo {volume} {37}},\ \bibinfo
		{pages} {3557} (\bibinfo {year} {1988})}\BibitemShut {NoStop}%
	\bibitem [{\citenamefont {Akara-pipattana}\ and\ \citenamefont
		{Evnin}(2022)}]{akara2022random}%
	\BibitemOpen
	\bibfield  {author} {\bibinfo {author} {\bibfnamefont {P.}~\bibnamefont
			{Akara-pipattana}}\ and\ \bibinfo {author} {\bibfnamefont {O.}~\bibnamefont
			{Evnin}},\ }\bibfield  {title} {\bibinfo {title} {Random matrices with row
			constraints and eigenvalue distributions of graph laplacians},\ }\href@noop
	{} {\bibfield  {journal} {\bibinfo  {journal} {Journal of Physics A:
				Mathematical and Theoretical}\ } (\bibinfo {year} {2022})}\BibitemShut
	{NoStop}%
	\bibitem [{\citenamefont {Efetov}(1999)}]{efetov1999supersymmetry}%
	\BibitemOpen
	\bibfield  {author} {\bibinfo {author} {\bibfnamefont {K.}~\bibnamefont
			{Efetov}},\ }\href@noop {} {\emph {\bibinfo {title} {Supersymmetry in
				disorder and chaos}}}\ (\bibinfo  {publisher} {Cambridge university press},\
	\bibinfo {year} {1999})\BibitemShut {NoStop}%
	\bibitem [{\citenamefont {Azaele}\ and\ \citenamefont
		{Maritan}(2024)}]{azaele2024generalized}%
	\BibitemOpen
	\bibfield  {author} {\bibinfo {author} {\bibfnamefont {S.}~\bibnamefont
			{Azaele}}\ and\ \bibinfo {author} {\bibfnamefont {A.}~\bibnamefont
			{Maritan}},\ }\bibfield  {title} {\bibinfo {title} {Generalized dynamical
			mean field theory for non-gaussian interactions},\ }\href@noop {} {\bibfield
		{journal} {\bibinfo  {journal} {Physical Review Letters}\ }\textbf {\bibinfo
			{volume} {133}},\ \bibinfo {pages} {127401} (\bibinfo {year}
		{2024})}\BibitemShut {NoStop}%
	\bibitem [{\citenamefont {Akemann}\ \emph {et~al.}(2021)\citenamefont
		{Akemann}, \citenamefont {Byun},\ and\ \citenamefont
		{Kang}}]{akemann2021non}%
	\BibitemOpen
	\bibfield  {author} {\bibinfo {author} {\bibfnamefont {G.}~\bibnamefont
			{Akemann}}, \bibinfo {author} {\bibfnamefont {S.-S.}\ \bibnamefont {Byun}},\
		and\ \bibinfo {author} {\bibfnamefont {N.-G.}\ \bibnamefont {Kang}},\
	}\bibfield  {title} {\bibinfo {title} {A non-hermitian generalisation of the
			marchenko--pastur distribution: from the circular law to multi-criticality},\
	}in\ \href@noop {} {\emph {\bibinfo {booktitle} {Annales Henri
				Poincar{\'e}}}},\ Vol.~\bibinfo {volume} {22}\ (\bibinfo {organization}
	{Springer},\ \bibinfo {year} {2021})\ pp.\ \bibinfo {pages}
	{1035--1068}\BibitemShut {NoStop}%
	\bibitem [{\citenamefont {Kanzieper}\ and\ \citenamefont
		{Singh}(2010)}]{kanzieper2010non}%
	\BibitemOpen
	\bibfield  {author} {\bibinfo {author} {\bibfnamefont {E.}~\bibnamefont
			{Kanzieper}}\ and\ \bibinfo {author} {\bibfnamefont {N.}~\bibnamefont
			{Singh}},\ }\bibfield  {title} {\bibinfo {title} {Non-hermitean wishart
			random matrices (i)},\ }\href@noop {} {\bibfield  {journal} {\bibinfo
			{journal} {Journal of mathematical physics}\ }\textbf {\bibinfo {volume}
			{51}} (\bibinfo {year} {2010})}\BibitemShut {NoStop}%
	\bibitem [{\citenamefont {Feinberg}\ and\ \citenamefont
		{Zee}(1997)}]{feinberg1997non}%
	\BibitemOpen
	\bibfield  {author} {\bibinfo {author} {\bibfnamefont {J.}~\bibnamefont
			{Feinberg}}\ and\ \bibinfo {author} {\bibfnamefont {A.}~\bibnamefont {Zee}},\
	}\bibfield  {title} {\bibinfo {title} {Non-hermitian random matrix theory:
			Method of hermitian reduction},\ }\href@noop {} {\bibfield  {journal}
		{\bibinfo  {journal} {Nuclear Physics B}\ }\textbf {\bibinfo {volume}
			{504}},\ \bibinfo {pages} {579} (\bibinfo {year} {1997})}\BibitemShut
	{NoStop}%
	\bibitem [{\citenamefont {Altland}\ and\ \citenamefont
		{Simons}(2010)}]{altlandsimons}%
	\BibitemOpen
	\bibfield  {author} {\bibinfo {author} {\bibfnamefont {A.}~\bibnamefont
			{Altland}}\ and\ \bibinfo {author} {\bibfnamefont {B.~D.}\ \bibnamefont
			{Simons}},\ }\href@noop {} {\emph {\bibinfo {title} {Condensed Matter Field
				Theory}}}\ (\bibinfo  {publisher} {Cambridge University Press},\ \bibinfo
	{year} {2010})\BibitemShut {NoStop}%
	\bibitem [{\citenamefont {Martin}\ \emph {et~al.}(1973)\citenamefont {Martin},
		\citenamefont {Siggia},\ and\ \citenamefont {Rose}}]{msr}%
	\BibitemOpen
	\bibfield  {author} {\bibinfo {author} {\bibfnamefont {P.~C.}\ \bibnamefont
			{Martin}}, \bibinfo {author} {\bibfnamefont {E.~D.}\ \bibnamefont {Siggia}},\
		and\ \bibinfo {author} {\bibfnamefont {H.~A.}\ \bibnamefont {Rose}},\
	}\bibfield  {title} {\bibinfo {title} {Statistical dynamics of classical
			systems},\ }\href@noop {} {\bibfield  {journal} {\bibinfo  {journal} {Phys.
				Rev. A}\ }\textbf {\bibinfo {volume} {8}},\ \bibinfo {pages} {423} (\bibinfo
		{year} {1973})}\BibitemShut {NoStop}%
	\bibitem [{\citenamefont {Janssen}(1976)}]{janssen1976lagrangean}%
	\BibitemOpen
	\bibfield  {author} {\bibinfo {author} {\bibfnamefont {H.-K.}\ \bibnamefont
			{Janssen}},\ }\bibfield  {title} {\bibinfo {title} {On a lagrangean for
			classical field dynamics and renormalization group calculations of dynamical
			critical properties},\ }\href@noop {} {\bibfield  {journal} {\bibinfo
			{journal} {Zeitschrift f{\"u}r Physik B Condensed Matter}\ }\textbf {\bibinfo
			{volume} {23}},\ \bibinfo {pages} {377} (\bibinfo {year} {1976})}\BibitemShut
	{NoStop}%
	\bibitem [{\citenamefont {Dominicis}(1976)}]{dominicis1976techniques}%
	\BibitemOpen
	\bibfield  {author} {\bibinfo {author} {\bibfnamefont {C.~d.}\ \bibnamefont
			{Dominicis}},\ }\bibfield  {title} {\bibinfo {title} {Techniques de
			renormalisation de la th{\'e}orie des champs et dynamique des ph{\'e}nomenes
			critiques},\ }in\ \href@noop {} {\emph {\bibinfo {booktitle} {J. Phys.,
				Colloq}}},\ Vol.~\bibinfo {volume} {37}\ (\bibinfo {year} {1976})\ p.\
	\bibinfo {pages} {247}\BibitemShut {NoStop}%
	\bibitem [{\citenamefont {Kamenev}\ and\ \citenamefont
		{Andreev}(1999)}]{kamenev1999electron}%
	\BibitemOpen
	\bibfield  {author} {\bibinfo {author} {\bibfnamefont {A.}~\bibnamefont
			{Kamenev}}\ and\ \bibinfo {author} {\bibfnamefont {A.}~\bibnamefont
			{Andreev}},\ }\bibfield  {title} {\bibinfo {title} {Electron-electron
			interactions in disordered metals: Keldysh formalism},\ }\href@noop {}
	{\bibfield  {journal} {\bibinfo  {journal} {Physical Review B}\ }\textbf
		{\bibinfo {volume} {60}},\ \bibinfo {pages} {2218} (\bibinfo {year}
		{1999})}\BibitemShut {NoStop}%
	\bibitem [{\citenamefont {Altland}\ and\ \citenamefont
		{Kamenev}(2000)}]{altland2000wigner}%
	\BibitemOpen
	\bibfield  {author} {\bibinfo {author} {\bibfnamefont {A.}~\bibnamefont
			{Altland}}\ and\ \bibinfo {author} {\bibfnamefont {A.}~\bibnamefont
			{Kamenev}},\ }\bibfield  {title} {\bibinfo {title} {Wigner-dyson statistics
			from the keldysh $\sigma$-model},\ }\href@noop {} {\bibfield  {journal}
		{\bibinfo  {journal} {Physical review letters}\ }\textbf {\bibinfo {volume}
			{85}},\ \bibinfo {pages} {5615} (\bibinfo {year} {2000})}\BibitemShut
	{NoStop}%
	\bibitem [{\citenamefont {Altieri}\ \emph {et~al.}(2021)\citenamefont
		{Altieri}, \citenamefont {Roy}, \citenamefont {Cammarota},\ and\
		\citenamefont {Biroli}}]{altieri2021properties}%
	\BibitemOpen
	\bibfield  {author} {\bibinfo {author} {\bibfnamefont {A.}~\bibnamefont
			{Altieri}}, \bibinfo {author} {\bibfnamefont {F.}~\bibnamefont {Roy}},
		\bibinfo {author} {\bibfnamefont {C.}~\bibnamefont {Cammarota}},\ and\
		\bibinfo {author} {\bibfnamefont {G.}~\bibnamefont {Biroli}},\ }\bibfield
	{title} {\bibinfo {title} {Properties of equilibria and glassy phases of the
			random lotka-volterra model with demographic noise},\ }\href@noop {}
	{\bibfield  {journal} {\bibinfo  {journal} {Physical Review Letters}\
		}\textbf {\bibinfo {volume} {126}},\ \bibinfo {pages} {258301} (\bibinfo
		{year} {2021})}\BibitemShut {NoStop}%
	\bibitem [{\citenamefont {Ros}\ \emph {et~al.}(2019)\citenamefont {Ros},
		\citenamefont {Ben~Arous}, \citenamefont {Biroli},\ and\ \citenamefont
		{Cammarota}}]{ros2019complex}%
	\BibitemOpen
	\bibfield  {author} {\bibinfo {author} {\bibfnamefont {V.}~\bibnamefont
			{Ros}}, \bibinfo {author} {\bibfnamefont {G.}~\bibnamefont {Ben~Arous}},
		\bibinfo {author} {\bibfnamefont {G.}~\bibnamefont {Biroli}},\ and\ \bibinfo
		{author} {\bibfnamefont {C.}~\bibnamefont {Cammarota}},\ }\bibfield  {title}
	{\bibinfo {title} {Complex energy landscapes in spiked-tensor and simple
			glassy models: Ruggedness, arrangements of local minima, and phase
			transitions},\ }\href@noop {} {\bibfield  {journal} {\bibinfo  {journal}
			{Physical Review X}\ }\textbf {\bibinfo {volume} {9}},\ \bibinfo {pages}
		{011003} (\bibinfo {year} {2019})}\BibitemShut {NoStop}%
	\bibitem [{\citenamefont {Verbaarschot}\ and\ \citenamefont
		{Zirnbauer}(1985)}]{verbaarschot1985critique}%
	\BibitemOpen
	\bibfield  {author} {\bibinfo {author} {\bibfnamefont {J.}~\bibnamefont
			{Verbaarschot}}\ and\ \bibinfo {author} {\bibfnamefont {M.}~\bibnamefont
			{Zirnbauer}},\ }\bibfield  {title} {\bibinfo {title} {Critique of the replica
			trick},\ }\href@noop {} {\bibfield  {journal} {\bibinfo  {journal} {Journal
				of Physics A: Mathematical and General}\ }\textbf {\bibinfo {volume} {18}},\
		\bibinfo {pages} {1093} (\bibinfo {year} {1985})}\BibitemShut {NoStop}%
	\bibitem [{\citenamefont {Fyodorov}\ \emph {et~al.}(1998)\citenamefont
		{Fyodorov}, \citenamefont {Sommers},\ and\ \citenamefont
		{Khoruzhenko}}]{fyodorov1998universality}%
	\BibitemOpen
	\bibfield  {author} {\bibinfo {author} {\bibfnamefont {Y.~V.}\ \bibnamefont
			{Fyodorov}}, \bibinfo {author} {\bibfnamefont {H.-J.}\ \bibnamefont
			{Sommers}},\ and\ \bibinfo {author} {\bibfnamefont {B.~A.}\ \bibnamefont
			{Khoruzhenko}},\ }\bibfield  {title} {\bibinfo {title} {Universality in the
			random matrix spectra in the regime of weak non-hermiticity},\ }in\
	\href@noop {} {\emph {\bibinfo {booktitle} {Annales de l'IHP Physique
				th{\'e}orique}}},\ Vol.~\bibinfo {volume} {68}\ (\bibinfo {year} {1998})\
	pp.\ \bibinfo {pages} {449--489}\BibitemShut {NoStop}%
	\bibitem [{\citenamefont {Fyodorov}\ \emph {et~al.}(1997)\citenamefont
		{Fyodorov}, \citenamefont {Khoruzhenko},\ and\ \citenamefont
		{Sommers}}]{fyodorov1997almost}%
	\BibitemOpen
	\bibfield  {author} {\bibinfo {author} {\bibfnamefont {Y.~V.}\ \bibnamefont
			{Fyodorov}}, \bibinfo {author} {\bibfnamefont {B.~A.}\ \bibnamefont
			{Khoruzhenko}},\ and\ \bibinfo {author} {\bibfnamefont {H.-J.}\ \bibnamefont
			{Sommers}},\ }\bibfield  {title} {\bibinfo {title} {Almost-hermitian random
			matrices: eigenvalue density in the complex plane},\ }\href@noop {}
	{\bibfield  {journal} {\bibinfo  {journal} {Physics Letters A}\ }\textbf
		{\bibinfo {volume} {226}},\ \bibinfo {pages} {46} (\bibinfo {year}
		{1997})}\BibitemShut {NoStop}%
	\bibitem [{\citenamefont {Dhesi}\ and\ \citenamefont
		{Jones}(1990)}]{dhesi1990asymptotic}%
	\BibitemOpen
	\bibfield  {author} {\bibinfo {author} {\bibfnamefont {G.~S.}\ \bibnamefont
			{Dhesi}}\ and\ \bibinfo {author} {\bibfnamefont {R.~C.}\ \bibnamefont
			{Jones}},\ }\bibfield  {title} {\bibinfo {title} {Asymptotic corrections to
			the wigner semicircular eigenvalue spectrum of a large real symmetric random
			matrix using the replica method},\ }\href@noop {} {\bibfield  {journal}
		{\bibinfo  {journal} {Journal of Physics A: Mathematical and General}\
		}\textbf {\bibinfo {volume} {23}},\ \bibinfo {pages} {5577} (\bibinfo {year}
		{1990})}\BibitemShut {NoStop}%
	\bibitem [{\citenamefont {Br\'ezin}\ and\ \citenamefont
		{Zee}(1994)}]{brezin1994correlation}%
	\BibitemOpen
	\bibfield  {author} {\bibinfo {author} {\bibfnamefont {E.}~\bibnamefont
			{Br\'ezin}}\ and\ \bibinfo {author} {\bibfnamefont {A.}~\bibnamefont {Zee}},\
	}\bibfield  {title} {\bibinfo {title} {Correlation functions in disordered
			systems},\ }\href@noop {} {\bibfield  {journal} {\bibinfo  {journal}
			{Physical Review E}\ }\textbf {\bibinfo {volume} {49}},\ \bibinfo {pages}
		{2588} (\bibinfo {year} {1994})}\BibitemShut {NoStop}%
	\bibitem [{\citenamefont {'t~Hooft}(1974)}]{HOOFT1974461}%
	\BibitemOpen
	\bibfield  {author} {\bibinfo {author} {\bibfnamefont {G.}~\bibnamefont
			{'t~Hooft}},\ }\bibfield  {title} {\bibinfo {title} {A planar diagram theory
			for strong interactions},\ }\href@noop {} {\bibfield  {journal} {\bibinfo
			{journal} {Nuclear Physics B}\ }\textbf {\bibinfo {volume} {72}},\ \bibinfo
		{pages} {461} (\bibinfo {year} {1974})}\BibitemShut {NoStop}%
	\bibitem [{\citenamefont {Br{\'e}zin}\ \emph {et~al.}(1978)\citenamefont
		{Br{\'e}zin}, \citenamefont {Itzykson}, \citenamefont {Parisi},\ and\
		\citenamefont {Zuber}}]{brezin1978planar}%
	\BibitemOpen
	\bibfield  {author} {\bibinfo {author} {\bibfnamefont {E.}~\bibnamefont
			{Br{\'e}zin}}, \bibinfo {author} {\bibfnamefont {C.}~\bibnamefont
			{Itzykson}}, \bibinfo {author} {\bibfnamefont {G.}~\bibnamefont {Parisi}},\
		and\ \bibinfo {author} {\bibfnamefont {J.-B.}\ \bibnamefont {Zuber}},\
	}\bibfield  {title} {\bibinfo {title} {Planar diagrams},\ }\href@noop {}
	{\bibfield  {journal} {\bibinfo  {journal} {Communications in Mathematical
				Physics}\ }\textbf {\bibinfo {volume} {59}},\ \bibinfo {pages} {35} (\bibinfo
		{year} {1978})}\BibitemShut {NoStop}%
	\bibitem [{\citenamefont {Kuczala}(2019)}]{kuczala2019dynamics}%
	\BibitemOpen
	\bibfield  {author} {\bibinfo {author} {\bibfnamefont {A.}~\bibnamefont
			{Kuczala}},\ }\href@noop {} {\emph {\bibinfo {title} {Dynamics and
				Information Processing in Recurrent Networks}}}\ (\bibinfo  {publisher}
	{University of California, San Diego},\ \bibinfo {year} {2019})\BibitemShut
	{NoStop}%
	\bibitem [{\citenamefont {Wigner}(1958{\natexlab{b}})}]{Wigner1958}%
	\BibitemOpen
	\bibfield  {author} {\bibinfo {author} {\bibfnamefont {E.~P.}\ \bibnamefont
			{Wigner}},\ }\bibfield  {title} {\bibinfo {title} {On the distribution of the
			roots of certain symmetric matrices},\ }\href@noop {} {\bibfield  {journal}
		{\bibinfo  {journal} {Annals of Mathematics}\ }\textbf {\bibinfo {volume}
			{67}},\ \bibinfo {pages} {325} (\bibinfo {year}
		{1958}{\natexlab{b}})}\BibitemShut {NoStop}%
	\bibitem [{\citenamefont {Bray}\ and\ \citenamefont
		{Moore}(1982)}]{bray1982eigenvalue}%
	\BibitemOpen
	\bibfield  {author} {\bibinfo {author} {\bibfnamefont {A.}~\bibnamefont
			{Bray}}\ and\ \bibinfo {author} {\bibfnamefont {M.}~\bibnamefont {Moore}},\
	}\bibfield  {title} {\bibinfo {title} {On the eigenvalue spectrum of the
			susceptibility matrix for random spin systems},\ }\href@noop {} {\bibfield
		{journal} {\bibinfo  {journal} {Journal of Physics C: Solid State Physics}\
		}\textbf {\bibinfo {volume} {15}},\ \bibinfo {pages} {L765} (\bibinfo {year}
		{1982})}\BibitemShut {NoStop}%
	\bibitem [{\citenamefont {Tao}(2013)}]{tao2013outliers}%
	\BibitemOpen
	\bibfield  {author} {\bibinfo {author} {\bibfnamefont {T.}~\bibnamefont
			{Tao}},\ }\bibfield  {title} {\bibinfo {title} {Outliers in the spectrum of
			iid matrices with bounded rank perturbations},\ }\href@noop {} {\bibfield
		{journal} {\bibinfo  {journal} {Probability Theory and Related Fields}\
		}\textbf {\bibinfo {volume} {155}},\ \bibinfo {pages} {231} (\bibinfo {year}
		{2013})}\BibitemShut {NoStop}%
	\bibitem [{\citenamefont {Bunin}(2017)}]{bunin}%
	\BibitemOpen
	\bibfield  {author} {\bibinfo {author} {\bibfnamefont {G.}~\bibnamefont
			{Bunin}},\ }\bibfield  {title} {\bibinfo {title} {Ecological communities with
			{L}otka-{V}olterra dynamics},\ }\href@noop {} {\bibfield  {journal} {\bibinfo
			{journal} {Physical Review E}\ }\textbf {\bibinfo {volume} {95}},\ \bibinfo
		{pages} {042414} (\bibinfo {year} {2017})}\BibitemShut {NoStop}%
	\bibitem [{\citenamefont {Galla}(2018)}]{galla2018dynamically}%
	\BibitemOpen
	\bibfield  {author} {\bibinfo {author} {\bibfnamefont {T.}~\bibnamefont
			{Galla}},\ }\bibfield  {title} {\bibinfo {title} {Dynamically evolved
			community size and stability of random lotka-volterra ecosystems (a)},\
	}\href@noop {} {\bibfield  {journal} {\bibinfo  {journal} {Europhysics
				Letters}\ }\textbf {\bibinfo {volume} {123}},\ \bibinfo {pages} {48004}
		(\bibinfo {year} {2018})}\BibitemShut {NoStop}%
	\bibitem [{\citenamefont {Garcia~Lorenzana}\ and\ \citenamefont
		{Altieri}(2022)}]{garcialorenzana2022competitive}%
	\BibitemOpen
	\bibfield  {author} {\bibinfo {author} {\bibfnamefont {G.}~\bibnamefont
			{Garcia~Lorenzana}}\ and\ \bibinfo {author} {\bibfnamefont {A.}~\bibnamefont
			{Altieri}},\ }\bibfield  {title} {\bibinfo {title} {Well-mixed lotka-volterra
			model with random strongly competitive interactions},\ }\href@noop {}
	{\bibfield  {journal} {\bibinfo  {journal} {Phys. Rev. E}\ }\textbf {\bibinfo
			{volume} {105}},\ \bibinfo {pages} {024307} (\bibinfo {year}
		{2022})}\BibitemShut {NoStop}%
	\bibitem [{\citenamefont {Slanina}(2011)}]{slanina2011equivalence}%
	\BibitemOpen
	\bibfield  {author} {\bibinfo {author} {\bibfnamefont {F.}~\bibnamefont
			{Slanina}},\ }\bibfield  {title} {\bibinfo {title} {Equivalence of replica
			and cavity methods for computing spectra of sparse random matrices},\
	}\href@noop {} {\bibfield  {journal} {\bibinfo  {journal} {Physical Review
				E}\ }\textbf {\bibinfo {volume} {83}},\ \bibinfo {pages} {011118} (\bibinfo
		{year} {2011})}\BibitemShut {NoStop}%
	\bibitem [{\citenamefont {Slanina}(2012)}]{slanina2012localization}%
	\BibitemOpen
	\bibfield  {author} {\bibinfo {author} {\bibfnamefont {F.}~\bibnamefont
			{Slanina}},\ }\bibfield  {title} {\bibinfo {title} {Localization of
			eigenvectors in random graphs},\ }\href@noop {} {\bibfield  {journal}
		{\bibinfo  {journal} {The European Physical Journal B}\ }\textbf {\bibinfo
			{volume} {85}},\ \bibinfo {pages} {1} (\bibinfo {year} {2012})}\BibitemShut
	{NoStop}%
	\bibitem [{\citenamefont {Girko}(1985)}]{girko1985circular}%
	\BibitemOpen
	\bibfield  {author} {\bibinfo {author} {\bibfnamefont {V.~L.}\ \bibnamefont
			{Girko}},\ }\bibfield  {title} {\bibinfo {title} {Circular law},\ }\href@noop
	{} {\bibfield  {journal} {\bibinfo  {journal} {Theory of Probability \& Its
				Applications}\ }\textbf {\bibinfo {volume} {29}},\ \bibinfo {pages} {694}
		(\bibinfo {year} {1985})}\BibitemShut {NoStop}%
	\bibitem [{\citenamefont {Burda}\ \emph
		{et~al.}(2010{\natexlab{a}})\citenamefont {Burda}, \citenamefont {Jarosz},
		\citenamefont {Livan}, \citenamefont {Nowak},\ and\ \citenamefont
		{Swiech}}]{burda2010eigenvalues}%
	\BibitemOpen
	\bibfield  {author} {\bibinfo {author} {\bibfnamefont {Z.}~\bibnamefont
			{Burda}}, \bibinfo {author} {\bibfnamefont {A.}~\bibnamefont {Jarosz}},
		\bibinfo {author} {\bibfnamefont {G.}~\bibnamefont {Livan}}, \bibinfo
		{author} {\bibfnamefont {M.~A.}\ \bibnamefont {Nowak}},\ and\ \bibinfo
		{author} {\bibfnamefont {A.}~\bibnamefont {Swiech}},\ }\bibfield  {title}
	{\bibinfo {title} {Eigenvalues and singular values of products of rectangular
			gaussian random matrices},\ }\href@noop {} {\bibfield  {journal} {\bibinfo
			{journal} {Physical Review E—Statistical, Nonlinear, and Soft Matter
				Physics}\ }\textbf {\bibinfo {volume} {82}},\ \bibinfo {pages} {061114}
		(\bibinfo {year} {2010}{\natexlab{a}})}\BibitemShut {NoStop}%
	\bibitem [{\citenamefont {Burda}\ \emph
		{et~al.}(2010{\natexlab{b}})\citenamefont {Burda}, \citenamefont {Janik},\
		and\ \citenamefont {Waclaw}}]{burda2010spectrum}%
	\BibitemOpen
	\bibfield  {author} {\bibinfo {author} {\bibfnamefont {Z.}~\bibnamefont
			{Burda}}, \bibinfo {author} {\bibfnamefont {R.~A.}\ \bibnamefont {Janik}},\
		and\ \bibinfo {author} {\bibfnamefont {B.}~\bibnamefont {Waclaw}},\
	}\bibfield  {title} {\bibinfo {title} {Spectrum of the product of independent
			random gaussian matrices},\ }\href@noop {} {\bibfield  {journal} {\bibinfo
			{journal} {Physical Review E—Statistical, Nonlinear, and Soft Matter
				Physics}\ }\textbf {\bibinfo {volume} {81}},\ \bibinfo {pages} {041132}
		(\bibinfo {year} {2010}{\natexlab{b}})}\BibitemShut {NoStop}%
	\bibitem [{\citenamefont {O'Rourke}\ and\ \citenamefont
		{Soshnikov}(2011)}]{o2011products}%
	\BibitemOpen
	\bibfield  {author} {\bibinfo {author} {\bibfnamefont {S.}~\bibnamefont
			{O'Rourke}}\ and\ \bibinfo {author} {\bibfnamefont {A.}~\bibnamefont
			{Soshnikov}},\ }\bibfield  {title} {\bibinfo {title} {Products of independent
			non-hermitian random matrices},\ }\href@noop {} {\  (\bibinfo {year}
		{2011})}\BibitemShut {NoStop}%
	\bibitem [{\citenamefont {Hubbard}(1959)}]{hubbard}%
	\BibitemOpen
	\bibfield  {author} {\bibinfo {author} {\bibfnamefont {J.}~\bibnamefont
			{Hubbard}},\ }\bibfield  {title} {\bibinfo {title} {Calculation of partition
			functions},\ }\href@noop {} {\bibfield  {journal} {\bibinfo  {journal}
			{Physical Review Letters}\ }\textbf {\bibinfo {volume} {3}},\ \bibinfo
		{pages} {77} (\bibinfo {year} {1959})}\BibitemShut {NoStop}%
	\bibitem [{\citenamefont {Cui}\ \emph {et~al.}(2020)\citenamefont {Cui},
		\citenamefont {Rocks},\ and\ \citenamefont {Mehta}}]{cui2020perturbative}%
	\BibitemOpen
	\bibfield  {author} {\bibinfo {author} {\bibfnamefont {W.}~\bibnamefont
			{Cui}}, \bibinfo {author} {\bibfnamefont {J.~W.}\ \bibnamefont {Rocks}},\
		and\ \bibinfo {author} {\bibfnamefont {P.}~\bibnamefont {Mehta}},\ }\bibfield
	{title} {\bibinfo {title} {The perturbative resolvent method: Spectral
			densities of random matrix ensembles via perturbation theory},\ }\href@noop
	{} {\bibfield  {journal} {\bibinfo  {journal} {arXiv preprint
				arXiv:2012.00663}\ } (\bibinfo {year} {2020})}\BibitemShut {NoStop}%
	\bibitem [{\citenamefont {Silva}\ and\ \citenamefont
		{Metz}(2022)}]{silva2022analytic}%
	\BibitemOpen
	\bibfield  {author} {\bibinfo {author} {\bibfnamefont {J.~D.}\ \bibnamefont
			{Silva}}\ and\ \bibinfo {author} {\bibfnamefont {F.~L.}\ \bibnamefont
			{Metz}},\ }\bibfield  {title} {\bibinfo {title} {Analytic solution of the
			resolvent equations for heterogeneous random graphs: spectral and
			localization properties},\ }\href@noop {} {\bibfield  {journal} {\bibinfo
			{journal} {Journal of Physics: Complexity}\ }\textbf {\bibinfo {volume}
			{3}},\ \bibinfo {pages} {045012} (\bibinfo {year} {2022})}\BibitemShut
	{NoStop}%
	\bibitem [{\citenamefont {Rogers}(2010)}]{rogers2010universal}%
	\BibitemOpen
	\bibfield  {author} {\bibinfo {author} {\bibfnamefont {T.}~\bibnamefont
			{Rogers}},\ }\bibfield  {title} {\bibinfo {title} {Universal sum and product
			rules for random matrices},\ }\href@noop {} {\bibfield  {journal} {\bibinfo
			{journal} {Journal of mathematical physics}\ }\textbf {\bibinfo {volume}
			{51}},\ \bibinfo {pages} {093304} (\bibinfo {year} {2010})}\BibitemShut
	{NoStop}%
	\bibitem [{\citenamefont {Mirlin}\ and\ \citenamefont
		{Fyodorov}(1991)}]{mirlin1991universality}%
	\BibitemOpen
	\bibfield  {author} {\bibinfo {author} {\bibfnamefont {A.}~\bibnamefont
			{Mirlin}}\ and\ \bibinfo {author} {\bibfnamefont {Y.~V.}\ \bibnamefont
			{Fyodorov}},\ }\bibfield  {title} {\bibinfo {title} {Universality of level
			correlation function of sparse random matrices},\ }\href@noop {} {\bibfield
		{journal} {\bibinfo  {journal} {Journal of Physics A: Mathematical and
				General}\ }\textbf {\bibinfo {volume} {24}},\ \bibinfo {pages} {2273}
		(\bibinfo {year} {1991})}\BibitemShut {NoStop}%
	\bibitem [{\citenamefont {Br{\'e}zin}\ and\ \citenamefont
		{Zee}(1993)}]{brezin1993universality}%
	\BibitemOpen
	\bibfield  {author} {\bibinfo {author} {\bibfnamefont {E.}~\bibnamefont
			{Br{\'e}zin}}\ and\ \bibinfo {author} {\bibfnamefont {A.}~\bibnamefont
			{Zee}},\ }\bibfield  {title} {\bibinfo {title} {Universality of the
			correlations between eigenvalues of large random matrices},\ }\href@noop {}
	{\bibfield  {journal} {\bibinfo  {journal} {Nuclear Physics B}\ }\textbf
		{\bibinfo {volume} {402}},\ \bibinfo {pages} {613} (\bibinfo {year}
		{1993})}\BibitemShut {NoStop}%
	\bibitem [{\citenamefont {Nowak}\ and\ \citenamefont
		{Tarnowski}(2018)}]{nowak2018probing}%
	\BibitemOpen
	\bibfield  {author} {\bibinfo {author} {\bibfnamefont {M.~A.}\ \bibnamefont
			{Nowak}}\ and\ \bibinfo {author} {\bibfnamefont {W.}~\bibnamefont
			{Tarnowski}},\ }\bibfield  {title} {\bibinfo {title} {Probing
			non-orthogonality of eigenvectors in non-hermitian matrix models:
			diagrammatic approach},\ }\href@noop {} {\bibfield  {journal} {\bibinfo
			{journal} {Journal of High Energy Physics}\ }\textbf {\bibinfo {volume}
			{2018}},\ \bibinfo {pages} {1} (\bibinfo {year} {2018})}\BibitemShut
	{NoStop}%
	\bibitem [{\citenamefont {Allesina}\ \emph
		{et~al.}(2015{\natexlab{b}})\citenamefont {Allesina}, \citenamefont {Grilli},
		\citenamefont {Barab{\'a}s}, \citenamefont {Tang}, \citenamefont {Aljadeff},\
		and\ \citenamefont {Maritan}}]{allesina2015predicting}%
	\BibitemOpen
	\bibfield  {author} {\bibinfo {author} {\bibfnamefont {S.}~\bibnamefont
			{Allesina}}, \bibinfo {author} {\bibfnamefont {J.}~\bibnamefont {Grilli}},
		\bibinfo {author} {\bibfnamefont {G.}~\bibnamefont {Barab{\'a}s}}, \bibinfo
		{author} {\bibfnamefont {S.}~\bibnamefont {Tang}}, \bibinfo {author}
		{\bibfnamefont {J.}~\bibnamefont {Aljadeff}},\ and\ \bibinfo {author}
		{\bibfnamefont {A.}~\bibnamefont {Maritan}},\ }\bibfield  {title} {\bibinfo
		{title} {Predicting the stability of large structured food webs},\
	}\href@noop {} {\bibfield  {journal} {\bibinfo  {journal} {Nature
				communications}\ }\textbf {\bibinfo {volume} {6}},\ \bibinfo {pages} {1}
		(\bibinfo {year} {2015}{\natexlab{b}})}\BibitemShut {NoStop}%
\end{thebibliography}
\end{document}


\title{A path integral approach to sparse non-Hermitian random matrices\\ \medskip \medskip{\Large--- Supplemental Material ---}}
\author{Joseph W. Baron}
\email{joseph-william.baron@phys.ens.fr}
\affiliation{Laboratoire de Physique de l’Ecole Normale Sup\`{e}rieure, ENS, Universit\'{e} PSL, CNRS, Sorbonne Universit\'{e}, Universit\'{e} de Paris, F-75005 Paris, France}

\maketitle
\vspace{-2em}
\tableofcontents
\newpage

\section{Overview}
This document contains additional information about the diagrammatic formalism and details of the calculations, in particular, the results for sparse matrices. Also included are some additional results for non-Gaussian matrices, products of random matrices and block-structured matrices, which we use to demonstrate the flexibility of the path-integral method.

First, we give a more detailed introduction to the general method presented in the main text. In Section \ref{appendix:msrjd}, we comment on how the universality of the results is apparent from the path integral approach. We then give a pedagogical motivation and explanation of the diagrammatic formalism in Sections \ref{appendix:feynman} and \ref{appendix:distributive}, using the case of dense random matrices as a concrete example. 

Then, we go on to provide some details of the calculations of the sparse corrections that involve `ribbon' diagrams. In Section \ref{appendix:firstorder}, we describe how to obtain the results in Eqs.~(40--43) from Eq.~(39) of the main text. Then, we discuss in Section \ref{appendix:outlier} how the introduction of a non-zero value of $\mu$ gives rise to additional diagrams and changes the eigenvalue spectrum. In Section \ref{appendix:secondorder}, we show how one can sum additional $O(1/p^2)$ diagrams to obtain Eq.~(51) of the main text.

Finally, we present some additional results for other ensembles of random matrix. In Section \ref{appendix:nongaussian}, we discuss ensembles with non-negligible higher-order moments, and demonstrate the link to sparse matrices. We derive results for matrices with elements drawn from a truncated Cauchy distribution, and we also address the case where we allow the null entries of the sparse matrices examined in the main text to fluctuate about $a_{ij} = 0$. Additionally, in Sections \ref{appendix:marchenkopastur} and \ref{appendix:blockstructure}, we show how products of random matrices and block-structured random matrices (respectively) can be handled using the path-integral approach.

\section{A note on Universality}
\label{appendix:msrjd}

In this section, we argue that the expression for the disorder-averaged generating functional in Eq.~(18) of the main text is valid not only for matrices $\underline{\underline{a}}$ with Gaussian entries, but for any distribution with higher-order moments that decay sufficiently quickly with $N$. We proceed along similar lines to an argument made in Ref. \cite{baron2022eigenvalues}. 

More specifically, for the case of a dense random matrix, we show that if higher-order moments decay more quickly than $1/N$, they do not contribute to the response function in the thermodynamic limit and consequently they do not affect the eigenvalue spectrum. Similar arguments carry over to the sparse case. 

More precisely, taking the disorder average of Eq.~(14) of the main text, one obtains
\begin{align}
	&\overline{Z[\psi, h]} = \int D[x, \hat x] \exp\left[i \sum_{i,a}\int dt \psi^a_i x^{a\star}_i + \psi^{a\star}_i x^{a}_i\right] \nonumber \\
	&\times\exp \left[ i\sum_{i,a} \int dt \, \left\{\hat x^{a\star}_i \left(\dot x_i^a + \sum_{b, j}(\mathcal{H}_0^{-1})^{ab}_{ij} x^b_j - h_j^a \right) + \mathrm{c.c.}\right\}\right] \nonumber \\
	&\times\overline{\prod_{i<j}\exp \left[- i \int dt \,  \sum_{ab} (\hat x^{a\star}_i \mati_{ij}^{ab} x^{b}_j + \hat x^{a\star}_j \mati_{ji}^{ab} x^{b}_i  + \mathrm{c.c.}) \right]} , \label{genfunct}
\end{align}

Noting that only transpose pairs of elements $a_{ij}$ and $a_{ji}$ are correlated, we can consider each combination $(i,j)$ separately. One expands the exponential to obtain
\begin{align}
	&\overline{\exp \left[- i \int dt \,  \sum_{ab} (\hat x^{a\star}_i \mati_{ij}^{ab} x^{b}_j + \hat x^{a\star}_j \mati_{ji}^{ab} x^{b}_i  + \mathrm{c.c.}) \right]} = 1 - i \int dt \,  \sum_{ab} \overline{(\hat x^{a\star}_i \mati_{ij}^{ab} x^{b}_j + \hat x^{a\star}_j \mati_{ji}^{ab} x^{b}_i  + \mathrm{c.c.})} \nonumber \\
	&- \frac{1}{2!} \int dt dt' \,  \sum_{aba'b'}  \overline{[\hat x^{a\star}_i \mati_{ij}^{ab} x^{b}_j + \hat x^{a\star}_j \mati_{ji}^{ab} x^{b}_i  + \mathrm{c.c.}] [\hat x^{a'\star}_i \mati_{ij}^{a'b'} x^{b'}_j + \hat x^{a'\star}_j \mati_{ji}^{a'b'} x^{b'}_i + \mathrm{c.c.}]} + \cdots\label{expanded}
\end{align}
For $a\neq b$ and $a' \neq b'$, we have
\begin{align}
	\overline{ \mati_{ij}^{ab} \mati_{i'j'}^{a'b'}} = \delta_{aa'}\delta_{bb'} \frac{\sigma^2}{N}\left[\delta_{ii'}\delta_{jj'}+\Gamma \delta_{ij'}\delta_{i' j}  \right] + \delta_{ba'}\delta_{ab'} \frac{\sigma^2}{N}\left[\Gamma\delta_{ii'}\delta_{jj'}+ \delta_{ij'}\delta_{i' j}  \right] .
\end{align}
One sees that if the higher moments of $a_{ij}$ decay more quickly than $1/N$ then, for large $N$, we can truncate the expansion in Eq.~(\ref{expanded}) at second order in $\mati$. We can then reexponentiate to obtain 
\begin{align}
	&\overline{Z[\psi, h]} = \int D[x, \hat x] \exp\left[i \sum_{i,a}\int dt \psi^a_i x^{a\star}_i + \psi^{a\star}_i x^{a}_i\right] \nonumber \\
	&\times\exp \left[ i\sum_{i,a} \int dt \, \left\{\hat x^{a\star}_i \left(\dot x_i^a + \sum_{b, j}(\mathcal{H}_0^{-1})^{ab}_{ij} x^b_j - h_j^a \right) + \mathrm{c.c.}\right\}\right] \nonumber \\ 
	&\times\exp\left\{- \frac{1}{2!2} \sum_{ij} \int dt dt' \,  \sum_{aba'b'}  \overline{[\hat x^{a\star}_i \mati_{ij}^{ab} x^{b}_j + \hat x^{a\star}_j \mati_{ji}^{ab} x^{b}_i  + \mathrm{c.c.}] [\hat x^{a'\star}_i \mati_{ij}^{a'b'} x^{b'}_j + \hat x^{a'\star}_j \mati_{ji}^{a'b'} x^{b'}_i + \mathrm{c.c.}]} \right\},\label{fullgenfunct}
\end{align}
In the following section, we explain why the matrices $\underline{\underline{\mati}}$ can be replaced with $\matj$, and also why the complex conjugate terms can be neglected, to obtain Eq.~(19) of the main text.

If we were to keep the terms that are subleading in $1/N$ in the above expansion, they would only give rise to corrections to the interaction term of the action that were also subleading in $N$. We thus can thus see readily that the higher-order moments do not contribute to the calculation of the response functions, and therefore the eigenvalue spectrum. We consider the case where all higher-order moments are of the order $1/N$, for which the preceding argument does not apply, in Section \ref{appendix:nongaussian} of this document.

For sparse matrices on the other hand, the higher-order moments of the distribution $\pi(\cdot, \cdot)$ very much affect the eigenvalue spectrum, since they do not scale with powers of $1/N$. However, the results that we obtain for sparse matrices are universal in the following sense. If, for example, we choose to approximate the eigenvalue spectrum with an expression that is accurate to first order in $1/p$, then all matrix ensembles with the same values of $\mu$, $\Gamma$, $\Gamma_4^{(1)}$, $\Gamma_4^{(2)}$, and $\Gamma_4^{(3)}$ will have the same expressions for the boundary, density and outlier [see Eqs.~(40), (42), (44) and (46)].

\section{Detailed introduction to the diagrammatic formalism. Example: dense random matrices} \label{appendix:feynman}
The evaluation of the series in Eq.~(22) of the main text using Wick's theorem at first appears a daunting task. It would certainly seem at first glance that the large collection of separate Wick pairings of the dynamic variables would be unmanageable. However, we show here that upon careful consideration, most terms can be neglected in the thermodynamic limit, and a great simplification occurs. The identification of which terms survive amounts to constructing a set of Feynman rules. Indeed, we show that the terms that do survive can be represented by so-called planar `rainbow' diagrams in the dense case discussed in Section IV of the main text. The following considerations and the rules that we derive also apply directly to the calculation of the sparse corrections.

In what follows, we largely follow the discussion detailed in Ref. \cite{hertz2016path}, which developed a diagrammatic representation for dynamical systems with non-linear terms. We extend this approach to our case, incorporating also the topological reasoning of t' Hooft \cite{HOOFT1974461} (see also \cite{brezin1978planar}), which allows one to find efficiently the terms that survive in the limit $N \to \infty$.

We wish to find all of the Wick pairings in the series in Eq.~(22) of the main text that we can neglect and find an efficient way of identifying the set of surviving terms for $N\to \infty$. We begin with the expression for the interaction term of the action in Eq.~(\ref{fullgenfunct})
\begin{align}
	S_\mathrm{int} &= - \frac{1}{2!2} \sum_{ij} \int dT dT' \,  \sum_{aba'b'}  \overline{[\hat x^{a\star}_i \mati_{ij}^{ab} x^{b}_j + \hat x^{a\star}_j \mati_{ji}^{ab} x^{b}_i  + \mathrm{c.c.}] [\hat x^{a'\star}_i \mati_{ij}^{a'b'} x^{b'}_j + \hat x^{a'\star}_j \mati_{ji}^{a'b'} x^{b'}_i + \mathrm{c.c.}]},\nonumber \\
	&=- \frac{1}{2!} \sum_{ij, i',j'} \int dT dT' \,  \sum_{aba'b'}  \overline{[\hat x^{a\star}_i \mati_{ij}^{ab} x^{b}_j + \mathrm{c.c.}] [\hat x^{a'\star}_{i'} \mati_{i'j'}^{a'b'} x^{b'}_{j'}  + \mathrm{c.c.}]}\label{fullsint}
\end{align}
where the dynamic variables in the first square bracket have time coordinate $T$ and those in the second have time coordinate $T'$. It will become apparent during the following discourse why the expression in Eq.~(\ref{fullsint}) above can be replaced with the expression in Eq.~(19) of the main text. 

Let us consider the first-order term in $S_{\mathrm{int}}$ [i.e. $r=1$] in Eq.~(22) of the main text
\begin{align}
	&-\frac{i}{1!}\langle x^c_k(t) \hat x^{d\star}_l(t') S_{\mathrm{int}}\rangle_0 \nonumber \\
	&= \frac{i}{1! 2!}\left\langle  x^c_k(t) \hat x^{d\star}_l(t')\overline{\left\{ \sum_{i,j,a,b} \int dT \, \hat x^{a\star}_i  \mati_{ij}^{ab} x^b_j   + \mathrm{c.c.} \right\}\left\{ \sum_{i',j',a',b'} \int dT' \, \hat x^{a'\star}_{i'}  \mati_{i'j'}^{a'b'} x^{b'}_{j'}   + \mathrm{c.c.} \right\}} \right\rangle_0 \nonumber \\
	&= \frac{(-i)^3}{1! 2!} \sum_{i,j,a,b} \sum_{i',j',a',b'} \int dT \int dT'\Bigg[\overline{\mati_{ij}^{ab} \mati_{i'j'}^{a'b'}}\left\langle  x^c_k(t) \hat x^{d\star}_l(t')  \hat x^{a\star}_i   x^b_j    \, \hat x^{a'\star}_{i'}   x^{b'}_{j'}   \right\rangle_0  \nonumber \\
	&+ \overline{\mati_{ij}^{ab} \mati_{i'j'}^{a'b'}}\left\langle  x^c_k(t) \hat x^{d\star}_l(t')  \hat x^{a}_i   x^{b\star}_j    \, \hat x^{a'\star}_{i'}   x^{b'}_{j'}     \right\rangle_0 + \overline{\mati_{ij}^{ab} \mati_{i'j'}^{a'b'}}\left\langle  x^c_k(t) \hat x^{d\star}_l(t')  \hat x^{a\star}_i   x^b_j    \, \hat x^{a'}_{i'}   x^{b'\star}_{j'}    \right\rangle_0 \nonumber \\
	&+ \overline{\mati_{ij}^{ab} \mati_{i'j'}^{a'b'}}\left\langle  x^c_k(t) \hat x^{d\star}_l(t')  \hat x^{a}_i   x^{b\star}_j    \, \hat x^{a'}_{i'}   x^{b'\star}_{j'}  \right\rangle_0 \Bigg]. \label{firstordersint}
\end{align}
We see that we obtain four averages of the dynamic variables here, which are to be evaluated using Wick's theorem. In other words, the average of a product of the dynamic variables is equal to the sum of all possible combinations of the variables averaged in pairs. This applies in our case since the bare action $S_0$ is quadratic in the dynamic variables. See Ref. \cite{hertz2016path} for a more in-depth discussion of this point. 

First, we note that in order for a particular Wick pairing to be non-zero, all `hatted' variables must be paired with `unhatted' variables. This is because 
\begin{align}
	-i\langle \hat x^{d\star}_l(t')   \hat x^{a\star}_i(T) \rangle_0 = \frac{\delta \langle 1 \rangle_0}{\delta h^d_l(t') \delta h_i^a(T)} = 0, 
\end{align}
with analogous expressions being valid for the complex conjugate terms, while pairings of unhatted with hatted variables give rise to factors of the bare response function, which we can evaluate readily [see Eqs.~(9) and (23) of the main text]
\begin{align}
	\lim_{\eta \to 0} \overline{(\hat K_0)_{ij}^{ab}} &= \lim_{\eta\to 0}\mathcal{L}_t\left[ -i\langle x^a_i(t) \hat x^{b\star}_j(0)  \rangle_0 \right]=  (\mathcal{H}_0)_{ij}^{ab} .
\end{align}
Further, we note also that
\begin{align}
	-i\langle x^{c}_k(t)   \hat x^{a}_i(T) \rangle &= \frac{\delta \langle x^c_k(t) \rangle}{\delta h^{a\star}_i(T)} = 0, \nonumber \\
	-i\langle x^{b\star}_j(T) \hat x^{a}_i(T) \rangle &= \frac{\delta \langle x^{b\star}_j(T) \rangle}{\delta h^{a\star}_i(T)} = 0.
\end{align}
That is, the dynamic variables $x^c_k(t)$ are analytic functions of $h_i^a(t)$ and the equal-time response function is zero. With the combination of these observations, we therefore see that only the first of the four averages in Eq.~(\ref{firstordersint}) survives, since there is no way to Wick-pair the other terms to produce a non-vanishing contribution.

This is true of terms that are higher-order in $S_{\mathrm{int}}$ as well. This means that, in general, we can ignore the complex conjugate terms that appear in $S_{\mathrm{int}}$ when calculating the response functions that we desire. That is, the only non-vanishing term in Eq.~(\ref{firstordersint}) is
\begin{align}
	O_1 \equiv \frac{1}{1! 2 } \int dT dT'\sum_{i, j, i', j'}(-i)^3 \overline{\mati^{ab}_{ij} \mati^{a'b'}_{i'j'}}\left\langle  x^c_k(t) \hat x^{d\star}_l(t')  \hat x^{a\star}_i(T)   x^b_j(T)    \, \hat x^{a'\star}_{i'}(T')   x^{b'}_{j'} (T')  \right\rangle_0 ,
\end{align}
Let us now consider more carefully the average $\langle \cdot \rangle_0$, which we evaluate using Wick's theorem. There are $5!!$ possible Wick pairings that we potentially have to deal with. However, as argued above, only Wick pairings that involve solely hatted-unhatted pairs survive. We therefore have a drastic simplification, and the only surviving Wick pairings are
\begin{align}
	O_1 &= \frac{1}{1!2} \int dT dT'\sum_{i, j, i', j'}(-i)^3 \overline{\mati^{ab}_{ij} \mati^{a'b'}_{i'j'}}\Bigg[\left\langle  x^c_k(t)   \hat x^{a\star}_i(T) \right\rangle_0   \left\langle x^b_j(T)    \, \hat x^{a'\star}_{i'}(T') \right\rangle_0 \left\langle   x^{b'}_{j'} (T') \hat x^{d\star}_l(t') \right\rangle_0 \nonumber \\
	&+ \left\langle  x^c_k(t)  \hat x^{a'\star}_{i'}(T')  \right\rangle_0   \left\langle  x^{b'}_{j'} (T')  \hat x^{a\star}_i(T)  \, \right\rangle_0 \left\langle  x^b_j(T)  \hat x^{d\star}_l(t') \right\rangle_0 \Bigg] .
\end{align}
We note that we cannot have a pairing containing $\left\langle  x^c_k(t) \hat x^{d\star}_l(t') \right\rangle_0 $ due to time ordering. Such a pairing would require there to be a factor of the bare response function $K_0(T,T')$ for which $T<T'$, which would evaluate to nil. Diagrammatically, this would correspond to two disconnected loops, which always evaluate to nil \cite{hertz2016path}.

Now, we consider the disorder-averaged object $ \overline{\mati^{ab}_{ij} \mati^{a'b'}_{i'j'}}$. From the definition in Eq.~(9) of the main text, for $a\neq b$ and $a' \neq b'$, we have
\begin{align}
	\overline{ \mati_{ij}^{ab} \mati_{i'j'}^{a'b'}} =& \delta_{aa'}\delta_{bb'} \frac{\sigma^2}{N}\left[\delta_{ii'}\delta_{jj'}+\Gamma \delta_{ij'}\delta_{i' j}  \right] \nonumber \\
	&+ \delta_{ba'}\delta_{ab'} \frac{\sigma^2}{N}\left[\Gamma\delta_{ii'}\delta_{jj'}+ \delta_{ij'}\delta_{i' j}  \right] .
\end{align}
We see that either we have the constraint $i = i'$ and $j = j'$, or $i = j'$ and $j = i'$. Because the bare resolvent is diagonal in the lower indices, the former constraint gives rise to a sub-leading contribution in $1/N$, whereas the latter constraint correctly pairs `hatted' variables with `unhatted' variables (so to speak) in the summation over lower indices, cancelling the factors of $N$ that appear in the denominator. This same reasoning again carries over to all higher orders in $S_\mathrm{int}$ (and indeed to the ribbon diagrams for the sparse correction). We can therefore effectively consider $\overline{ \mati_{ij}^{ab} \mati_{i'j'}^{a'b'}} \propto \delta_{ij'}\delta_{ji'}$. Diagrammatically, this amounts to ignoring diagrams with twisted double arcs \cite{brezin1994correlation}. We now introduce the $2 \times 2$ matrix $\matj$ (denoted in calligraphic font without the double underline) 
\begin{align}
	\matj = 
	\begin{bmatrix}
		0 & a_{12} \\
		a_{21} & 0
	\end{bmatrix},\label{jdef}
\end{align}
where $a_{12}$ and $a_{21}$ have the statistics given in Eq.~(17) of the main text. In other words, for $a\neq b$ and $a'\neq b'$ we have 
\begin{align}
	\overline{\matj^{ab} (\matj^\dagger)^{a'b'}} = \delta_{aa'}\delta_{bb'} \sigma^2/N + \delta_{ab'}\delta_{ba'} \Gamma \sigma^2/N.
\end{align}
We note that $\matj$ is a non-Hermitan matrix, in contrast to $\underline{\underline{\mati}}$. We have therefore shown that we can replace $\overline{ \mati_{ij}^{ab} \mati_{i'j'}^{a'b'}}$ with $\overline{\matj^{ab} (\matj^\dagger)^{a'b'}}$, and one thus arrives at the simplified expression for the interaction term $S_\mathrm{int}$ in Eq.~(19) of the main text. Using this simplification, we find for the first order term
\begin{align}
	O_1 &= \frac{1}{1!2} \int dT dT'\sum_{i, j}(-i)^3 \overline{\matj^{ab} (\matj^\dagger)^{a'b'}}\Bigg[\left\langle  x^c_k(t)   \hat x^{a\star}_i(T) \right\rangle_0   \left\langle x^b_j(T)    \, \hat x^{a'\star}_{j}(T') \right\rangle_0 \left\langle   x^{b'}_{i} (T') \hat x^{d\star}_l(t') \right\rangle_0 \nonumber \\
	&+ \left\langle  x^c_k(t)  \hat x^{a'\star}_{j}(T)  \right\rangle_0   \left\langle  x^{b'}_{i} (T)  \hat x^{a\star}_i(T')  \, \right\rangle_0 \left\langle  x^b_j(T')  \hat x^{d\star}_l(t') \right\rangle_0 \Bigg] .
\end{align}
One notes that the only difference between the first and second terms in the square brackets above is the time ordering. Because the internal times are integrated over, both terms above are equal, and thus cancel the factor of $2$ in the denominator. So, we have succeeded in finding the non-vanishing contribution up to leading order in $S_\mathrm{int}$
\begin{align}
	&O_1= -\frac{i}{1!}  \langle x^c_k(t) \hat x^{d\star}_l(t') S_{\mathrm{int}}\rangle_0 \nonumber \\
	&= \sum_{a_1,b_1, a'_1, b'_1}  \int_{T_1>T'_1} dT_1 dT'_1\sum_{ i_1, j_1}(-i)^3 \Bigg\{ \nonumber \\
	&\times\Bigg[\left\langle  x^c_k(t)   \hat x^{a_1\star}_{i_1}(T_1) \right\rangle_0 \overline{ \matj^{a_1b_1} \left\langle x^{b_1}_{j_1}(T_1)    \, \hat x^{a_1'\star}_{j_1}(T'_1) \right\rangle_0  (\matj^\dagger)^{a'_1b'_1}}\left\langle   x^{b'_1}_{i_1} (T'_1) \hat x^{d\star}_l(t') \right\rangle_0 \Bigg]\Bigg\} ,
\end{align}
where we have now given the indices a subscript `1' in anticipation of the pattern of higher-order terms. This surviving term can be represented diagrammatically as
\begin{figure}[H]
	\centering 
	\includegraphics[scale = 0.5]{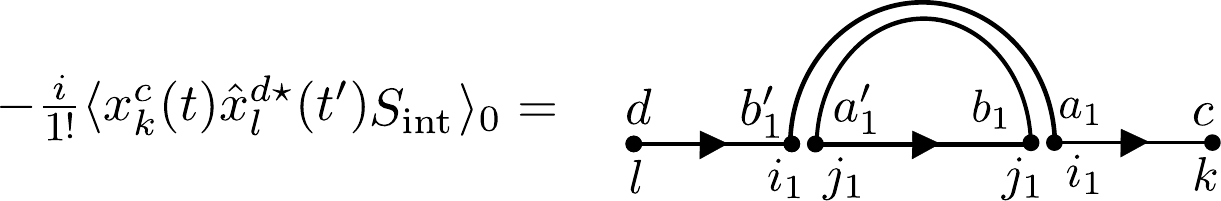}
	\captionsetup{labelformat=empty}
\end{figure}

The above diagram should be interpreted as follows (see also Ref. \cite{baron2022eigenvalues}):  A dot on the left-hand end of a directed edge represents an $\hat x$-type variable, and a dot on the right-hand end of a directed edge represents an $x$-type variable. Pairs of dots positioned together have the same time coordinate, and each pair of dots carries a matrix factor $\matj$ (on the right-hand side of an arc) or $\matj^\dagger$ (on the left-hand side). Double arcs connect pairs of $\matj$ matrices that are disorder-averaged together. The $x$ and $\hat x$ variables connected by a single arc are also constrained to have the same lower indices. Points connected by horizontal edges are Wick-paired together (averaged with respect to the bare action), and thus evaluate to the bare response function. Because $(K_0)^{ab}_{ij}(t,t') = 0$ for $t<t'$, the time coordinate of an $x$-type variable must always be greater than that of an $\hat x$-type variable, hence the directionality of the edges. Finally, all internal (i.e. not corresponding to the nodes at either end of the diagram) times and indices are summed/integrated over.

These representations are known as `rainbow' diagrams, and they have been obtained previously by other methods \cite{JANIK1997603,  feinberg1997non, brezin1994correlation}. Only rainbow diagrams that are `planar' (i.e. with no crossing arcs) survive in the thermodynamic limit \cite{HOOFT1974461}, which arise from the bare resolvent matrix $(\mathcal{H}_0)_{ij}^{ab}$ being diagonal in the indices $i$ and $j$. We give a motivation for this statement below.

To see that the contribution depicted diagrammatically above is indeed non-vanishing for the purposes of calculating the resolvent that we desire [see main text Eqs.~(13) and (22)], we compute
\begin{align}
		& \mathcal{L}_t \left\{ -\frac{i}{1! N } \sum_k \langle x^c_k(t) \hat x^{d\star}_k(0) S_{\mathrm{int}}\rangle_0 \right\}(\eta) \nonumber \\
		&= \frac{1}{N}\sum_{a_1,b_1, a'_1, b'_1}  \sum_{k, i_1, j_1}  \Bigg[(\mathcal{H}_0)_{k, i_1}^{c a_1} \overline{ \matj^{a_1b_1} (\mathcal{H}_0)_{j_1, j_1}^{b_1 a_1'} (\matj^\dagger)^{a'_1b'_1}} (\mathcal{H}_0)_{i_1,k}^{b_1' d} \Bigg] \nonumber \\
		&= \Bigg[ N\sum_{a_1,b_1, a'_1, b'_1}(\mathcal{H}_0)^{c a_1} \overline{ \matj^{a_1b_1} (\mathcal{H}_0)^{b_1 a_1'} (\matj^\dagger)^{a'_1b'_1}} (\mathcal{H}_0)^{b_1' d} \Bigg] \left[ \frac{1}{N^2} \sum_{k, i_1, j_1} \delta_{k,i_1} \delta_{j_1, j_1} \delta_{i_1, k} \right] \nonumber \\
		&= N\sum_{a_1,b_1, a'_1, b'_1}(\mathcal{H}_0)^{c a_1} \overline{ \matj^{a_1b_1} (\mathcal{H}_0)^{b_1 a_1'} (\matj^\dagger)^{a'_1b'_1}} (\mathcal{H}_0)^{b_1' d} ,
\end{align} 
which is of the order $N^0$ (since $\overline{ \matj^{a_1b_1}(\matj^\dagger)^{a'_1b'_1}}$). We note that we have used the convolution theorem for Laplace transforms above. We have also used the notation for $\mathcal{H}_0$ without lower indices to correspond to the $2\times 2$ matrix defined in Eq.~(9) of the main text. All of the terms that we neglected in the discourse above would have been subleading $1/N$, which we neglect in the thermodynamic limit, or simply outright vanishing.

To solidify the focus on planar diagrams, let us consider a higher-order term in $S_{\mathrm{int}}$ from the sum in Eq.~(22). The second order term in $S_\mathrm{int}$ has the following surviving terms (neglecting terms that pair unhatted dynamic variables together)
\begin{align}
	&-\frac{i}{2! }\langle x^c_k(t) \hat x^{d\star}_l(t') S_{\mathrm{int}}^2\rangle_0 \nonumber \\
	&= \sum_{a_1,b_1, a'_1, b'_1}  \int_{T_1>T'_1} dT_1 dT'_1\sum_{i_1, j_1} \sum_{a_2,b_2, a'_2, b'_2}  \int_{T_2>T'_2} dT_2 dT'_2\sum_{i_2, j_2}(-i)^6\Bigg\{ \overline{\matj_1^{a_1b_1} (\matj_1^\dagger)^{a'_1b'_1}} \;\overline{\matj_2^{a_2b_2} (\matj_2^\dagger)^{a'_2b'_2}} \nonumber \\
	& \Bigg[\left\langle  x^c_k(t)   \hat x^{a_1\star}_{i_1}(T_1) \right\rangle_0  \left\langle x^{b_1}_{j_1}(T_1)    \, \hat x^{a_1'\star}_{j_1}(T'_1) \right\rangle_0   \left\langle x^{b_1}_{i_1}(T_1')    \, \hat x^{a_2\star}_{i_2}(T_2) \right\rangle_0 \left\langle x^{b_2}_{j_2}(T_2)    \, \hat x^{a_2'\star}_{j_2}(T'_2) \right\rangle_0   \left\langle   x^{b'_2}_{i_2} (T'_2) \hat x^{d\star}_l(t') \right\rangle_0  \nonumber \\
	&+ \left\langle  x^c_k(t)   \hat x^{a_1\star}_{i_1}(T_1) \right\rangle_0  \left\langle x^{b_1}_{j_1}(T_1)    \,  \hat x^{a_2\star}_{i_2}(T_2) \right\rangle_0 \left\langle x^{b_2}_{j_2}(T_2)  \,  \hat x^{a_2'\star}_{j_2}(T'_2) \right\rangle_0  \left\langle   x^{b'_2}_{i_2} (T'_2)  \, \hat x^{a_1'\star}_{j_1}(T'_1) \right\rangle_0    \left\langle x^{b_1'}_{i_1}(T_1') \hat x^{d\star}_l(t') \right\rangle_0 \nonumber \\
	&+\left\langle  x^c_k(t)   \hat x^{a_1\star}_{i_1}(T_1) \right\rangle_0  \left\langle x^{b_1}_{j_1}(T_1)    \, \hat x^{a_2\star}_{i_2}(T_2) \right\rangle_0 \left\langle x^{b_2}_{j_2}(T_2)   \, \hat x^{a_1'\star}_{j_1}(T'_1) \right\rangle_0   \left\langle x^{b_1}_{i_1}(T_1')   \, \hat x^{a_2'\star}_{j_2}(T'_2) \right\rangle_0   \left\langle   x^{b'_2}_{i_2} (T'_2) \hat x^{d\star}_l(t') \right\rangle_0
	\Bigg] \Bigg\}  , \label{secondordersurviving}
\end{align}
and we see that there is symmetry between the times labelled $1$ and $2$ (which has cancelled the factor of $2!$) and symmetry between dashed and undashed times (which has cancelled a factor of $2^2$). We note that due to this symmetry, the specific labelling of the vertices in the diagrams is irrelevant. The numbers of ways of ordering the times always cancels the appropriate multiplicative factor, and so the only salient feature of a diagram is its topology (as discussed in more detail below). From now on, we drop the labelling of the vertices.

Only the first two of these Wick pairings survive in the thermodynamic limit. This can be seen simply by observing 
\begin{align}
	&\frac{1}{N}\sum_{k}\mathcal{L}_t\left\{-\frac{i}{2! }\langle x^c_k(t) \hat x^{d\star}_k(t') S_{\mathrm{int}}^2\rangle_0\right\} \nonumber \\
	&= \left[N^2\sum_{a_1,b_1, a'_1, b'_1}  \sum_{a_2,b_2, a'_2, b'_2}  \left[(\mathcal{H}_0)^{c a_1} \overline{\matj^{a_1b_1}_1 (\mathcal{H}_0)^{b_1 a_1'} (\matj^\dagger_1)^{a'_1b'_1} (\mathcal{H}_0)^{b_1' a_2}\;\matj_2^{a_2b_2} (\mathcal{H}_0)^{b_2 a_2'} (\matj_2^\dagger)^{a'_2b'_2}} (\mathcal{H}_0)^{b_2' d} \right] \right]\nonumber \\
	& \times\left[\frac{1}{N^3} \sum_{k, i_1, j_1, i_2, j_2} \left( \delta_{k,i_1} \delta_{j_1,j_1} \delta_{i_1,i_2} \delta_{j_2,j_2} \delta_{j_2,k}\right)	\right] \nonumber \\ 
	&+\left[N^2\sum_{a_1,b_1, a'_1, b'_1}  \sum_{a_2,b_2, a'_2, b'_2}  \left[(\mathcal{H}_0)^{c a_1} \overline{\matj_1^{a_1b_1} (\mathcal{H}_0)^{b_1 a_2}  \matj^{a_2b_2}_2 (\mathcal{H}_0)^{b_2 a_2'} (\matj_2^\dagger)^{a'_2b'_2} (\mathcal{H}_0)^{b_2' a_1'} (\matj_1^\dagger)^{a'_1b'_1}} (\mathcal{H}_0)^{b_1' d}  \right] \right]\nonumber \\
	& \times\left[\frac{1}{N^3} \sum_{k, i_1, j_1, i_2, j_2} \left(  \delta_{k,i_1} \delta_{j_1,i_2} \delta_{j_2,j_2} \delta_{i_2,j_1}\delta_{i_1,k} \right)	\right] \nonumber \\
	&+\left[N^2\sum_{a_1,b_1, a'_1, b'_1}  \sum_{a_2,b_2, a'_2, b'_2}  \left[(\mathcal{H}_0)^{c a_1} \overline{\matj_1^{a_1b_1} (\mathcal{H}_0)^{b_1 a_2} \matj_2^{a_2 b_2} (\mathcal{H}_0)^{b_2 a_1'}\; (\matj_1^\dagger)^{a_1' b_1} (\mathcal{H}_0)^{b_1 a_2'} (\matj_2^\dagger)^{a'_2b'_2}} (\mathcal{H}_0)^{b_2' d} \right] \right]\nonumber \\
	& \times\left[\frac{1}{N^3} \sum_{k, i_1, j_1, i_2, j_2} \left(  \delta_{k,i_1} \delta_{j_1,i_2} \delta_{j_2,j_1} \delta_{i_1,j_2} \delta_{i_2,k}\right)	\right],
\end{align} 
where we see that the first two products of Kronecker deltas evaluate to $1$, whereas the final set gives $1/N^2$. Note that we have introduced two sets of statistically independent $\matj$-type matrices above, so that $\overline{\matj_1^{ab} \matj_2^{a'b'}} = 0$ for all combinations of upper indices.

The real advantage of the diagrammatic representation is in identifying those Wick pairings that vanish in the same way that the third pairing in Eq.~(\ref{secondordersurviving}) did. The Wick pairings in Eq.~(\ref{secondordersurviving}) can be represented diagrammatically as 
\begin{figure}[H]
	\centering 
	\includegraphics[scale = 0.45]{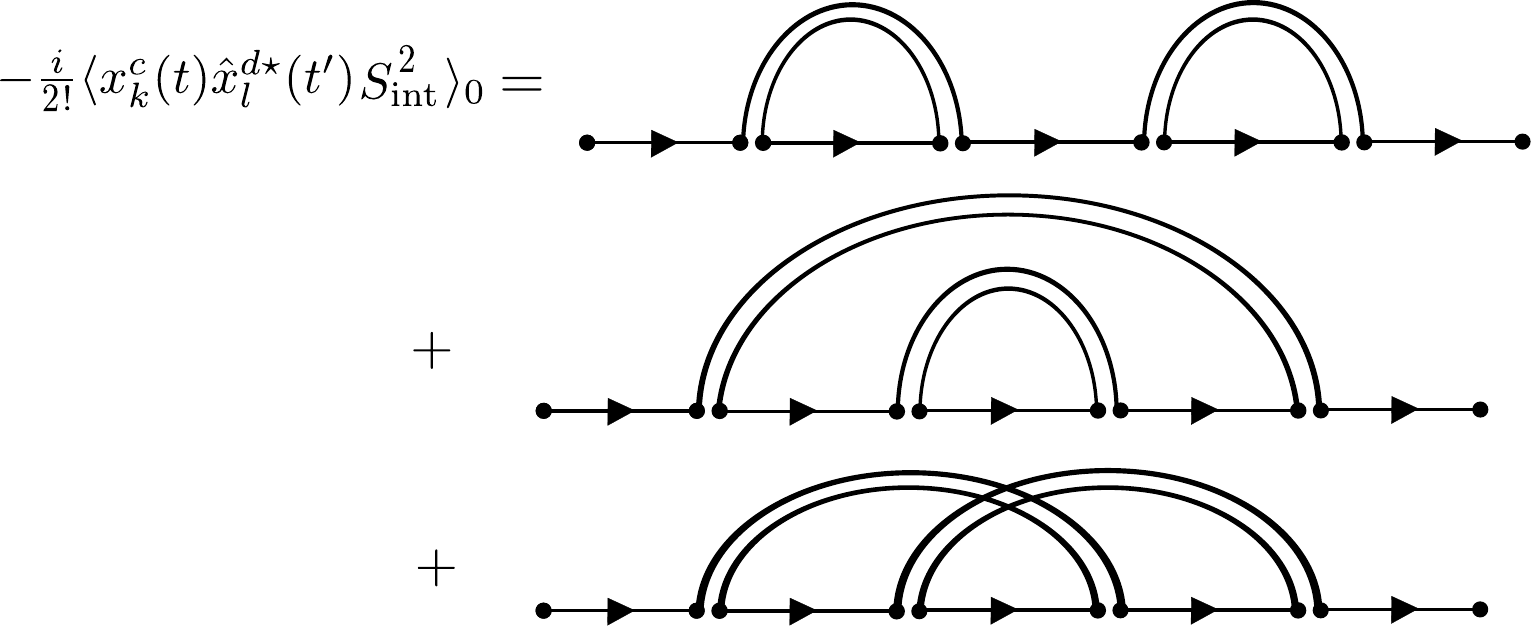}
	\captionsetup{labelformat=empty}
\end{figure}
These first two digrams each have three disconnected sets of directed edges, which corresponds to three factors of $\sum_l \delta_{ll}$. This cancels the factor of $N^{-3}$ when finding $N^{-1}\mathrm{Tr}\mathcal{H}$. In contrast, the third term in Eq.~(\ref{secondordersurviving}) is represented by a diagram whose directed edges are all connected by arcs. This means that one obtains only a single factor of $\sum_{l} \delta_{ll}$ after summing over all other indices. One thus finds that this diagram is an $O(N^{-2})$ contribution to the sum $N^{-1}\mathrm{Tr}\mathcal{H}$.

We thus see that the `non-planar' diagram gives a contribution that vanishes in the limit $N\to \infty$ and only the planar diagrams survive. We also saw that a factor of $1/(2! 2^2)$ cancelled due to time ordering. 

To summarise, we have so far argued that the following simplifying rules apply generally: 
\begin{enumerate}
	\item The only Wick pairings we need to consider pair solely hatted and unhatted dynamic variables.
	\item We can make the replacement $\overline{ \mati_{ij}^{ab} \mati_{i'j'}^{a'b'}} \to \overline{\matj^{ab} (\matj^\dagger)^{a'b'}} \delta_{i, j'}\delta_{ji'}$.
	\item We can ignore the complex conjugate terms in the action when calculating the response functions $\overline{\delta x_i^a(t)/\delta h_j^b(t')}$.
	\item The only non-vanishing Wick pairings for $N\to \infty$ correspond to planar diagrams.
	\item The number of combinations of Wick pairings that are equivalent up to time ordering always exactly cancels a prefactor, allowing us to discard the labelling of the internal nodes in the Feynman diagrams.
\end{enumerate}

Let us argue in more detail that points (iv) and (v) above indeed apply generally. When calculating $N^{-1} \sum_i \mathcal{H}_{ii}^{ab}$, the term in the series in Eq.~(22) containing $S_{\mathrm{int}}^r$ gives rise to $2r + 1$ factors of the response function, $2r$ internal times and is multiplied by a factor $1/(2^r N^{(r+1)} r!)$. 

As we have exemplified, the factor of $N^{-(r+1)}$ is cancelled when we perform the sums of the lower indices when calculating $N^{-1} \sum_i \mathcal{H}_{ii}^{ab}$. The $S_{\mathrm{int}}^r$ term gives rise to $2r+1$ factors of the bare response function, which is represented diagrammatically as $2r+1$ directed edges, and $r$ factors of $N\overline{\mathcal{J} \mathcal{J}^\dagger}$, which correspond to $r$ double arcs. Each set of disconnected edges corresponds to a factor of $N$ when we sum over the lower indices. The arcs must connect at least $r$ edges to another edge, so the maximum number of sets of disconnected edges is $r+1$, which is satisfied for planar diagrams. When the arcs cross, we obtain fewer disconnected sets of edges. This is why only planar diagrams survive in the thermodynamic limit, since only they cancel the factor of $N^{-(r+1)}$ when one sums over the lower indices.

The remaining factor of $1/(2^r r!)$ is cancelled due to time-ordering. More specifically, each pair of dashed and undashed internal times (corresponding to each factor of $S_{\mathrm{int}}$) can be reversed. This is also equivalent to switching the lower indices, say $i_1$ and $j_1$. This cancels the factor of $2^r$. The order of the \textit{first appearances} of the pairs of internal times can also be arranged in any way. This corresponds to putting the pairs of indices $(i_1, j_1)$, $(i_2, j_2)$, ..., in any order. This in turns cancels the factor of $r!$. So, we can neglect the multiplicative factor of $1/(2^r r!)$ and drop the labelling of the vertices in the diagrams.

One therefore sees that the sum in main text Eq.~(22) can be evaluated in the thermodynamic limit by considering the set of all planar rainbow diagrams, where we consider diagrams that can be obtained from one another simply by changing the labels of the vertices (while maintaining the topology) as being `the same'. Diagrams that are `the same' should not contribute to the sum more than once. Exactly the same principle applies to the ribbon diagrams for the sparse correction.

As a final example, we find the following non-vanishing diagrams for the third-order term
\begin{figure}[H]
	\centering 
	\includegraphics[scale = 0.45]{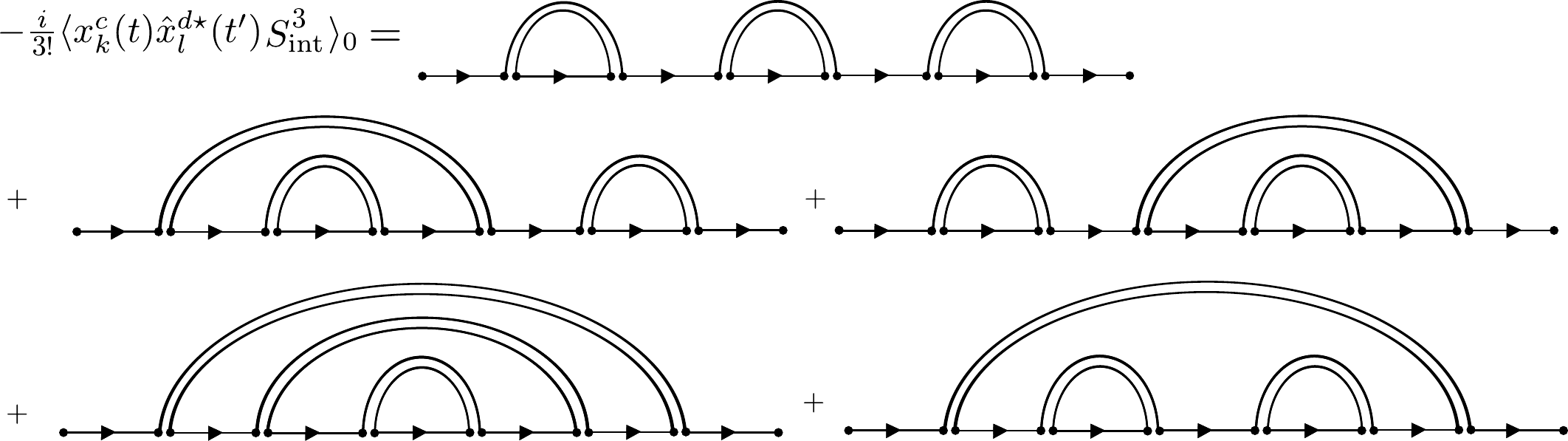}
	\captionsetup{labelformat=empty}
\end{figure}
These diagrams give us 
\begin{align}
	&\frac{1}{N} \sum_k \mathcal{L}_t \left\{-\frac{i}{3!}\left\langle x^c_k(t) \hat x_k^{d \star}(t') S_\mathrm{int}^3 \right\rangle_0 \right\} \nonumber \\  
	&=N^3\overline {\mathh_0^{ca_1} \matj_1^{a_1 b_1} \mathh_0^{b_1 a_1'} (\matj_1^\dagger)^{a_1' b_1'} \mathh_0^{b_1' a_2} \matj_2^{a_2 b_2} \mathh_0^{b_2 a_2'} (\matj_2^\dagger)^{a_2' b_2'} \mathh_0^{b_2' a_3} \matj_3^{a_3 b_3} \mathh_0^{b_3 a_3'} (\matj_3^\dagger)^{a_3' b_3'} \mathh_0^{b_3' d}} \nonumber \\
	&+N^3\overline {\mathh_0^{ca_1} \matj_1^{a_1 b_1} \mathh_0^{b_1 a_2} \matj_2^{a_2 b_2} \mathh_0^{b_2 a_2'} (\matj_2^\dagger)^{a_2' b_2'}  \mathh_0^{b_2' a_1'}  (\matj_1^\dagger)^{a_1' b_1'}  \mathh_0^{b_1' a_3} \matj_3^{a_3 b_3} \mathh_0^{b_3 a_3'} (\matj_3^\dagger)^{a_3' b_3'} \mathh_0^{b_3' d}} \nonumber \\
	&+N^3\overline {\mathh_0^{ca_1} \matj_1^{a_1 b_1} \mathh_0^{b_1 a_1'} (\matj_1^\dagger)^{a_1' b_1'} \mathh_0^{b_1' a_2} \matj_2^{a_2 b_2} \mathh_0^{b_2 a_3} \matj_3^{a_3 b_3} \mathh_0^{b_3 a_3'} (\matj_3^\dagger)^{a_3' b_3'} \mathh_0^{b_3' a_2'} (\matj_2^\dagger)^{a_2' b_2'} \mathh_0^{b_2' d}} \nonumber \\
	&+N^3\overline {\mathh_0^{ca_1} \matj_1^{a_1 b_1} \mathh_0^{b_1 a_2} \matj_2^{a_2 b_2} \mathh_0^{b_2 a_3} \matj_3^{a_3 b_3}    \mathh_0^{b_3 a_3'}  (\matj_3^\dagger)^{a_3' b_3'} \mathh_0^{b_3' a_2'}(\matj_2^\dagger)^{a_2' b_2'} \mathh_0^{b_2' a_1'} (\matj_1^\dagger)^{a_1' b_1'}  \mathh_0^{b_1' d}} \nonumber \\
	&+N^3\overline {\mathh_0^{ca_1} \matj_1^{a_1 b_1} \mathh_0^{b_1 a_2} \matj_2^{a_2 b_2} \mathh_0^{b_2 a_2'} (\matj_2^\dagger)^{a_2' b_2'}  \mathh_0^{b_2' a_3}   \matj_3^{a_3 b_3} \mathh_0^{b_3 a_3'} (\matj_3^\dagger)^{a_3' b_3'} \mathh_0^{b_3' a_1'} (\matj_1^\dagger)^{a_1' b_1'} \mathh_0^{b_1' d}} ,
\end{align}
where the sum over repeated upper indices above is implied. We thus see how the formidable task of evaluating the series in Eq.~(22) of the main text simplifies to summing a series of planar diagrams, each of which can be evaluated fairly easily. The strategy for evaluating the full resulting series of planar diagrams is described in the next section.

\section{The distributive convention and recovering the elliptic law}\label{appendix:distributive}
We employ one additional diagrammatic convention to simplify the notation when we perform sums over many diagrams. We denote a sum of planar diagrams by an edge with a double arrow, accompanied by a label for identification purposes. For example, let us take the surviving planar diagrams for the second-order term above $-\frac{i}{2!} \langle x_k^c(t) \hat x_l^{d\star} S_\mathrm{int}^2\rangle_0$, for which we write
\begin{figure}[H]
	\centering 
	\includegraphics[scale = 0.45]{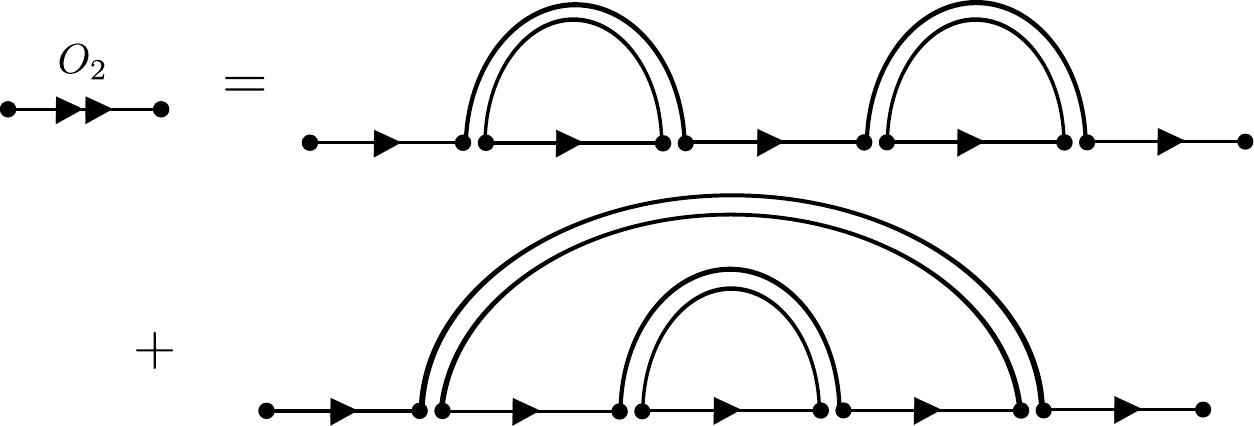}
	\captionsetup{labelformat=empty}
\end{figure}
When we draw an arc (or ribbon) over a double-arrowed edge, this is also to be interpreted as a sum of diagrams. Bearing in mind the distributivity of matrix multiplication over addition, this provides a consistent and meaningful notation. Precisely, for the example above we have
\begin{figure}[H]
	\centering 
	\includegraphics[scale = 0.45]{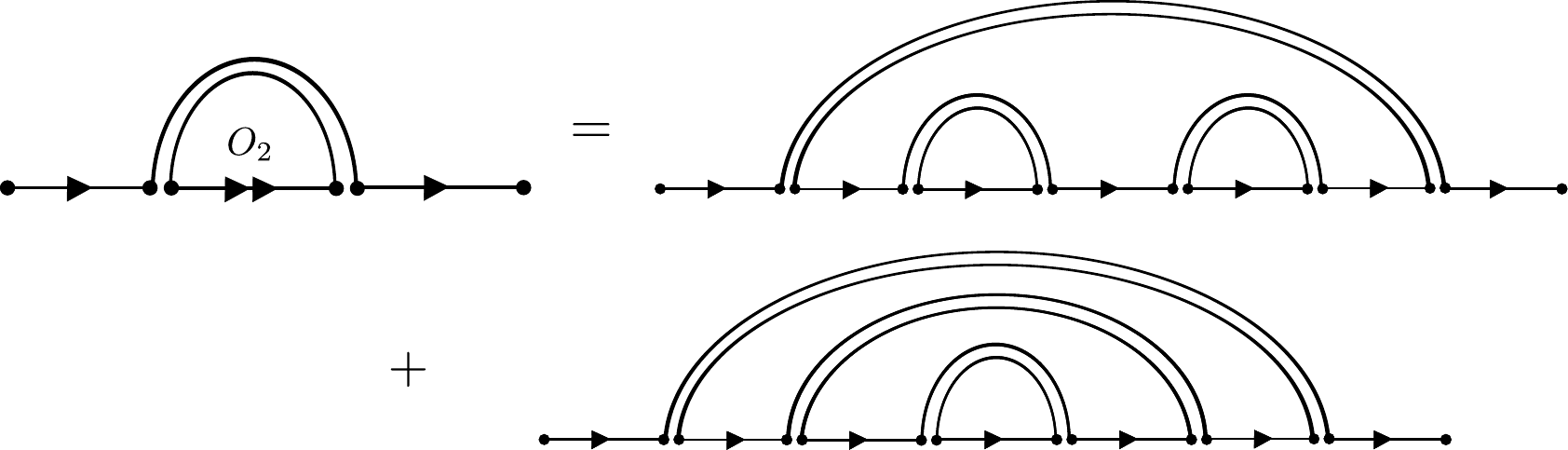}
	\captionsetup{labelformat=empty}
\end{figure}

We use this convention in Fig. 2 of the main text to sum the series of planar diagrams. We summarise the argument briefly here as to why the two series in Fig. 2 are same. 

Let us say that a diagram has $r$ `external arcs' if, by following a completely connected path of vertices from the leftmost vertex to the rightmost, we traverse $r$ arcs. We can categorise a general planar rainbow diagram by the number of external arcs that it has, since no arcs intersect. The full collection of diagrams with a single external arc, for example, can then be found by taking every planar diagram in the series, placing each of them inside a single arc, and attaching two directed edges to either side. Similar statements apply for diagrams with any number of external arcs. The complete series of planar diagrams can therefore be generated by summing together all of the sets of diagrams with $r = 1, 2, 3, \cdots$ external arcs, where under each arc is the sum of all diagrams in the series. In this statement, we have thus identified a self-similarity quality of the series, which allows us to perform the resummation.

Because of this argument, we thus see why the full series of rainbow diagrams can be represented by the simpler series involving $\mathh$ (the dressed resolvent) in main text Fig. 2. This simpler series is recognised to be geometric and is given by
\begin{align}
	&\lim_{\eta\to 0} N^{-1}\mathcal{H}_{kk}^{cd}= \lim_{\eta\to 0}N^{-1}\mathcal{L}_t\left[-i\langle x^c_k(t) \hat x^{d\star}_k(0)  \rangle_S \right](\eta) \nonumber \\
	&= (\mathcal{H}_0)^{cd}_{kk} + (\mathcal{H}_0)^{ca_1}_{k i_1}\Sigma^{a_1 b_1'} (\mathcal{H}_0)^{b_1'd}_{i_1k} + (\mathcal{H}_0)^{ca_1}_{k i_1} \Sigma^{a_1 b_1'} (\mathcal{H}_0)^{b_1' a_2}_{i_1 i_2} \Sigma^{a_2 b'_2} (\mathcal{H}_0)^{b_2'd}_{i_2 k} + \cdots , \label{selfconsistentseriesexplicit}
\end{align}
where sums over repeated indices above are implied, and we have defined
\begin{align}
	\Sigma^{a b'} \equiv \sum_{j, b, a'}\overline{ \matj^{ab} \mathh_{j j}^{b a'} (\matj^\dagger)^{a'b'} }.
\end{align}
Noting that $(\mathcal{H}_0)^{ab}_{ij} \propto \delta_{ij}$, we can thus sum the Eq.~(\ref{selfconsistentseriesexplicit}) to yield the self-consistent expression for the Hermitized resolvent in Eq. (26) of the main text.

\section{First-order in $1/p$ correction to the elliptic and semi-circular laws}\label{appendix:firstorder}
This section provides more detail on how the results for the modified elliptic and semi-circular laws in Eqs.~(40)--(43) are derived from the self-consistent expression for the Hermitized resolvent in Eq.~(39). 

\subsection{Finding the resolvent $C$}
We begin by finding two expressions for the resolvent $C$ -- one that is valid inside the bulk region to which most of the eigenvalues are confined, and one that is valid outside of this region. 

Using the notation in Eq.~(10) and the definitions in Eq.~(33), one obtains from Eq.~(39) of the main text
\begin{align}
	A&= \frac{1}{\Delta}\left\{\sigma^2 A + \frac{\sigma^4 A}{p} \left[\Gamma^{(1)}_4 AD + \Gamma^{(2)}_4 (B^2 + C^2) +\Gamma^{(3)}_4 BC \right] \right\} ,\nonumber \\
	B&= \frac{1}{\Delta}\left\{\omega-\Gamma \sigma^2 C - \frac{\sigma^4}{p}\left[2 \Gamma^{(2)}_4 ADC +\Gamma^{(3)}_4 (C^3 + ADB) \right] \right\}, \nonumber \\
	C&= \frac{1}{\Delta}\left\{ \omega^\star-\Gamma \sigma^2 B - \frac{\sigma^4}{p}\left[2 \Gamma^{(2)}_4 ADB +\Gamma^{(3)}_4 (B^3 + ADC) \right] \right\}, \nonumber \\
	D &= \frac{1}{\Delta}\left\{ \sigma^2 D + \frac{\sigma^4 D}{p} \left[\Gamma^{(1)}_4 AD + \Gamma^{(2)}_4 (B^2 + C^2) +\Gamma^{(3)}_4 BC \right] \right\} , \nonumber \\
	\frac{1}{\Delta} &= BC - AD. \label{simultsparsegeneral}
\end{align}
There are two solutions to this set of equations. In one case, we have $A = D= 0$, which yields in combination with the second and the last of Eqs.~(\ref{simultsparsegeneral})
\begin{align}
	C &= \frac{1}{\omega - \Gamma \sigma^2 C - \frac{\sigma^4}{p} \Gamma^{(3)}_4 C^3 }, \nonumber \\
	&\approx \frac{1}{\omega - \Gamma \sigma^2 C} + \frac{\sigma^4}{p} \Gamma^{(3)}_4 C^5 . \label{resolventoutside}
\end{align}
This expression is analytic in $\omega$, so it must correspond to the region of the complex plane in which the eigenvalue density is zero [see main text Eq.~(3)]. 

Alternatively, we can obtain the following expression from the equation for $D$ in Eqs.~(\ref{simultsparsegeneral})
\begin{align}
	AD = BC - \frac{1}{\sigma^2} + \frac{1}{p} \left[\Gamma^{(1)}_4\left(BC - \frac{1}{\sigma^2} \right) + \Gamma^{(2)}_4 (B^2 + C^2)  + \Gamma^{(3)}_4 BC  \right]. \label{adexpression}
\end{align}
We can therefore eliminate $AD$ to find (accurate to first order in $1/p$)
\begin{align}
	C &= \frac{\omega^\star - \Gamma \sigma^2 B}{\sigma^2}  - \frac{\sigma^2}{p} h, \nonumber \\
	B &= C^\star, \nonumber \\
	h &= \left( BC - \frac{1}{\sigma^2} \right) \left[(\Gamma^{(1)}_4 + \Gamma^{(3)}_4  ) C + 2 \Gamma^{(2)}_4 B  \right] + \Gamma^{(2)}_4 C \left(B^2 + C^2 \right) +\Gamma^{(3)}_4 B(B^2+C^2) .
\end{align}
Using the fact that $B = C^\star$, we arrive at an alternative expression for $C$
\begin{align}
	C = \frac{\omega^\star - \Gamma \omega}{\sigma^2 (1-\Gamma^2)} - \frac{\sigma^2}{p} \frac{1}{(1-\Gamma^2)} (h - \Gamma h^\star) . \label{resolventinside}
\end{align}
This expression is non-analytic in $\omega$, so it must correspond to the bulk region of the eigenvalue spectrum.

\subsection{Boundary of the support}
The boundary of the support of the eigenvalue spectrum is given by the set of values of $\omega$ for which both Eq.~(\ref{resolventoutside}) and Eq.~(\ref{resolventinside}) are simultaneously true. To simplify the problem of finding the boundary, we define a new set of coordinates. We let 
\begin{align}
	z = \frac{\omega - \Gamma \omega^\star}{(1 - \Gamma^2) } .
\end{align}
With this in mind, we find
\begin{align}
	C&= \frac{1}{z} + \frac{1}{p} \left[\sigma^4 \Gamma^{(3)}_4 C^5  - \Gamma\sigma^6 C^2 \frac{(h - \Gamma h^\star)}{(1-\Gamma^2)} \right], \nonumber \\
	C &= \frac{z^\star}{\sigma^2} - \frac{1}{p} \frac{\sigma^2}{(1-\Gamma^2)} (h - \Gamma h^\star).
\end{align}
Equating these expressions for $C$, we obtain (defining also $z = r_z e^{i \theta_z}$ and using that $r_z^2 \approx \sigma^2$ to zeroth order)
\begin{align}
	r_z^2 &= \sigma^2 + \frac{ \sigma^2}{p} \left[\Gamma^{(3)}_4 e^{-4i\theta_z} +\sigma^3\frac{(e^{i\theta_z} - \Gamma e^{-i\theta_z} ) }{(1-\Gamma^2)} (h - \Gamma h^\star)  \right] , \label{rcompact}
\end{align}
Similarly, we obtain (noting that $h$ only appears in the term proportional to $1/p$)
\begin{align}
	h &= \left( \frac{1}{r_z^2} - \frac{1}{\sigma^2} \right) \frac{r_z}{\sigma^2}\left[(\Gamma^{(1)}_4 + \Gamma^{(3)}_4  ) e^{-i\theta_z} + 2 \Gamma^{(2)}_4 e^{i\theta_z} \right] \nonumber \\
	&\,\,\,\,\,+ \frac{r_z^3}{\sigma^6}\left[\Gamma^{(2)}_4  \left(e^{i\theta_z}  + e^{-3i\theta_z}  \right) +\Gamma^{(3)}_4 e^{3i\theta_z} + \Gamma^{(3)}_4 e^{-i\theta_z}  \right], \nonumber \\
	&\approx  \frac{1}{\sigma^3} \left(e^{2i\theta_z}  + e^{-2i\theta_z}  \right)  \left[\Gamma^{(2)}_4  e^{-i\theta_z}   +\Gamma^{(3)}_4 e^{i\theta_z}  \right], \nonumber \\
	\Rightarrow h - \Gamma h^\star &= \frac{1}{\sigma^3} \left(e^{2i\theta_z}  + e^{-2i\theta_z}  \right)  \left[\Gamma^{(2)}_4 ( e^{-i\theta_z} - \Gamma e^{i\theta_z}  )  +\Gamma^{(3)}_4 (e^{i\theta_z} - \Gamma e^{-i\theta_z} )  \right]. 
\end{align}
Inserting this into Eq.~(\ref{rcompact}), we finally arrive at a parametric expression for the boundary of the support of the eigenvalue spectrum 
\begin{align}
	r_z^2  =& \sigma^2 + \frac{ \sigma^2}{p (1- \Gamma^2)} \Big\{(1+\Gamma^2) \Gamma_4^{(3)} - 2 \Gamma \Gamma_4^{(2)} \nonumber \\
	&+ \left[2 (1 + \Gamma^2)\Gamma_4^{(2)} - 4 \Gamma \Gamma_4^{(3)} \right]\cos2\theta_z + 2\left(\Gamma_4^{(3)} - \Gamma \Gamma_4^{(2)} \right) \cos 4 \theta_z \Big\} \label{boundarygeneral} ,
\end{align}
where one obtains the boundary in the original coordinates via
\begin{align}
	x &= (1+\Gamma) r_z \cos\theta_z , \nonumber \\
	y &= (1-\Gamma) r_z \sin\theta_z  . \label{inversetransformgeneral}
\end{align}
One notes that in the case $\Gamma = 0$, we have no need for the coordinate transformation and the above expressions reduce to 
\begin{align}
	r(\theta) &= \sigma\left\{1 + \frac{1}{2p}\left[ \Gamma_4^{(3)}  + 2 \Gamma_4^{(2)}\cos2\theta + 2 \Gamma_4^{(3)}\cos 4 \theta_z \right] \right\}. \label{boundarycircular}
\end{align}
To obtain an expression in cartesian coordinates from Eq.~(\ref{boundarygeneral}), one simply expands $\cos2\theta_z$ and $\cos4\theta_z$ in terms of $\cos\theta_z$ and $\sin\theta_z$ and then substitutes $x$ and $y$ using Eqs.~(\ref{inversetransformgeneral}). It is useful to note that in the term multiplying $1/p$, $r_z$ can be replaced with $\sigma$ and the error will be $O(1/p^2)$. One eventually finds the following expression that is valid to first order in $1/p$
\begin{align}
	\frac{x^2}{a^2} + \frac{y^2}{b^2} = \sigma^2 \left[ 1 + \frac{1}{p} \left( c -  \frac{d}{\sigma^4} \frac{x^2}{a^2} \frac{y^2}{b^2}\right)\right],  \label{ellipsemodcart}
\end{align}
where we have
\begin{align}
	a &= (1 + \Gamma) \left[ 1 + \frac{1}{p} \frac{(1 + \Gamma^2) \Gamma_4^{(2)} - 2 \Gamma \Gamma_4^{(3)}}{1-\Gamma^2} \right], \nonumber \\ 
	b &= (1 - \Gamma) \left[ 1 - \frac{1}{p} \frac{(1 + \Gamma^2) \Gamma_4^{(2)} - 2 \Gamma \Gamma_4^{(3)}}{1-\Gamma^2} \right], \nonumber \\ 
	c &= \frac{(3 + \Gamma^2) \Gamma_4^{(3)} - 4 \Gamma \Gamma_4^{(2)}}{1-\Gamma^2}, \nonumber \\
	d &= 16\frac{\Gamma_4^{(3)} - \Gamma \Gamma_4^{(2)}}{1-\Gamma^2}.
\end{align}
In the limit $p\to \infty$, we clearly recover the usual elliptic law for the dense case. To obtain the formula in Eq.~(40) of the main text, we first note that the value of $y$ at which $x=0$ on the modified ellipse is given by (to first order in $1/p$)
\begin{align}
	y_c = \sigma(1-\Gamma) + \frac{\sigma}{2p} \left[(3 + \Gamma) \Gamma_4^{(3)} - 2 (1+\Gamma) \Gamma_4^{(2)} \right] ,
\end{align}
and similarly, we have 
\begin{align}
	x_c = \sigma(1+\Gamma)  + \frac{\sigma}{2p} \left[ (3-\Gamma) \Gamma_4^{(3)} + 2 (1-\Gamma) \Gamma^{(2)}_4\right] . \label{xcmuzero}
\end{align}
One can show that (again to first order in $1/p$) one has
\begin{align}
	\frac{y_c^2}{x_c^2}  = \frac{b^2}{a^2} .
\end{align}
We can therefore rewrite for the expression in Eq.~(\ref{ellipsemodcart}) as
\begin{align}
	y_\pm(x) &= \pm y_c\sqrt{1 - \frac{x^2}{x_c^2}}\left[1 + \frac{1}{p}\frac{ x^2 d}{\sigma^2 a^2}\right]^{-1/2}. \label{ypmellipse}
\end{align}
This is the most convenient expression to use for plotting the modified ellipse, since both of the square root singularities at $x = x_c$ and $y = y_c$ are faithfully preserved.
\subsection{Density inside the support}
From the expression in Eq.~(\ref{resolventinside}), we can also find the eigenvalue density via Eq.~(3) of the main text. Differentiating, one finds
\begin{align}
	\frac{\partial C}{\partial \omega^\star} = \frac{1}{\sigma^2(1-\Gamma^2)} - \frac{\sigma^2}{p} \frac{1}{(1-\Gamma^2)} \left[\frac{\partial h}{\partial \omega^\star} - \Gamma\frac{\partial h^\star}{\partial \omega^\star} \right],
\end{align}
which subsequently yields
\begin{align}
	\rho(x, y) &= \frac{1}{\pi\sigma^2(1-\Gamma^2)} + \frac{1}{p} \frac{1}{\pi\sigma^2 (1-\Gamma^2)^2} \left[-4 \Gamma_4^{(2)} \Gamma  + (\Gamma_4^{(1)} +\Gamma_4^{(3)}) (1+\Gamma^2)  \right] \nonumber \\
	&- \frac{1}{p} \frac{2}{\pi\sigma^2 (1-\Gamma^2)^2} \left[ \Gamma_4^{(1)} (1 - \Gamma + \Gamma^2) + 3 \Gamma_4^{(2)} (1-\Gamma)^2 + \Gamma_4^{(3)} (2 - 5 \Gamma + 2 \Gamma^2)\right] \frac{x^2}{\sigma^2 (1+\Gamma)^2} \nonumber \\
	&- \frac{1}{p} \frac{2}{\pi\sigma^2 (1-\Gamma^2)^2} \left[ \Gamma_4^{(1)} (1 + \Gamma + \Gamma^2) - 3 \Gamma_4^{(2)} (1+\Gamma)^2 + \Gamma_4^{(3)} (2 + 5 \Gamma + 2 \Gamma^2)\right] \frac{y^2}{\sigma^2 (1-\Gamma)^2}. \label{generaldensityfull}
\end{align}

\subsection{Correction to the generalised Wigner semi-circle law}
To find Eq.~(42) of the main text, one must integrate Eq.~(\ref{generaldensityfull}) between the limits given by Eq.~(\ref{ypmellipse}).

Integrating Eq.~(\ref{generaldensityfull}) between $y_-(x)$ and $y_+(x)$, one obtains 
\begin{align}
	\rho_x = \frac{2}{\pi\sigma^2 (1-\Gamma^2)} \left\{ 1 + \frac{1}{p} \frac{1}{ (1-\Gamma^2)}\left[\alpha_0 - \alpha_x \frac{x^2}{\sigma^2(1+\Gamma)^2} - \alpha_y\frac{y_+^2}{3\sigma^2(1-\Gamma)^2} \right]\right\} y_+(x) , 
\end{align}
where we have
\begin{align}
	\alpha_0 &= -4 \Gamma_4^{(2)} \Gamma  + (\Gamma_4^{(1)} +\Gamma_4^{(3)}) (1+\Gamma^2) , \nonumber \\
	\alpha_x &= \Gamma_4^{(1)} (1 - \Gamma + \Gamma^2) + 3 \Gamma_4^{(2)} (1-\Gamma)^2 + \Gamma_4^{(3)} (2 - 5 \Gamma + 2 \Gamma^2), \nonumber \\
	\alpha_y &= \Gamma_4^{(1)} (1 + \Gamma + \Gamma^2) - 3 \Gamma_4^{(2)} (1+\Gamma)^2 + \Gamma_4^{(3)} (2 + 5 \Gamma + 2 \Gamma^2).
\end{align}

After expanding the expression that multiplies $\sqrt{x_c^2 - x^2}$ up to leading order in $1/p$ and simplifying the algebra, one eventually finds
\begin{align}
	\rho_x(x) =& \frac{2 \sqrt{x_c^2-x^2}}{\pi(1+\Gamma)^2\sigma^2} \Bigg( 1 +\frac{1}{p} \frac{1}{3(1+\Gamma)} \bigg\{ (\Gamma_4^{(1)}- \Gamma_4^{(3)})(1-\Gamma)   \nonumber \\
	&- 4 \left[(1-\Gamma)\Gamma_4^{(1)}+ 6 \Gamma_4^{(2)}(1-\Gamma) + 2 \Gamma_4^{(3)}(4-\Gamma)\right] \frac{x^2}{\sigma^2(1+\Gamma)^2}\bigg\} \Bigg).
\end{align}
Noting the expression for $x_c$ in Eq.~(41) of the main text, one thus arrives at the expression in Eq.~(42), which is valid up to $O(1/p)$.
. 
\subsection{Correction to the semi-circle law in the case $\Gamma =1$}\label{appendix:semicirclegamma1}
In the case $\Gamma = 1$, the eigenvalues are all real, and so the resolvent is an analytic function of $\omega$ along the real axis. This means that one can also use Eq.~(4) of the main text in combination with Eq.~(\ref{resolventoutside}) to obtain the density of eigenvalues along the real axis. We proceed along similar lines to Ref. \cite{kim1985density} (see Appendix A in particular of this reference). We do this mainly to verify the formula in Eq.~(42) of the main text.

Rearranging Eq.~(\ref{resolventoutside}) and setting $\Gamma = 1$, one obtains [noting $\Gamma_4^{(1)}=\Gamma_4^{(2)}=\Gamma_4^{(3)}$]
\begin{align}
	\sigma^2 C^2 - \omega C +1 + \frac{\Gamma_4^{(1)}}{p}  (\omega C - 1)^2 = 0 ,
\end{align}
where we have used the fact that to leading order in $1/p$ we have $\sigma^2 C^2 =  \omega C - 1$. Solving this quadratic  for $C$, one finds
\begin{align}
	C = \frac{1}{2 (\sigma^2 + \Gamma_4^{(1)} \omega^2/p)} \left[ \omega + \frac{2 \Gamma_4^{(1)} \omega}{p}  + \sqrt{\omega^2 - 4 \sigma^2 \left(1 + \frac{\Gamma_4^{(1)}}{p}\right)} \right] .
\end{align}
Noting that in this case $x_c^2 = 4 \sigma^2 (1 + \Gamma_4^{(1)}/p)$ [see Eq.~(41) of the main text], we obtain
\begin{align}
	C(x) \approx \frac{2}{x^2_c} \left[ 1 + \frac{\Gamma_4^{(1)}}{p} \left(1 - \frac{4 x^2}{x_c^2}\right)\right]\left[ x + \frac{2 \Gamma_4^{(1)} x}{p}  + \sqrt{x^2 - x_c^2} \right]. 
\end{align}
Using main text Eq.~(4), one then obtains
\begin{align}
	\rho_x(x) = \frac{2}{\pi x^2_c} \left[ 1 + \frac{\Gamma_4^{(1)}}{p} \left(1 - \frac{4 x^2}{x_c^2}\right)\right] \sqrt{x_c^2 - x^2 } .\label{semicirclesparsegeneral}
\end{align}	
We have thus succeeded in finding an approximation for the real eigenvalue density that is valid up to leading order in $1/p$. Crucially, this expression also preserves the critical point $x_c$ at which the eigenvalue density first becomes non-zero up to leading order in $1/p$. This expression can be seen readily to be in agreement with Eq.~(42) of the main text in the case $\Gamma = 1$.

\section{Outlier eigenvalue and modified bulk spectrum due to $\mu \neq 0$ }\label{appendix:outlier}
In this appendix, we discuss how the action is altered by the inclusion of a non-zero value of $\mu$, and we present the new kinds of Feynman diagram that come about in the calculation of the outlier eigenvalue.

Let us now compute the action of the system with $\mu \neq 0$ to first order in $1/p$. Letting $\mathcal{Z} = \mathcal{J}- \frac{1}{p}\mathcal{M}$ where $\mathcal{M}^{ab} = \mu$ for $a\neq b$ and zero otherwise, we obtain by averaging the generating functional in Eq.~(14) of the main text 
\begin{widetext}
	
	\begin{align}
		S_0 &= i  \sum_{i,a} \int dt \, \hat x^{a\star}_i \left(\dot x_i^a +\sum_{b, j}(\mathinv)^{ab}_{ij}  x^b_j  \right)    , \,\,\,\,\,\,\,\,\,\,	S_{0}^{\mathrm{off}} = -\frac{i}{N}  \int dt \, \hat x^{a\star}_i \mathcal{M}^{ab}  x^b_j    , \nonumber \\
		S_1 &= \frac{p}{2N} \frac{(-i)^2}{2!}\, {}^2C_1  \int dt_1 dt_2 \left\langle\hat x^{a\star}_i \mathcal{Z}^{ab} x^{b}_j  \hat x^{a'\star}_{j} (\mathcal{Z}^\dagger)^{a'b'} x^{b'}_{i} \right\rangle_\pi \nonumber \\
		S_2 &=\frac{p}{2N} \frac{(-i)^4}{4!} \, {}^4C_2  \int dt_1 \cdots dt_4 \left\langle\hat x^{a_1\star}_{i} \mathcal{Z}^{a_1b_1} x^{b_1}_{j}  \hat x^{a_2\star}_{j} (\mathcal{Z}^\dagger)^{a_2b_2} x^{b_2}_{i}\hat x^{a_3\star}_{i} \mathcal{Z}^{a_3b_3} x^{b_3}_{j}\hat x^{a_4\star}_{j} (\mathcal{Z}^\dagger)^{a_4b_4} x^{b_4}_{i}\right\rangle_\pi  \nonumber \\
		&+ \frac{1}{2N} \frac{1}{p} \frac{(-i)^2}{2!}\, {}^2C_1  \int dt_1 dt_2 \left\langle\hat x^{a\star}_i \mathcal{M}^{ab} x^{b}_j  \hat x^{a'\star}_{j} \mathcal{M}^{a'b'} x^{b'}_{i} \right\rangle_\pi \nonumber \\
		S_2^{(\mathrm{off})}&= \frac{1}{2N} \frac{(-i)^2}{2!1!}\, {}^2C_1  \int dt_1 dt_2 dt_3  \left\langle\hat x^{a\star}_i \mathcal{Z}^{ab} x^{b}_j  \hat x^{a'\star}_{j} (\mathcal{Z}^\dagger)^{a'b'} x^{b'}_{i} \hat x^{a''\star}_i \mathcal{M}^{a''b''} x^{b''}_j \right\rangle_\pi , \nonumber \\
		S_3 &= \cdots , \label{actionseriestermsmu}
	\end{align}
\end{widetext}
where sums over repeated indices are implied. In principle, if one considered a distribution $\pi(a_{ij}, a_{ji})$ for which the third moments were non-zero, this would also contribute to $S_2^{(\mathrm{off})}$, but we do not consider this here. 

Comparing the action in Eq.~(\ref{actionseriestermsmu}) with Eq.~(34) of the main text and noting that the statistics of $\mathcal{Z}$ here are the same as those of $\matj$ when $\mu = 0$, we see that to $O(1/p)$ there are essentially three differences in the action: 

(i) There is a new term $S_0^\mathrm{off}$ involving $\mathcal{M}$, which would also have come about if we were to simply add $\mu/N$ to every element of the matrix (including the null entries). We can think of this term as representing an effective rank-1 perturbation to the random matrix \cite{benaych2011eigenvalues, orourke, tao2013outliers}. This term does not contribute to the trace of the resolvent in the limit $N\to\infty$, and thus does not affect the bulk spectrum, but it is crucial for determining the outlier eigenvalue. 

(ii) There is a new term in $S_2$. This term does in fact modify the trace of the resolvent and so affects the bulk. It has exactly the same structure as the zeroth-order (in $1/p$) action contribution $S_1$, and would lead to similar rainbow diagrams if we performed a diagrammatic expansion. However, for the purposes of obtaining an $O(1/p)$ approximation to the eigenvalue spectrum, this term can simply be thought of as modifying $\Gamma$ and $\sigma^2$ as follows
\begin{align}
	\sigma^2 &\to \sigma^2 + \frac{\mu^2}{p} \nonumber \\
	\Gamma &\to \Gamma + \frac{1}{p}(1-\Gamma)\frac{\mu^2}{\sigma^2}  .  \label{statsmumod}
\end{align}
That is, we absorb this term into the $S_1$ contribution to the action, neglecting the $O(1/p^2)$ error that this entails. This enables us to see quite easily how the bulk spectrum is modified by the introduction of a non-zero value of $\mu$, as described in the main text.

(iii) Finally, we have identified a new contribution to the action $S_2^{(\mathrm{off})}$. Like the effective rank-1 perturbation of (i), it does not affect the sum of the diagonal elements of the resolvent in the thermodynamic limit. It does however contribute to the sum of \textit{all} the elements of the resolvent, which is important for calculating the outlier eigenvalue, as we will show below.

To obtain the $1/p$ modification to the bulk spectrum, it is only necessary to consider the change to $S_2$ that arises as described in point (ii) above. By inserting the substitution in Eqs.~(\ref{statsmumod}) into Eqs.~(40), (41) and (42) of the main text, we see that the explicit formulae for the boundary of the bulk region and the density of eigenvalues therein do not change, but the expressions for $x_c$ and $y_c$ do. One obtains, rather straightforwardly, the new expressions for $x_c$ and $y_c$ in Eq.~(44) of the main text.

Now, we consider the outlier eigenvalue, which comes about due to effective rank-1 perturbation that was highlighted above in point (i). Following Refs. \cite{orourke, benaych2011eigenvalues}, the outlier eigenvalue can thus be found by noting that it must satisfy axiomatically
\begin{align}
	\det\left[\lambda_{\mathrm{outlier}} \underline{\underline{\id}} - \underline{\underline{z}} - \frac{1}{N} \underline{\underline{\mu}} \right] = 0,\nonumber \\
	\Rightarrow \det\left[\underline{\underline{\id}} - \frac{1}{N} \underline{\underline{\mu}} \underline{\underline{C}} \right] = 0,
\end{align}
where $z_{ij} = a_{ij} - \mu/N$ and $\underline{\underline{C}} \equiv [\lambda_{\mathrm{outlier}}\id - \underline{\underline{z}}]^{-1}$. We use Sylvester's determinant identity to find
\begin{align}
	\frac{1}{N}\sum_{ij}C_{ij}(\lambda_{\mathrm{outlier}}) = \frac{1}{\mu} , \label{sylvester}
\end{align}
where, crucially, we are now summing over all elements of the resolvent matrix in order to obtain the outlier eigenvalue. As mentioned in point (iii) above, this means that we have to take into account additional diagrams. A similar phenomenon was noted in Ref. \cite{baron2022eigenvalues}, where additional correlations gave rise to new diagrams.

Let us now calculate the position of the outlier eigenvalue using Eq.~(\ref{sylvester}). To use this formula, we must obtain $\frac{1}{N}\sum_{ij}C_{ij}$ in the thermodynamic limit. We must therefore take into account the new Feynman diagrams that arise from $S_2^{(\mathrm{off})}$. For example, we have the following term that contributes at the order $O(1/p)$ 
\begin{figure}[H]
	\centering 
	\includegraphics[scale = 0.18]{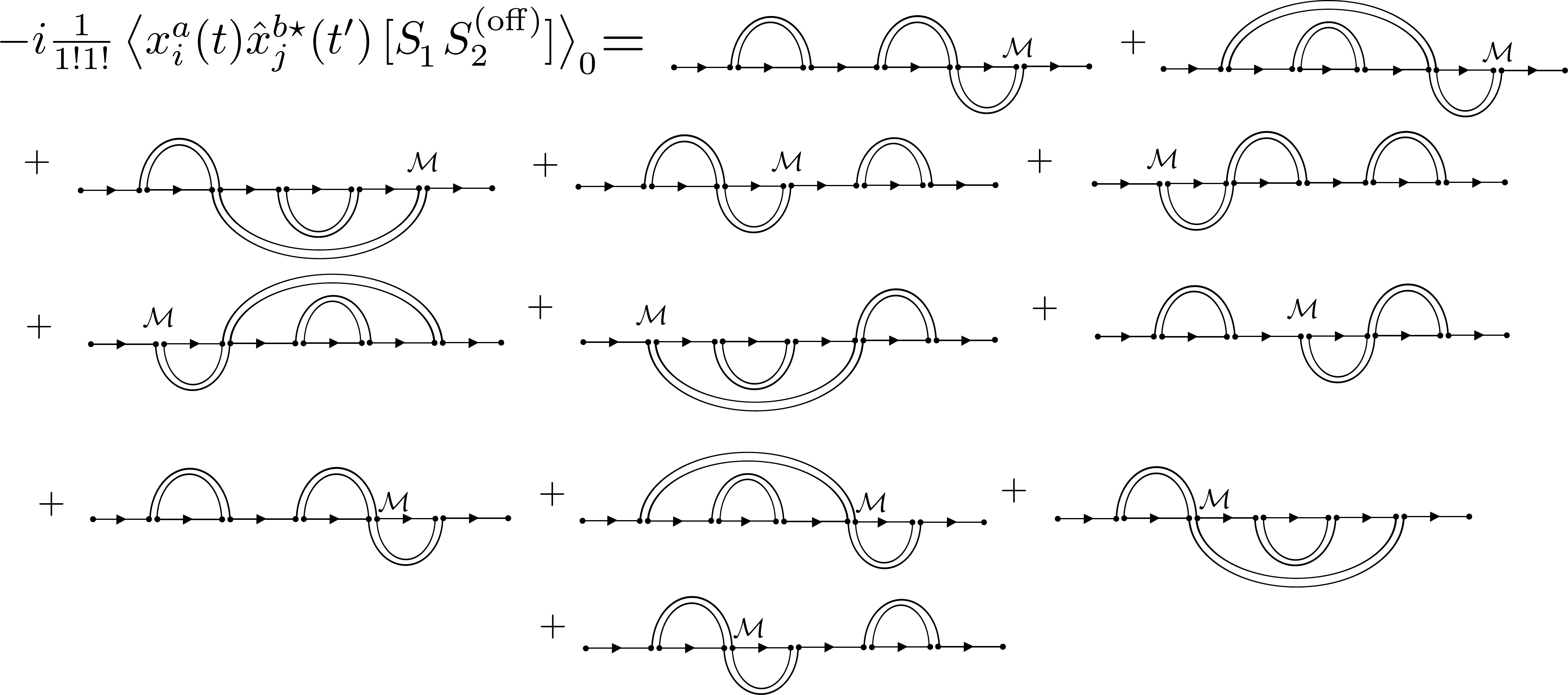}
	\captionsetup{labelformat=empty}
\end{figure}
We note that in the above diagrams, double points that carry a factor of $\mathcal{M}$ instead of $\mathcal{Z}$ are highlighted. For example, the first diagram on the right-hand side of the equality above should be understood to represent the Wick pairing 
\begin{align}
 & \frac{p^2}{N^2} (-i)^6\int_{\mathrm{(Time \, ordered)}} dT_1 dT_1' dT_2 dT_2'dT_2''\left\langle \mathcal{Z}^{a_1b_1} (\mathcal{Z}^\dagger)^{a_1'b_1'} \right\rangle_\pi \left\langle \mathcal{Z}^{a_2b_2} (\mathcal{Z}^\dagger)^{a_2'b_2'} \right\rangle_\pi \nonumber \\
&\times \left \langle \hat x_j^{b\star} (t') x_{i_1}^{a_1} \right\rangle_0 \left \langle \hat x_{j_1}^{b_1\star}  x_{j_1}^{a'_1} \right\rangle_0  \left\langle \hat x_{i_1}^{b'_1\star} x_{i_2}^{a_2} \right\rangle_0 \left\langle \hat x_{j_2}^{b_2\star}  x_{j_2}^{a'_2} \right\rangle_0  \left\langle \hat x_{i_2}^{b'_2\star}  x_{i_2}^{a''_2} \right\rangle_0\mathcal{M}^{a_2''b_2''} \left\langle \hat x_{j_2}^{b''_2\star}  x_{i}^{a}(t) \right\rangle_0
\end{align}
Further, the vertices at either end of the above diagrams are not constrained to have the same lower index (i.e. $i$ can be in general different to $j$). This is why these diagrams give rise to non-zero off-diagonal elements in $\mathcal{H}_{ij}^{ab}$ that affect only the outlier and not the bulk.

When we calculate $N^{-1} \sum_{ij} \mathcal{H}_{ij}$, the sum of the non-vanishing diagrams is performed in an analogous way to Fig. 4 of the main text. One finds ultimately 
\begin{align}
	\frac{1}{N}\sum_{ij}\mathcal{H}_{ij} = \mathcal{H} + \langle \mathcal{H}\mathcal{Z}\mathcal{H}\mathcal{Z}\mathcal{H}\mathcal{M}\mathcal{H}\rangle_\pi + \langle\mathcal{H}\mathcal{Z}\mathcal{H}\mathcal{M}\mathcal{H}\mathcal{Z}\mathcal{H}\rangle_\pi + \langle\mathcal{H}\mathcal{M}\mathcal{H}\mathcal{Z}\mathcal{H}\mathcal{Z}\mathcal{H} \rangle_\pi,
\end{align}
where we note that the off-diagonal contribution is of order $1/p$. Examining the $(2,1)$ component and taking the analytic solution $D = A = 0$ (which is valid outside the bulk where the outlier eigenvalue resides), we obtain
\begin{align}
	\frac{1}{N}\sum_{ij} C_{ij} &= C + \frac{1}{p} \mu (1 + 2 \Gamma) \sigma^2 C^4 . \label{simultoutlier1}
\end{align} 
We also have for the trace
\begin{align}
	C &= \frac{1}{\lambda_{\mathrm{outlier}} - \Gamma \sigma^2C - \frac{1}{p} \left[\mu^2 C+ \Gamma_4^{(3)}  \sigma^4 C^3\right] } . \label{simultoutlier2}
\end{align}
where we have used the substitution in Eq.~(\ref{statsmumod}), along with Eq.~(\ref{resolventoutside}) for the trace of the resolvent matrix. Using Eq.~(\ref{sylvester}) in combination with Eqs.~(\ref{simultoutlier1}) and (\ref{simultoutlier2}), we then obtain
\begin{align}
	C  \approx \frac{1}{\mu} - \frac{1}{p} \frac{(1+2\Gamma) \sigma^2}{\mu^3}. \label{cexpr}
\end{align}
Finally, the outlier eigenvalue can be found by substituting Eq.~(\ref{cexpr}) into Eqs.~(\ref{simultoutlier1}) and we arrive at Eq.~(46) of the main text.

In order to obtain a criterion for the validity of this expression for the outlier eigenvalue, we must examine one of the other blocks of the hermitised resolvent matrix. This is similar to the procedure that was used in Refs. \cite{baron2020dispersal,baron2022eigenvalues}. Let us examine the quantity $D/\eta \equiv \mathh_{22}/\eta$ in the limit $\eta \to 0$. This is given by 
\begin{align}
	\lim_{\eta\to 0}\frac{D}{\eta} = -\overline{\frac{1}{N}\mathrm{Tr} \left[(\omega^\star \underline{\underline{\id}}_N - \underline{\underline{a}}^T)(\omega\underline{\underline{\id}}_N - \underline{\underline{a}})\right]^{-1}} = -\overline{\frac{1}{N} \sum_{\nu} \frac{1}{\vert \omega - \lambda_\nu \vert^2}} \leq 0 .\label{conditiond}
\end{align}
One has from Eq.~(39) of the main text [using also the substitution $\sigma^2 \to \sigma^2 + \mu^2/p$ above and keeping only $O(1/p)$ terms]
\begin{align}
	\lim_{\eta \to 0} \frac{D}{\eta} &=\frac{-BC}{1-BC\left\{\sigma^2  + \frac{\sigma^4 }{p} \left[\frac{\mu^2}{\sigma^4} + \Gamma^{(2)}_4 (B^2 + C^2) +\Gamma^{(3)}_4 BC \right]\right\}} .
\end{align}
We see that in order for the inequality in Eq.~(\ref{conditiond}) to be satisfied, we must have (recalling that $B = C^\star$)
\begin{align}
	\frac{1}{BC} \geq \sigma^2  + \frac{\sigma^4 }{p} \left[\frac{\mu^2}{\sigma^4} + \Gamma^{(2)}_4 (B^2 + C^2) +\Gamma^{(3)}_4 BC \right]. \label{bccond}
\end{align}
This therefore provides us with a criterion for the expression for $C$ in Eq.~(\ref{cexpr}), and therefore the expression for the outlier eigenvalue, to be valid. Inserting Eq.~(\ref{cexpr}) into (\ref{bccond}), one thus obtains (ignoring higher-order terms in $1/p$)
\begin{align}
	\mu^2 + \frac{2}{p} (1+2\Gamma) \sigma^2 \geq \sigma^2 + \frac{\sigma^2}{ p}\left[\frac{(2 \Gamma_2 + \Gamma_3)}{\mu^2} + \frac{\mu^2}{\sigma^4} \right].
\end{align}
This is a quadratic inequality in $\mu^2$, which can be solved to yield the condition in Eq.~(47) of the main text.

\section{Summation of the diagrammatic series up to second order in $1/p$}\label{appendix:secondorder}

In this section, we discuss how one can sum the series of all non-vanishing diagrams up to $O(1/p^2)$ to obtain the expression for the resolvent in Eq.~(51) of the main text. This is done in an analogous way to how the first order diagrams in main text Fig. 4 were summed to find $\frac{1}{p} \mathcal{H}_2$.

The second-order terms are best summed in four separate series such that we define the sum of all second-order diagrams as $\mathh_3/p^2 = (\mathh_3^{(1)}+\mathh_3^{(2)}+\mathh_3^{(3)}+ \mathh_3^{(4)})/p^2$. The first three series are given by
\begin{figure}[H]
	\centering 
	\includegraphics[scale = 0.19]{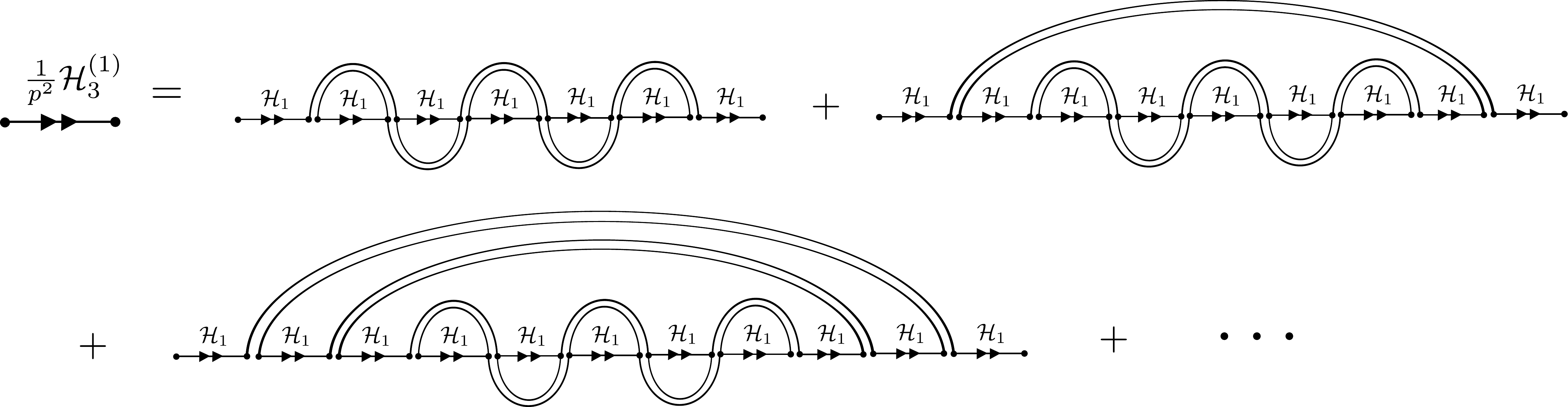}
	\captionsetup{labelformat=empty}
\end{figure}

\begin{figure}[H]
	\centering 
	\includegraphics[scale = 0.19]{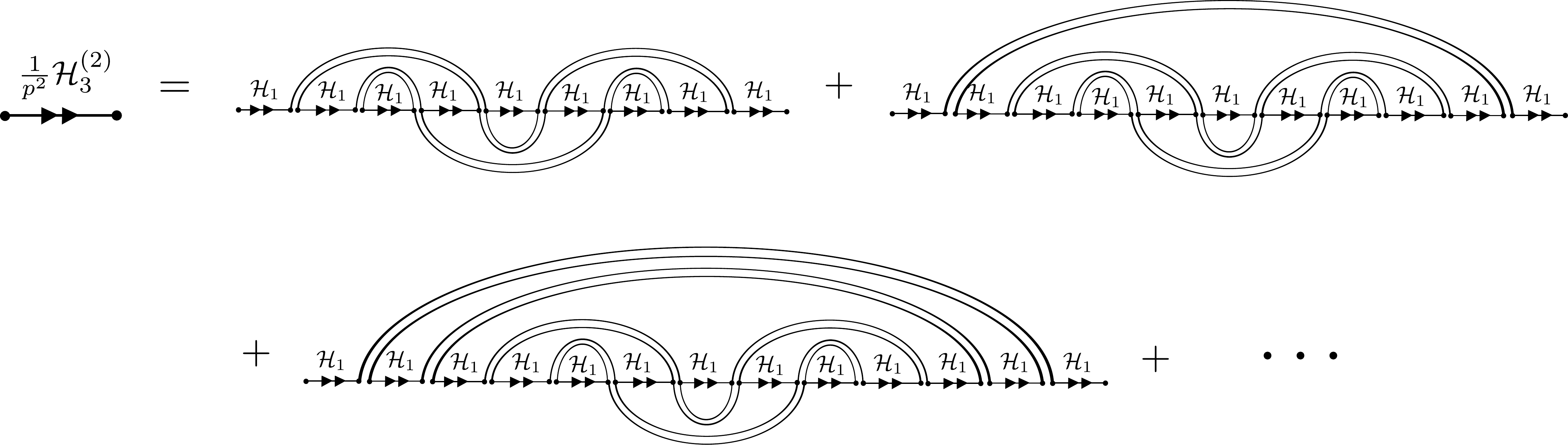}
	\captionsetup{labelformat=empty}
\end{figure}

\begin{figure}[H]
	\centering 
	\includegraphics[scale = 0.19]{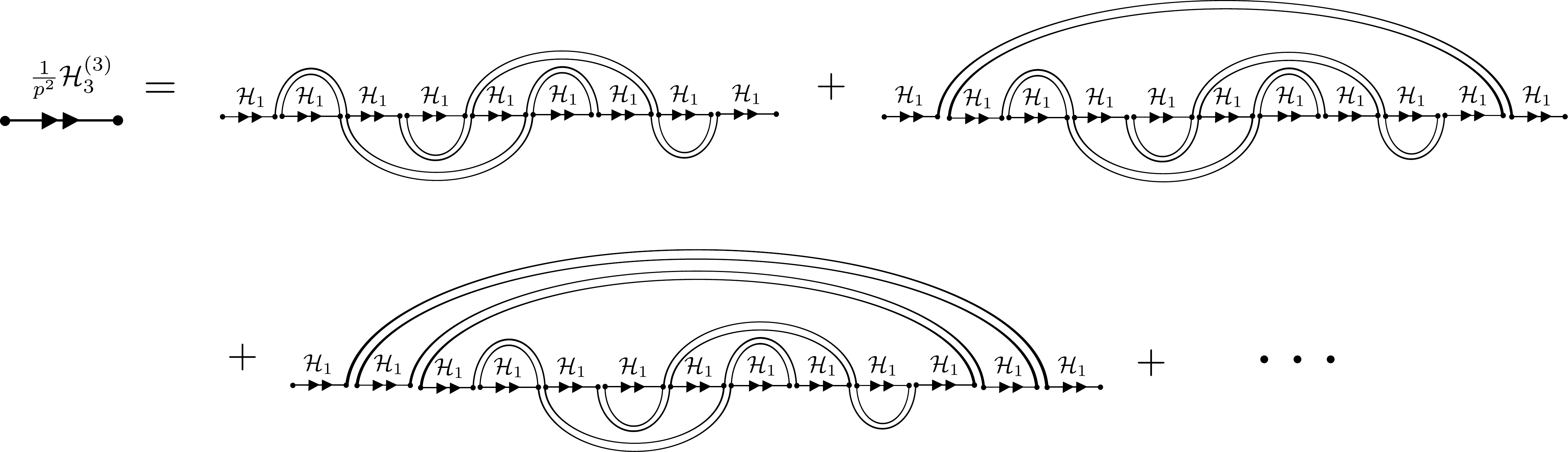}
	\captionsetup{labelformat=empty}
\end{figure}
We sum these diagrammatic series to obtain [c.f. Eq.~(38) of the main text]
\begin{align}
	\frac{1}{p^2}\mathcal{H}^{(1)}_{3}&= \frac{1}{p}\langle \mathcal{H}_1 \matj \mathcal{H}^{(1)}_{3} \matj \mathcal{H}_1 \rangle_\pi + p \langle \mathcal{H}_1 \matj \mathcal{H}_1 \matj \mathcal{H}_1 \matj\mathcal{H}_1 \matj \mathcal{H}_1  \matj \mathcal{H}_1  \matj \mathcal{H}_1 \rangle_\pi \nonumber \\
	\frac{1}{p^2}\mathcal{H}^{(2)}_{3}&= \frac{1}{p}\langle \mathcal{H}_1 \matj \mathcal{H}^{(2)}_{3} \matj \mathcal{H}_1 \rangle_\pi + p^2  \left\langle \mathcal{H}_1 \matj_1 \mathcal{H}_1 \matj_2 \mathcal{H}_1 \matj_2\mathcal{H}_1 \matj_1 \mathcal{H}_1  \matj_1 \mathcal{H}_1  \matj_2 \mathcal{H}_1 \matj_2 \mathcal{H}_1 \matj_1 \mathcal{H}_1 \right\rangle_\pi \nonumber \\
	\frac{1}{p^2}\mathcal{H}^{(3)}_{3}&= \frac{1}{p}\langle \mathcal{H}_1 \matj \mathcal{H}^{(3)}_{3} \matj \mathcal{H}_1 \rangle_\pi + p^2  \left\langle \mathcal{H}_1 \matj_1 \mathcal{H}_1 \matj_1 \mathcal{H}_1 \matj_2\mathcal{H}_1 \matj_2 \mathcal{H}_1  \matj_1 \mathcal{H}_1  \matj_1 \mathcal{H}_1 \matj_2 \mathcal{H}_1 \matj_2 \mathcal{H}_1  \right\rangle_\pi  , \label{secondordercorrections1}
\end{align}
where here we have defined the $2 \times 2$ matrices $\matj_1$ and $\matj_2$, which each individually have the same statistics as $\matj$ [defined in Eq.~(\ref{jdef}), but with elements drawn from $\pi(a_{12},a_{21})$], but are statistically independent of one another, so that $\langle \matj_1^{ab} \matj_2^{a'b'} \rangle_\pi = 0$ for all combinations of upper indices. 

The self-similarity of the remaining sub-series for the second-order terms is somewhat different to the others. We see that this series can be related to the square of the series for the first-order correction diagrams 
\begin{figure}[H]
	\centering 
	\includegraphics[scale = 0.18]{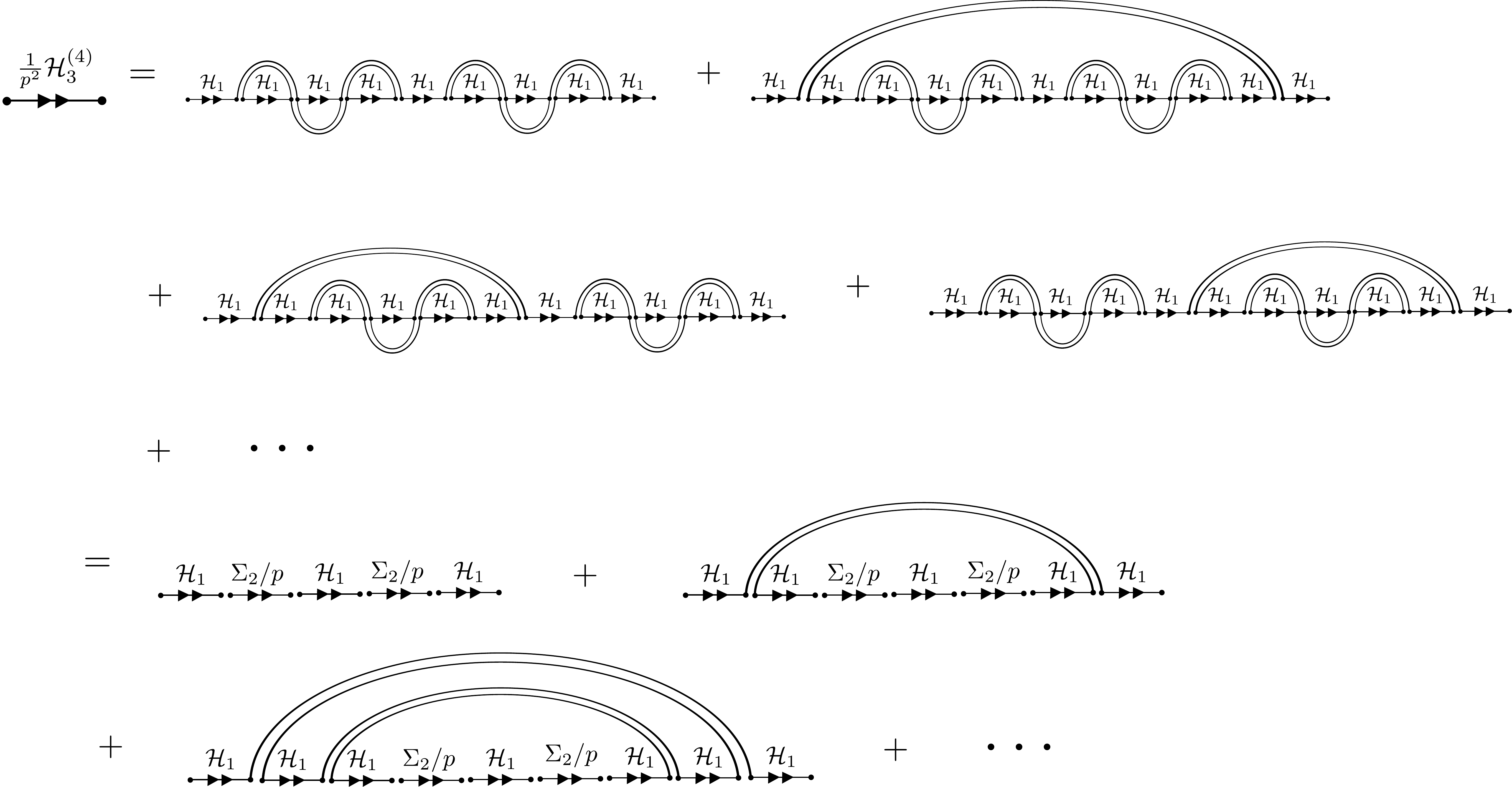}
	\captionsetup{labelformat=empty}
\end{figure}
where we have identified once again the series for $\mathh_2/p$ [see main text Fig. 4] and we have defined $\Sigma_2/p$ via 
\begin{figure}[H]
	\centering 
	\includegraphics[scale = 0.18]{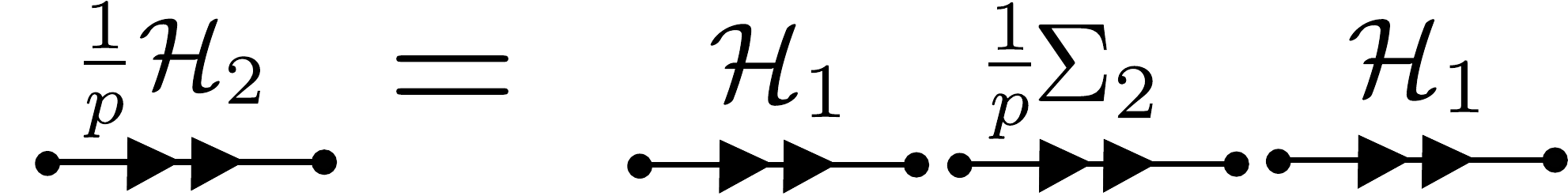}
	\captionsetup{labelformat=empty}
\end{figure}
The series for $\frac{1}{p^2}\mathcal{H}^{(4)}_{3}$ can therefore be summed as follows
\begin{align}
	\frac{1}{p^2}\mathcal{H}^{(4)}_{3}&= \frac{1}{p}\langle \mathcal{H}_1 \matj \mathcal{H}^{(4)}_{3} \matj^\dagger \mathcal{H}_1 \rangle_\pi +\frac{1}{p^2} \mathh_2\mathh_1^{-1}\mathh_2 . \label{secondordercorrections2}
\end{align}
We have thus successfully summed all of the second-order correction diagrams. Now, we wish to find a self-consistent expression for the full hermitised resolvent $\mathh$. We remind ourselves that up to first order we also have [see Eqs.~(37) and (38) of the main text]
\begin{align}
	\mathh_1 &= \left[\mathh_0^{-1} - p\langle \matj \mathh_1 \matj \rangle_\pi   \right]^{-1}, \nonumber \\
	\frac{1}{p}\mathcal{H}_{2}&= \langle \mathcal{H}_1 \matj \mathcal{H}_{2} \matj \mathcal{H}_1 \rangle_\pi + p \langle \mathcal{H}_1 \matj \mathcal{H}_1 \matj \mathcal{H}_1 \matj\mathcal{H}_1 \matj\mathcal{H}_1 \rangle_\pi .
\end{align}
It is helpful to consider the following quantity 
\begin{align}
	I &= \left[ \mathh_1^{-1} - \frac{1}{p} \Sigma_2 - \frac{1}{p^2}\Sigma_3\right]^{-1}, \nonumber \\
	\Rightarrow I \mathh_1^{-1} &= \id+  \frac{1}{p} I\Sigma_2 + \frac{1}{p^2}I\Sigma_3 , 
\end{align}
where if we consider the expansion up to $O(1/p^2)$ we find
\begin{align}
	I  &\approx \mathh_1+  \frac{1}{p} \left[\mathh_1 + \frac{1}{p} I\Sigma_2 \mathh_1\right]\Sigma_2\mathh_1 + \frac{1}{p^2}\mathh_1 \Sigma_3 \mathh_1 , \nonumber \\
	\Rightarrow I  &\approx \mathh_1+  \frac{1}{p} \mathh_1\Sigma_2\mathh_1   + \frac{1}{p^2}\left[\mathh_1\Sigma_2 \mathh_1\Sigma_2\mathh_1+\mathh_1 \Sigma_3 \mathh_1 \right] .
\end{align}
Making the substitutions
\begin{align}
	\Sigma_2 &= \mathh_1^{-1}\mathh_2 \mathh_1^{-1}, \nonumber \\
	\Sigma_3 &= \mathh_1^{-1}\left[\mathh_3^{(1)} + \mathh_3^{(2)} + \mathh_3^{(3)} + \mathh_3^{(4)}\right]\mathh_1^{-1} - \mathh_2\mathh_1^{-1}\mathh_2, \label{sigmassecondorder}
\end{align}
we then find using the expressions in Eqs.~(\ref{secondordercorrections1}) and (\ref{secondordercorrections2})
\begin{align}
	I  &\approx \mathh_1+  \frac{1}{p} \mathh_2   + \frac{1}{p^2}\left[\mathh_3^{(1)} + \mathh_3^{(2)} + \mathh_3^{(3)} + \mathh_3^{(4)} \right] \approx \mathh
\end{align}
Thus, we have 
\begin{align}
	\mathcal{H} \approx \left[ \mathh_1^{-1} - \frac{1}{p} \Sigma_2 - \frac{1}{p^2}\Sigma_3\right]^{-1}, \label{selfconsistentsecondorder}
\end{align}
where $\Sigma_2$ and $\Sigma_3$ are given by Eq.~(\ref{sigmassecondorder}). Using $	\mathh \approx  \,\,\, \mathh_1 + \frac{1}{p} \mathh_2 + \frac{1}{p^2} \left( \mathh_3^{(1)} + \mathh_3^{(2)} + \mathh_3^{(3)} + \mathh_3^{(4)} \right)$, we thus see that Eq.~(\ref{selfconsistentsecondorder}) is equivalent to the self-consistent equation for $\mathh$ in Eq.~(39) of the main text, which is valid up to second order in $1/p$.

\section{Ensembles with non-negligible higher-order moments}\label{appendix:nongaussian}

\subsection{General considerations}

In a certain sense, the ensemble of sparse matrices defined in Eq.~(32) of the main text can be thought of as a special case of a broader class of random matrix. In this section, we consider other random matrix ensembles where most elements have values close to zero, but there are a small number of larger entries per row. 

We show here that matrices whose elements are all drawn from a long-tailed distribution (specifically, we consider a Cauchy-Lorentz distribution) can be of this type, and can also be treated in a similar way to those of the main text. Despite the fact that both ensembles can be dealt with using the perturbative approach, there are some non-trivial differences in the resulting eigenvalue spectra. We also consider a fairly straightforward generalisation of the ensemble in Eq.~(32) of the main text, where we replace the null entries with random variables that fluctuate within $1/\sqrt{N}$ of zero.

Generally, we consider ensembles of all-to-all-connected random matrices whose entries have the following statistics
\begin{align}
\overline{ a_{ij} } &= \frac{\mu}{N}, \nonumber \\
	\overline{ (a_{ij}- \mu/N)^2} &= \frac{\sigma^2}{N}, \nonumber \\
	\overline{ (a_{ij}-\mu/N)(a_{ji}- \mu/N)} &= \frac{\Gamma \sigma^2}{N},\nonumber \\
	\overline{(a_{ij}-\mu/N)^4 } &= \frac{\Gamma^{(1)}_4 \sigma^4}{q N}, \nonumber \\
	\overline{ (a_{ij}-\mu/N)^3( a_{ji}-\mu/N) } &= \frac{\Gamma^{(2)}_4 \sigma^4}{q N}, \nonumber \\
	\overline{ (a_{ij}-\mu/N)^2 (a_{ji}-\mu/N)^2 } &= \frac{\Gamma^{(3)}_4 \sigma^4}{q N},  \label{loworderstats}
\end{align}
where all higher-order moments here scale as $1/N$ [c.f. Eq.~(33) of the main text]. We have included the factor $1/q$ to highlight the fact that in order for the perturbative approach to be useful, successively higher moments must be decreasing. The factor $1/q$ is primarily a placeholder that allows us to perform the perturbative expansion in a transparent way. 

One notes that the disorder averages in Eq.~(\ref{loworderstats}) are of the form $\overline{\cdots}$ [i.e. with respect to $P(a_{ij}, a_{ji})$] and not $\langle \cdot \rangle_\pi$. The ensembles considered in the main text would fall also be of the type discussed here with a redefinition of the parameters $\mu$, $\sigma^2$, and so on.

We consider here distributions $P(a_{ij}, a_{ji})$ where the third moments vanish. In principle, non-zero third moments could also be included using similar diagrams to those in Section \ref{appendix:outlier}. These terms only affect the position of the outlier eigenvalue, not the bulk. 

The corresponding action for the ensemble in Eq.~(\ref{loworderstats}) is (using the same definitions for $\mathcal{M}^{ab}$ and $\mathcal{Z}^{ab}$ as in Section \ref{appendix:outlier}) 
\begin{align}
	S_0 &= i  \sum_{i,a} \int dt \, \hat x^{a\star}_i \left(\dot x_i^a +\sum_{b, j}(\mathinv)^{ab}_{ij}  x^b_j  \right)    , \,\,\,\,\,\,\,\,\,\,	S_{0}^{\mathrm{off}} = -\frac{i}{N} \sum_{i,j,a,b} \int dt \, \hat x^{a\star}_i \mathcal{M}^{ab}  x^b_j    , \nonumber \\
	S_1 &= \frac{1}{2N} \frac{(-i)^2}{2!}\, {}^2C_1 \int dt dt' \; N\overline{\hat x^{a\star}_i \mathcal{Z}^{ab} x^{b}_j  \hat x^{a'\star}_{j} (\mathcal{Z}^\dagger)^{a'b'} x^{b'}_{i} } \nonumber \\
	S_2 &=\frac{1}{2N} \frac{(-i)^4}{4!} \, {}^4C_2  \int dt_1 \cdots dt_4 \; N\overline{\hat x^{a_1\star}_{i} \mathcal{Z}^{a_1b_1} x^{b_1}_{j}  \hat x^{a_2\star}_{j} (\mathcal{Z}^\dagger)^{a_2b_2} x^{b_2}_{i}\hat x^{a_3\star}_{i} \mathcal{Z}^{a_3b_3} x^{b_3}_{j}\hat x^{a_4\star}_{j} (\mathcal{Z}^\dagger)^{a_4b_4} x^{b_4}_{i}} , \nonumber \\
	S_3 &= \cdots , \label{actionseriestermsnongaussian}
\end{align}
again, sums over repeated indices here are implied. 

The only difference between Eq.~(\ref{actionseriestermsmu}) and Eq.~(\ref{actionseriestermsnongaussian}) are the terms involving $\mathcal{M}_{ij}^{ab}$. In Eq.~(\ref{actionseriestermsnongaussian}) the only term that is affected by the value of $\mu$ is $S_0^{\mathrm{off}}$. This means that a non-zero value of $\mu$ here serves only to introduce a single outlier eigenvalue. It does not alter the bulk spectrum as in the case introduced in the main text. This is because, in the main text, the non-zero mean only affected the non-zero entries (i.e. the rare large entries), whereas here it affects all entries.

Truncating the series at first order in $1/q$, we can easily see that the bulk spectrum ought to obey the same modified elliptical and semi-circle laws as in the main text. That is, most of the eigenvalues are confined to the region enclosed by
\begin{align}
	\frac{x^2}{x_c^2}+ \frac{y^2}{y_c^2} = 1 - \frac{16}{q} \frac{(\Gamma_4^{(3)}- \Gamma \Gamma_4^{(2)})}{(1-\Gamma^2)} \frac{x^2}{x_c^2} \frac{y^2}{y_c^2}\label{ellipsenongauss}
\end{align}
where we have identified the rightmost and topmost eigenvalues of the modified ellipse (respectively)
\begin{align}
	x_c &= \sigma(1+\Gamma)  + \frac{\sigma}{2 q} \left[ (3-\Gamma) \Gamma_4^{(3)} + 2 (1-\Gamma) \Gamma^{(2)}_4\right] \nonumber \\
	y_c &= \sigma(1-\Gamma) + \frac{\sigma}{2 q} \left[(3 + \Gamma) \Gamma_4^{(3)} - 2 (1+\Gamma) \Gamma_4^{(2)} \right], \label{xcnongauss}
\end{align}
and we also have the modified semi-circle law for the density of the real parts of the eigenvalues
\begin{align}
	\rho_x(x) &=  \frac{2}{\pi x_c^2} \left\{ 1 + \frac{\beta}{q} \frac{\sigma}{x_c} \left[1 - 4\frac{x^2}{x_c^2} \right]\right\} \sqrt{x_c^2 - x^2},  \label{generalizedwignernongauss}
\end{align}
where again
\begin{align}
	\beta &= \frac{1}{3}  \left[ (1-\Gamma) \Gamma_4^{(1)} + 6 (1-\Gamma) \Gamma_4^{(2)} + 2 (4-\Gamma) \Gamma_4^{(3)}\right].
\end{align}
These expressions are tested in Figs. \ref{fig:cauchysymmetric} and \ref{fig:denseplussparse} below. We also derive some additional results that are valid to second order in $1/p$ for the more specific cases below. 

\subsection{Example: Dense random matrices with i.i.d. Cauchy-distributed entries}
Let us consider the relatively simple example of a dense matrix whose elements are each independently selected from the following truncated Cauchy-Lorentz distribution
\begin{align}
	P(a_{ij}) = \frac{\epsilon}{\epsilon^2 + (a_{ij} -\mu/N)^2} \frac{1}{\arctan\left( \frac{w}{\epsilon}\right)} \Theta\left( w + \frac{\mu}{N} - \vert a_{ij} \vert\right) , \label{cauchydistribution}
\end{align}
where $\Theta(\cdot)$ is the Heaviside function, and we choose $\epsilon = \frac{\sigma \pi \sqrt{q}}{2 N}$ and $w = \frac{\sigma}{\sqrt{q}}$. The choice of scaling for these parameters with $q$ and $N$ gives rise to statistics of the form in Eq.~(\ref{loworderstats}) for large $N$
\begin{align}
	\overline{ a_{ij} } &= \frac{\mu}{N}, \nonumber \\
	\overline{ (a_{ij}- \mu/N)^2} &= \frac{\sigma^2}{N}, \nonumber \\
	\overline{(a_{ij}-\mu/N)(a_{ji}- \mu/N)} &= 0,\nonumber \\
	\overline{ (a_{ij}-\mu/N)^4 } &= \frac{ \gamma_1 \sigma^4}{ q N}, \nonumber \\
	\overline{ (a_{ij}-\mu/N)^3( a_{ji}-\mu/N) } &= 0, \nonumber \\
	\overline{ (a_{ij}-\mu/N)^2 (a_{ji}-\mu/N)^2 } &= O\left(\frac{1}{N^2}\right), \nonumber \\
	\overline{ (a_{ij} - \mu/N)^6} &=  \frac{\gamma_2\sigma^6}{ q^2 N}, \cdots \label{cauchystats}
\end{align}
where here $\gamma_r = 1/(2r +1)$ and we see that the only non-negligible moments are of the form $\overline{ (a_{ij}-\mu/N)^r }$. In principle, we could introduce non-trivial correlations by considering instead a multivariate student t-distribution, but the simple Cauchy distribution will suffice here. In Section \ref{appendix:symmetricnongauss}, we examine the case of a symmetric matrix (where $a_{ij}$ and $a_{ji}$ are maximally correlated).

By considering the cumulative distribution function corresponding to Eq.~(\ref{cauchydistribution}), one arrives at a rather succinct formula for sampling the Cauchy-distributed matrix elements
\begin{align}
	a_{ij} = \epsilon \tan\left[(2 r_{ij} -1) \arctan\left( \frac{w}{\epsilon}\right) \right] + \frac{\mu}{N} ,
\end{align}
where $r_{ij} \in [0,1]$ is a uniform random number.

Since the action in Eq.~(\ref{actionseriestermsnongaussian}) is essentially identical to that considered in the main text in Section VII (or Section \ref{appendix:secondorder} in this document), we can use the same diagrammatic expansion to arrive at [c.f. Eq.~(51) of the main text]
\begin{align}
	\mathh &\approx  \Bigg\{\mathh_0^{-1} - N\overline{ \matz \mathh \matz^\dagger } -  N\overline{ \matz \mathcal{H} \matz^\dagger \mathcal{H} \matz\mathcal{H} \matz^\dagger  } -  N\overline{ \matz \mathcal{H} \matz^\dagger \mathcal{H} \matz\mathcal{H} \matz^\dagger\mathcal{H}  \matz \mathcal{H} \matz^\dagger } \nonumber \\
	&- N^2\overline{  \matz_1 \mathcal{H} \matz_2 \mathcal{H} \matz_2^\dagger \mathcal{H} \matz_1^\dagger \mathcal{H}  \matz_1 \mathcal{H}  \matz_2 \mathcal{H} \matz_2^\dagger \mathcal{H} \matz_1^\dagger  } -  N^2\overline{  \matz_1 \mathcal{H} \matz_1^\dagger \mathcal{H} \matz_2\mathcal{H} \matz_2^\dagger \mathcal{H} \matz_1 \mathcal{H}  \matz_1^\dagger \mathcal{H} \matz_2 \mathcal{H} \matz_2^\dagger } \Bigg\}^{-1}, \label{secondorderselfconsistentnongaussian}
\end{align}
where we note once again that all disorder averages here are taken with respect to $P(\cdot , \cdot)$ rather than $\pi(\cdot , \cdot)$ and that the entries of $\mathcal{J}_1$ are to be taken as statistically independent of those of $\mathcal{J}_2$. Carrying out the disorder averages, one finds for $\eta \to 0$
\begin{align}
	A &= -\frac{A \sigma^2}{\Delta} \left\{1 + \frac{\gamma_1 \sigma^2}{q} AD + \frac{1}{q^2} AD \sigma^4 \left[ AD \gamma_2 + B^2 C^2 \gamma_1^2 \sigma^2 +  A^2 D^2 \gamma_1^2\sigma^2 (1+\gamma_1^2) \right]  \right\}, \nonumber \\
	B &= \frac{1}{\Delta} \left[ - \omega + \frac{\gamma_1}{q^2} A^2 D^2 B^2 C\right], \nonumber \\
	C &= \frac{1}{\Delta} \left[ - \omega^\star + \frac{\gamma_1}{q^2} A^2 D^2 B C^2 \right], \nonumber \\
	D &= -\frac{D \sigma^2}{\Delta} \left\{1 + \frac{\gamma_1 \sigma^2}{q} AD + \frac{1}{q^2} AD \sigma^4 \left[ AD \gamma_2 + B^2 C^2 \gamma_1^2 \sigma^2 +  A^2 D^2 \gamma_1^2\sigma^2 (1+\gamma_1^2) \right]  \right\}, \nonumber \\
	\frac{1}{\Delta} &= AD -BC . \label{disorderaveragednongaussian}
\end{align}
As usual, there are two solutions for $C(\omega, \omega^\star)$. One solution, valid outside the bulk region, is for the case $A = D = 0$
\begin{align}
	C(\omega) = \frac{1}{\omega}. \label{nongaussres1}
\end{align}
Inside the bulk region, we can instead solve for $AD \neq 0$ to find up to $O(1/q^2)$
\begin{align}
	&AD \approx BC -\frac{1}{\sigma^2} + \frac{1}{q} \gamma_1 \left(BC -\frac{1}{\sigma^2}\right) \nonumber \\
	&+ \frac{1}{q^2} \left(BC -\frac{1}{\sigma^2}\right) \left\{ \gamma_1^2  + \sigma^4 \gamma_1^2 B^2 C^2 + \sigma^2 \left(BC -\frac{1}{\sigma^2}\right) \left[\gamma_2 - \gamma_1^2+ \gamma_1^2 \sigma^2 \left(BC -\frac{1}{\sigma^2}\right) (1 + \gamma_1^2) \right]\right\} . 
\end{align}
We thus find the following expression, which is accurate up to second order in $1/q$, and is valid inside the bulk region
\begin{align}
	C(\omega, \omega^\star) &\approx \frac{\omega^\star}{\sigma^2} + \frac{\gamma_1\omega^\star}{q \sigma^2}  \left( 1 - \frac{\vert \omega\vert^2}{\sigma^2}\right) \nonumber \\
	&+ \frac{\omega^\star}{\sigma^2 q^2} \left( 1 -  \frac{\vert \omega\vert^2}{\sigma^2}\right)^2 \left[3 \gamma_1^2 \left( 1 - \frac{\vert \omega\vert^2}{\sigma^2}\right) -\gamma_2 - \gamma_1^4 \left( 1 - \frac{\vert \omega\vert^2}{\sigma^2}\right) \right] . \label{nongaussres2}
\end{align}
One notes that all higher order terms in $1/q$ in Eq.~(\ref{nongaussres2}) are proportional to $\left( 1 - \frac{\vert \omega\vert^2}{\sigma^2}\right)$. This owes to the fact all terms in the expression for $D$ in Eq.~(\ref{disorderaveragednongaussian}) are proportional to $AD$, which in turn is because the only moments of the matrix elements that are non-negligible are of the form $\overline{(a_{ij}-\mu/N)^r}$ [c.f. Eq.~(\ref{simultsparsegeneral})]. Similarly, one can see that the expression in Eq.~(\ref{nongaussres1}) should be valid to all orders in $1/q$, since all higher order terms will also be proportional to $AD$ in the expressions for $B$ and $C$ in Eq.~(\ref{disorderaveragednongaussian}).

We can thus see that, to all orders in $1/q$, when we solve Eqs.~(\ref{nongaussres1}) and (\ref{nongaussres2}) simultaneously to find the boundary of the bulk of the eigenvalue spectrum, all the higher-order terms in $1/q$ vanish, and we obtain the simple circular law
\begin{align}
	\vert \omega \vert^2 = \sigma^2. \label{circlenongauss}
\end{align}
This ought to apply generally to any distribution that has statistics like those in Eq.~(\ref{cauchystats}) for which a perturbative treatment is valid. However, one notes that if we were to study a non-Gaussian distribution for which $a_{ij}$ and $a_{ji}$ were correlated, such as the multivariate student's t-distribution, the elliptical law might not be as robust as this, and there would be corrections due to the higher-order moments. This is exemplified in the following subsection where we consider the Hermitian case $a_{ij} = a_{ji}$.

\begin{figure}[H]
	\centering 
	\includegraphics[scale = 0.45]{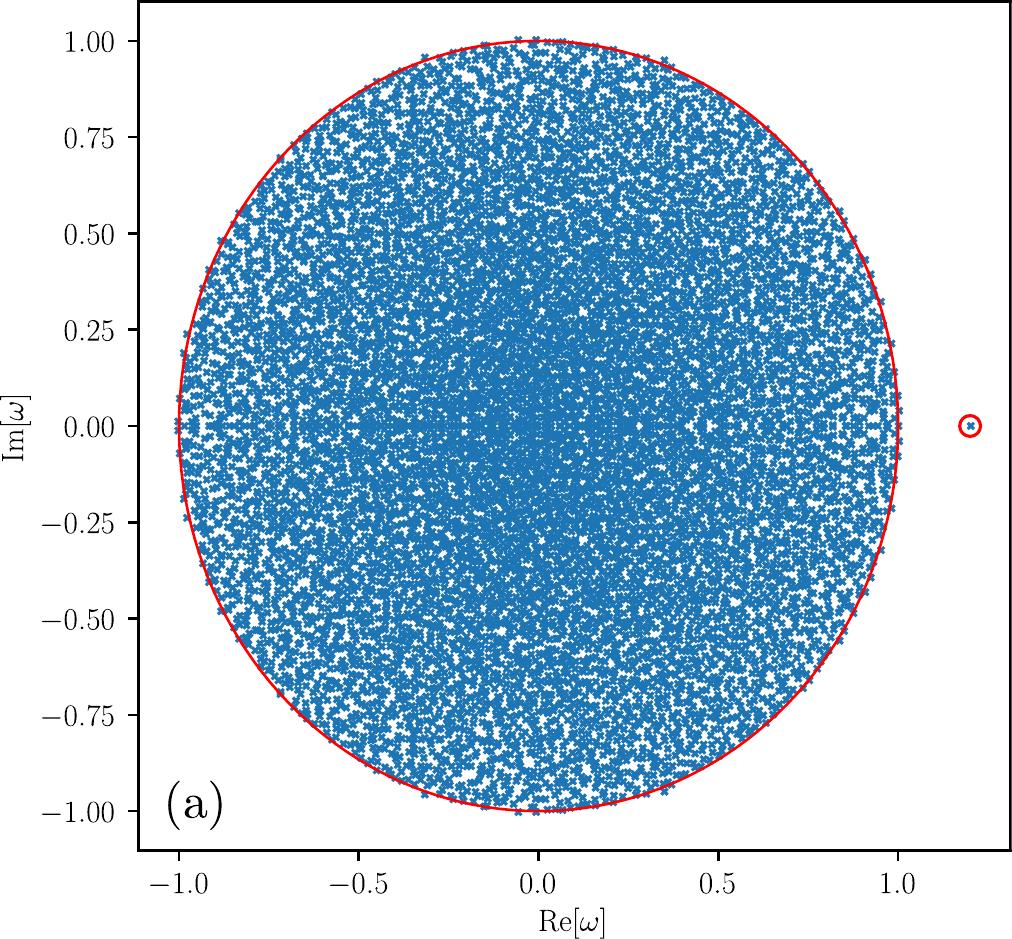}\hspace{2mm}
	\includegraphics[scale = 0.5]{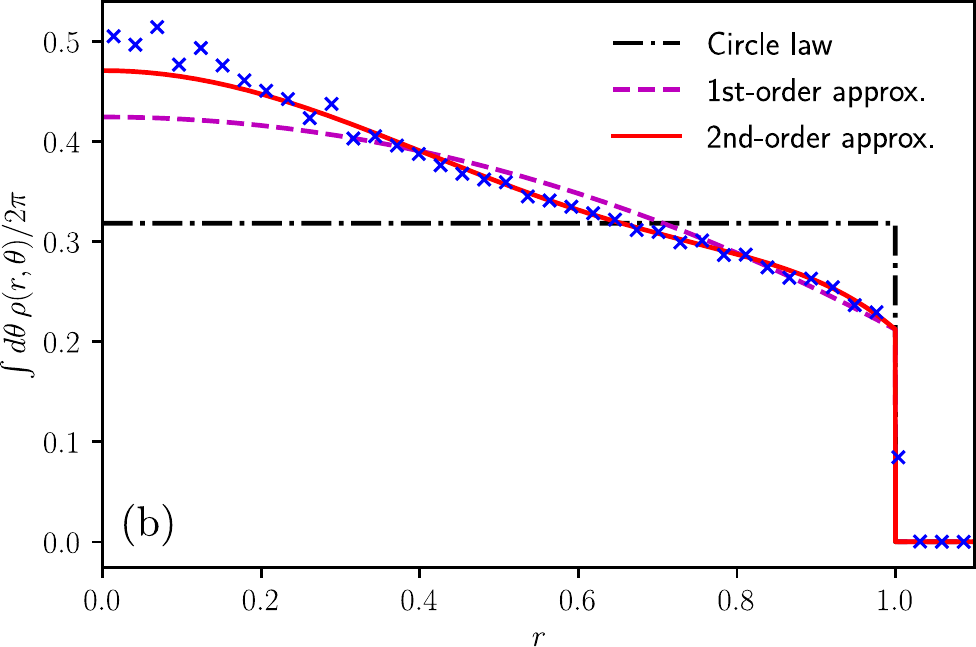}
	\caption{Eigenvalue spectra of random matrices with i.i.d. entries drawn from a truncated Cauchy distribution. Panel (a): Verification of Eqs.~(\ref{circlenongauss}) and Eq.~(\ref{outliernongauss}) for a single matrix. Panel (b): Verifying the second-order approximation for the radial density in Eq.~(\ref{densitynongauss}), averaging over 10 realisations. In both panels, matrices with parameters $N = 10000$, $q = 1$ and $\sigma = 1$ were used.}\label{fig:cauchy}
\end{figure}

Despite the boundary of the bulk of the eigenvalue spectrum being unaffected by higher-order moments in this case, the density inside the support of the spectrum certainly is. One obtains
\begin{align}
	\rho(r) &= \frac{1}{\pi} \frac{\partial C}{\partial \omega^\star}\nonumber \\
	&\approx \frac{1}{\pi \sigma^2} \bigg\{ 1 + \frac{\gamma_1}{q} \left(1 - 2\frac{r^2}{\sigma^2}\right) \nonumber \\
	&+ \frac{1}{q^2} \left[3 \gamma_1^2 + \gamma_1^4 - \gamma_2 + (4 \gamma_2 - 6 \gamma_1^4 - 18 \gamma_1^2) r^2 + (24 \gamma_1^2 + 9 \gamma_1^4 - 3 \gamma_2)r^4 - (12 \gamma_1^2 + 4 \gamma_1^4)r^6  \right] \bigg\}, \label{densitynongauss}
\end{align} 
where $r = \sqrt{\vert \omega \vert^2}$. Like the boundary of the bulk, the position of the outlier eigenvalue is similarly unaffected by the presence of higher order moments in this case. Following the reasoning in Section \ref{appendix:outlier}, one simply obtains
\begin{align}
	C(\lambda_{outlier}) &= \frac{1}{\mu}, \nonumber \\
	\Rightarrow \lambda_{outlier} &= \mu , \label{outliernongauss}
\end{align}
where we have used the expression for $C(\omega)$ in Eq.~(\ref{nongaussres1}). The expressions in Eqs.~(\ref{circlenongauss}), (\ref{densitynongauss}) and (\ref{outliernongauss}) are tested against the results of computer diagonalisation in Fig. \ref{fig:cauchy}.

\subsection{Example: Symmetric dense random matrices with Cauchy-distributed entries}\label{appendix:symmetricnongauss}
We now contrast the previous example with the case where the elements are again drawn from the distribution in Eq.~(\ref{cauchydistribution}), but where we constrain $a_{ij} = a_{ji}$. In this instance, we instead have 
\begin{align}
	\overline{ a_{ij} } &= \frac{\mu}{N}, \nonumber \\
	\overline{ (a_{ij}- \mu/N)^2} &= \overline{ (a_{ij}-\mu/N)(a_{ji}- \mu/N)} =\frac{\sigma^2}{N}, \nonumber \\
	\overline{ (a_{ij}-\mu/N)^4 } &= \overline{ (a_{ij}-\mu/N)^3( a_{ji}-\mu/N) } = \overline{ (a_{ij}-\mu/N)^2 (a_{ji}-\mu/N)^2 }=\frac{ \gamma_1 \sigma^4}{ q N}, \nonumber \\
	\overline{ (a_{ij} - \mu/N)^6} &=  \frac{\gamma_2\sigma^6}{ q^2 N}, \cdots \label{cauchystatssymmetric}
\end{align}
Comparing with Eq.~(\ref{loworderstats}), we can read off $\Gamma = 1$ and $\Gamma_4^{(1)} = \Gamma_4^{(2)} = \Gamma_4^{(3)} = \gamma_1 = 1/3$. From Eq.~(\ref{xcnongauss}) we see immediately that the support of the eigenvalue spectrum is modified by the higher-order moments, in contrast with the i.i.d. case of the previous subsection. 

\begin{figure}[H]
	\centering 
	\includegraphics[scale = 0.55]{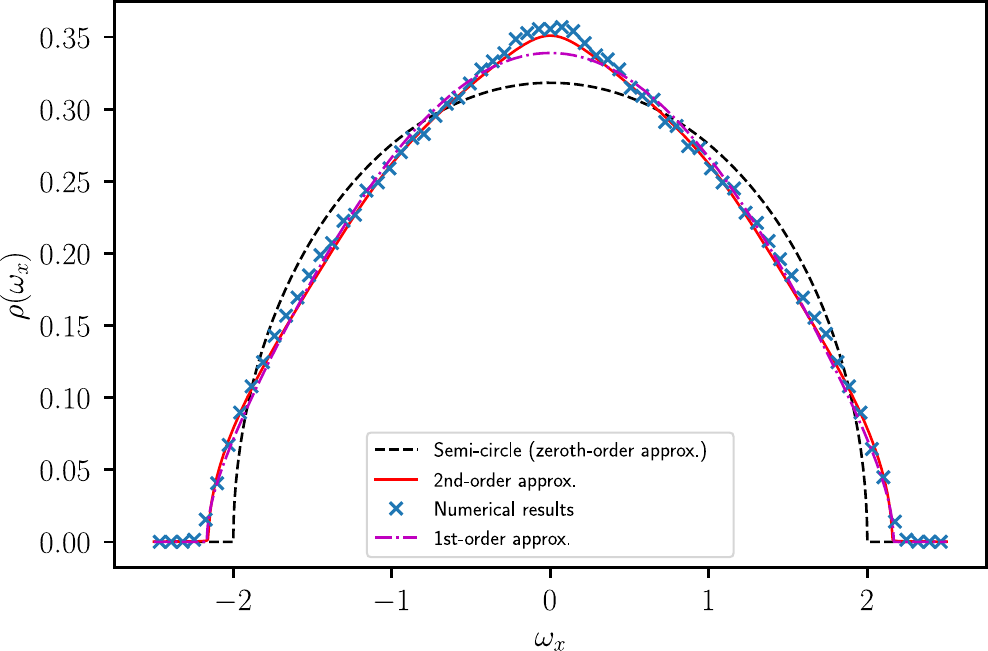}
	\caption{Eigenvalue spectrum of a symmetric random matrix with entries drawn from a truncated Cauchy distribution. The theory lines are the 1st-order approximation in Eq.~(\ref{generalizedwignernongauss}) and the second-order approximation derived from Eq.~(\ref{resgamma1nongauss}). A single matrix with parameters $N = 10000$, $q = 2$, $\mu = 0$ and $\sigma = 1$ was used.}\label{fig:cauchysymmetric}
\end{figure}

Performing a similar analysis to the previous section, one obtains a single analytic expression for the resolvent
\begin{align}
	C \approx \left[\omega - \sigma^2 C - \frac{\gamma_1 \sigma^4}{q} C^3 - \frac{1}{q^2} (\gamma_2 \sigma^6 C^5 + 2 \gamma_1^2 \sigma^8 C^7)  \right]^{-1}. \label{resgamma1nongauss}
\end{align}
If we were to solve this expression in a manner similar to Section \ref{appendix:semicirclegamma1}, we would arrive at an expression consistent with that in Eq.~(\ref{generalizedwignernongauss}). In Fig. \ref{fig:cauchysymmetric}, we show the result of solving Eq.~(\ref{resgamma1nongauss}) numerically and using the formula in Eq.~(4) of the main text to obtain the real eigenvalue density. Indeed, we see in this case that the both the extent of the support and the density within the support are modified heavily by the non-Gaussian corrections. 

\subsection{Example: Sparse matrices with small additional fluctuations}
Finally, we consider the case where most entries in a row fluctuate within $\pm\sigma/\sqrt{N}$ of zero (as would be the case for a dense random matrix), and we allow for the possibility of, on average, $p$ elements per row that are of order $N^0$. In other words, we modify the ensemble given in Eq.~(32) of the main text by allowing what were null entries to fluctuate around zero. 

\begin{figure}[H]
	\centering 
	\includegraphics[scale = 0.48]{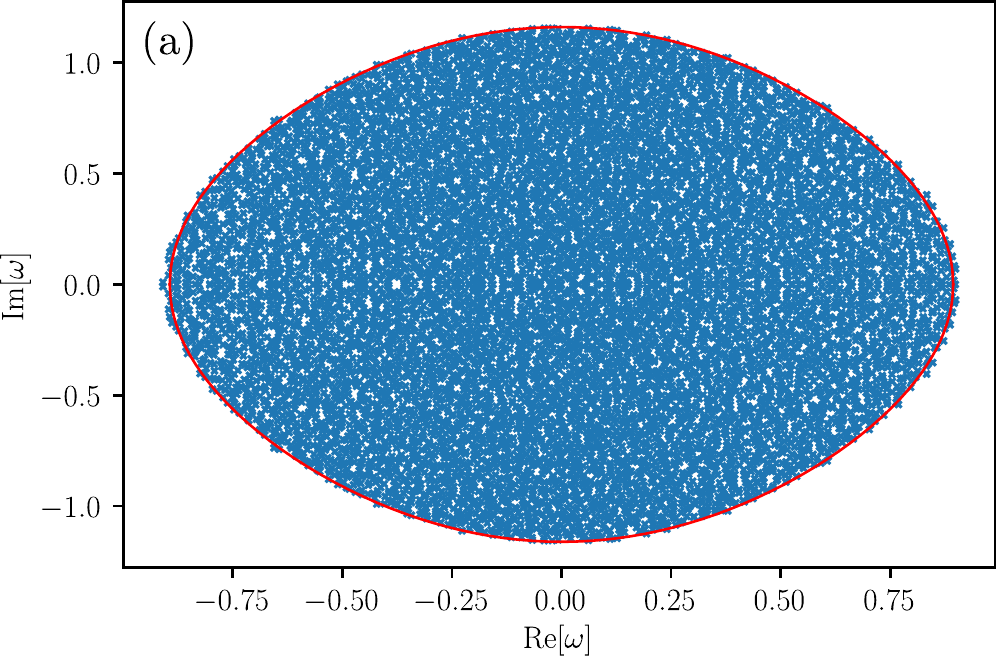}\hspace{2mm}
	\includegraphics[scale = 0.48]{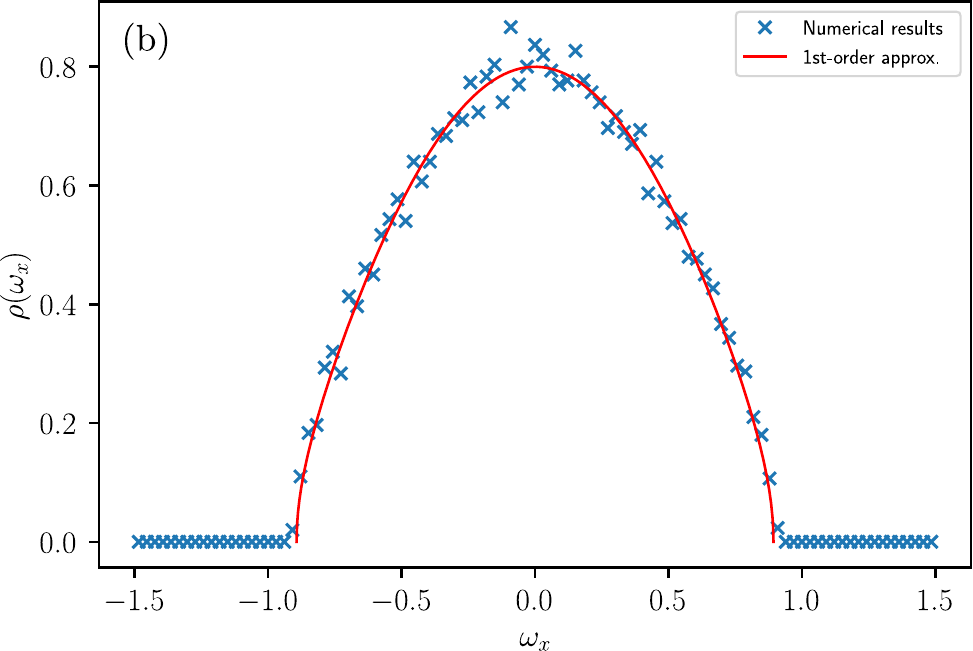}
	\caption{Eigenvalue spectrum of a dense matrix with elements drawn from Eq.~(\ref{denseplussparsedef}). Panel (a): Verification of the modified elliptic law in Eq.~(\ref{ellipsenongauss}) using the parameters in Eq.~(\ref{parameterconversionnongauss}). Panel (b): Verification of the modified semi-circle law in Eq.~(\ref{densitynongauss}). In both panels, a single matrix drawn using the distribution in Eqs.~(\ref{uniformdist}) and a joint Gaussian distribution for $\pi(a_{ij}, a_{ji})$, with parameters $N = 10000$, $p = 20$ and $\sigma_\tau = \sigma_\pi = 1/\sqrt{2}$, $\Gamma_\tau = -0.8$ and $\Gamma_\pi = 0.5$ was used.}\label{fig:denseplussparse}
\end{figure}

Specifically, we consider the ensemble
\begin{align}
	P(a_{ij}, a_{ji}) = \left( 1 - \frac{p}{N}\right) \tau(a_{ij}, a_{ji}) + \frac{p}{N} \pi(a_{ij}, a_{ji}), \label{denseplussparsedef}
\end{align}
where $\tau(a_{ij}, a_{ji})$ and $\pi(a_{ij}, a_{ji})$ are to be specified. We suppose that the distribution $\tau(a_{ij}, a_{ji})$ has statistics that obey
\begin{align}
	\langle a_{ij}\rangle_\tau &= \frac{\mu_\tau}{N}, \nonumber \\
	\langle (a_{ij}-\mu_\tau/N)^2\rangle_\tau &= \frac{\sigma_\tau^2}{N}, \nonumber \\
	\langle (a_{ij}-\mu_\tau/N)(a_{ji}-\mu_\tau/N)\rangle_\tau &= \frac{\Gamma_\tau\sigma_\tau^2}{N},
\end{align}
and we imagine that all higher-order moments decay more quickly than $1/N$. For the purposes of this section, we denote the statistics of $\pi(a_{ij}, a_{ji})$ as
\begin{align}
	\left\langle a_{ij} \right\rangle_\pi &= \frac{\mu_\pi}{p}, \nonumber \\
	\left\langle (a_{ij}- \mu_\pi/p)^2\right\rangle_\pi &= \frac{\sigma^2_\pi}{p}, \nonumber \\
	\left\langle (a_{ij}-\mu_\pi/p)(a_{ji}- \mu_\pi/p)\right\rangle_\pi &= \frac{\Gamma_\pi \sigma^2_\pi}{p},\nonumber \\
	\langle (a_{ij}-\mu_\pi/p)^4 \rangle_\pi &= \frac{\Gamma^{(1)}_{4,\pi} \sigma^4_\pi}{p^2 }, \nonumber \\
	\langle (a_{ij}-\mu_\pi/p)^3( a_{ji}-\mu_\pi/p) \rangle_\pi &= \frac{\Gamma^{(2)}_{4,\pi} \sigma_\pi^4}{p^2}, \nonumber \\
	\langle (a_{ij}-\mu/N)^2 (a_{ji}-\mu/N)^2 \rangle_\pi &= \frac{\Gamma^{(3)}_{4,\pi} \sigma^4_\pi}{p^2}.
\end{align}
One can then simply use the formulae the general formulae in Eqs.~(\ref{ellipsenongauss}), (\ref{xcnongauss}) and (\ref{generalizedwignernongauss}) by identifying
\begin{align}
	\sigma^2 &= \sigma^2_\tau + \sigma^2_\pi, \nonumber \\
	\Gamma &= \frac{\Gamma_\tau\sigma^2_\tau + \Gamma_\pi\sigma^2_\pi}{\sigma^2_\tau + \sigma^2_\pi} , \nonumber \\
	\Gamma^{(r)}_4 &= \frac{\Gamma_{4,\pi}^{(r)} \sigma_\pi^4}{(\sigma^2_\tau + \sigma^2_\pi)^2}, \nonumber \\
	q &= p . \label{parameterconversionnongauss}
\end{align}
As an example, we choose a correlated uniform distribution for $\tau(a_{ij}, a_{ji})$ so that
\begin{align}
	\tau(a_{ij}, a_{ji}) &= \sqrt{\frac{N}{12 \sigma^2}} \Theta\left( a_{ij} - \sqrt{\frac{3 \sigma^2}{N}}\right) \Theta\left( a_{ij} + \sqrt{\frac{3 \sigma^2}{N}}\right) \nonumber \\
	&\times \left[\frac{(1+\Gamma)}{2}\delta(a_{ij}- a_{ji}) + \frac{(1-\Gamma)}{2}\delta(a_{ij}+ a_{ji}) \right], \label{uniformdist}
\end{align}
and a Gaussian distribution for $\pi(a_{ij}, a_{ji})$. We verify  Eqs.~(\ref{ellipsenongauss}), (\ref{xcnongauss}) and (\ref{generalizedwignernongauss})  in Fig. \ref{fig:denseplussparse}, using the identification of parameters in Eq.~(\ref{parameterconversionnongauss}).

\section{Generalized Marchenko-Pastur law}\label{appendix:marchenkopastur}
We now use the dynamic approach to recover some results for products of random matrices \cite{akemann2021non,kanzieper2010non}. We also go on to recover some results for block-structured random matrices in Section \ref{appendix:blockstructure}. In Ref. \cite{cui2020perturbative}, the eigenvalue spectra of Wischart products of random matrices (i.e. matrices of the form $M = CC^T$) were analysed. Here, we take inspiration from their approach and generalise it for the case of arbitrary products of asymmetric random matrices using the path integral approach.

Consider a matrix of the the form $M_{ij} = \sum_k a_{ik}b_{kj}$, where the entries $a_{ik}$ and $b_{kj}$ are random variables and may have correlations. The matrix $\underline{\underline{a}}$ has dimensions $N \times M$ and the matrix $\underline{\underline{b}}$ has dimensions $M\times N$. We suppose the ratio $\alpha = M/N$ remains finite as $N\to \infty$ and we assume $\alpha \leq1$. In a similar fashion to Eqs.~(11) of the main text, we construct a dynamical system as follows
\begin{align}
	\dot x^1_i &= - \omega x_i^2 + \sum_{j = 1}^N a_{ij} y^2_j + h_i^1, \nonumber \\
	\dot x^2_i &= - \omega^\star x_i^1 + \sum_{j = 1}^N b^T_{ij} y^1_j + h_i^2 , \nonumber \\
	\dot y^1_i &= - y_i^2 + \sum_{j = 1}^N b_{ij} x^2_j + g_i^1, \nonumber \\
	\dot y^2_i &= - y_i^1 + \sum_{j = 1}^N a^T_{ij} x^1_j + g_i^2 .
\end{align}
We see once again that the Laplace transforms of the disorder-averaged response functions give the resolvent we desire
\begin{align}
	\frac{1}{N}\sum_{i}\lim_{\eta \to 0}\mathcal{L}_t\left\{ \overline{\frac{\delta x^2_i(t)}{h^1_i(t_0)}}\right\}(\eta) = \frac{1}{N} \sum_i \overline{\left[\omega  \underline{\underline{\id}} - \underline{\underline{a}} \underline{\underline{b}} \right]^{-1}_{ii}} . \label{resolventxx}
\end{align}
If we now use the alternative definitions [c.f. Eq.~(9) of the main text]
\begin{align}
	\mathinv &\equiv \begin{bmatrix}
		0 & \omega \underline{\underline{\id}}_{N} & 0 & 0 \\
		\omega^\star \underline{\underline{\id}}_{N} & 0 &0 &0 \\
		0 & 0 & 0 & \underline{\underline{\id}}_{M} \\
		0 & 0 & \underline{\underline{\id}}_{M} & 0
	\end{bmatrix} , \nonumber \\
	\mati &\equiv \begin{bmatrix}
		0& 0& 0& \underline{\underline{a}} \\
		0 & 0 & \underline{\underline{b}}^\dagger & 0\\
		0 & \underline{\underline{b}} & 0 & 0 \\ 
		\underline{\underline{a}}^\dagger & 0 & 0 & 0
	\end{bmatrix}, \nonumber \\
	\mathh &\equiv 	\left\langle \left[\eta \underline{\underline{\id}}_{2(N+M)} - \underline{\underline{\mati}} + \underline{\underline{\mathcal{H}}}_0^{-1} \right]^{-1} \right\rangle \nonumber \\
	&\equiv
	\begin{bmatrix}
		\underline{\underline{A}}_{xx} & \underline{\underline{B}}_{xx} & \underline{\underline{A}}_{xy} & \underline{\underline{B}}_{xy} \\
		\underline{\underline{C}}_{xx} & \underline{\underline{D}}_{xx} & \underline{\underline{C}}_{xy} & \underline{\underline{D}}_{xy} \\
		\underline{\underline{A}}_{yx} & \underline{\underline{B}}_{yx} & \underline{\underline{A}}_{yy} & \underline{\underline{B}}_{yy} \\
		\underline{\underline{C}}_{yx} & \underline{\underline{D}}_{yx} & \underline{\underline{C}}_{yy} & \underline{\underline{D}}_{yy} \\
	\end{bmatrix},\label{matrixdefs4}
\end{align}
with $\underline{\underline{H}} = \underline{\underline{\mathcal{H}}}_0^{-1}- \underline{\underline{\mati}}$ as before, we see that we can write the generating functional of this $2(N+M)$-component process in exactly the same way as Eq.~(14) of the main text. As a result, we find that the Hermitised resolvent for this random matrix ensemble also satisfies Eq.~(26), but with $\mathinv$, $\matj$, and $\mathh$ being $4\times4$ matrices corresponding to those in Eq.~(\ref{matrixdefs4}). For example
\begin{align}
	\matj &\equiv \begin{bmatrix}
		0& 0& 0& a_{12} \\
		0 & 0 & b_{21} & 0\\
		0 & b_{12} & 0 & 0 \\ 
		a_{21} & 0 & 0 & 0
	\end{bmatrix}.
\end{align}
We note that the resolvent we desire in Eq.~(\ref{resolventxx}) is given by $C_{xx} \equiv\frac{1}{N}\mathrm{Tr}C_{xx}$. 

\subsection{Products of square matrices with correlations}
\begin{figure}[H]
	\centering 
	\includegraphics[scale = 0.5]{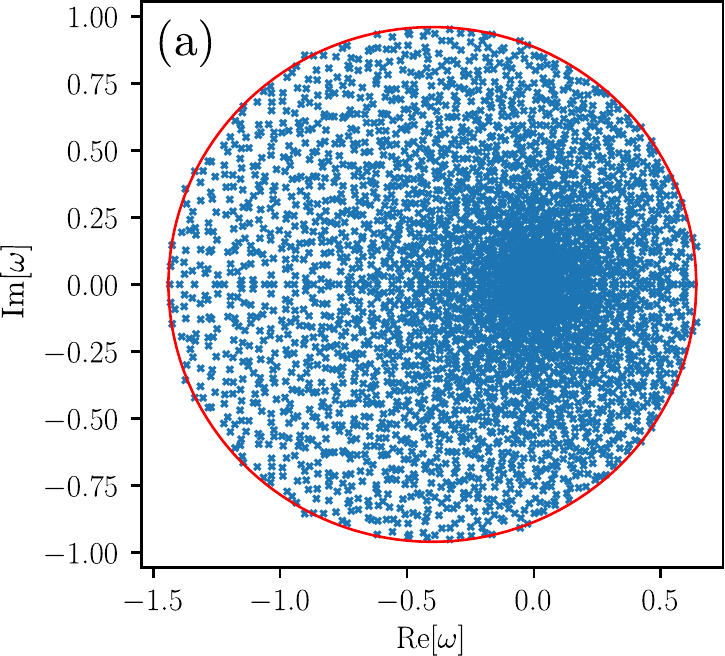}\hspace{2mm}
	\includegraphics[scale = 0.5]{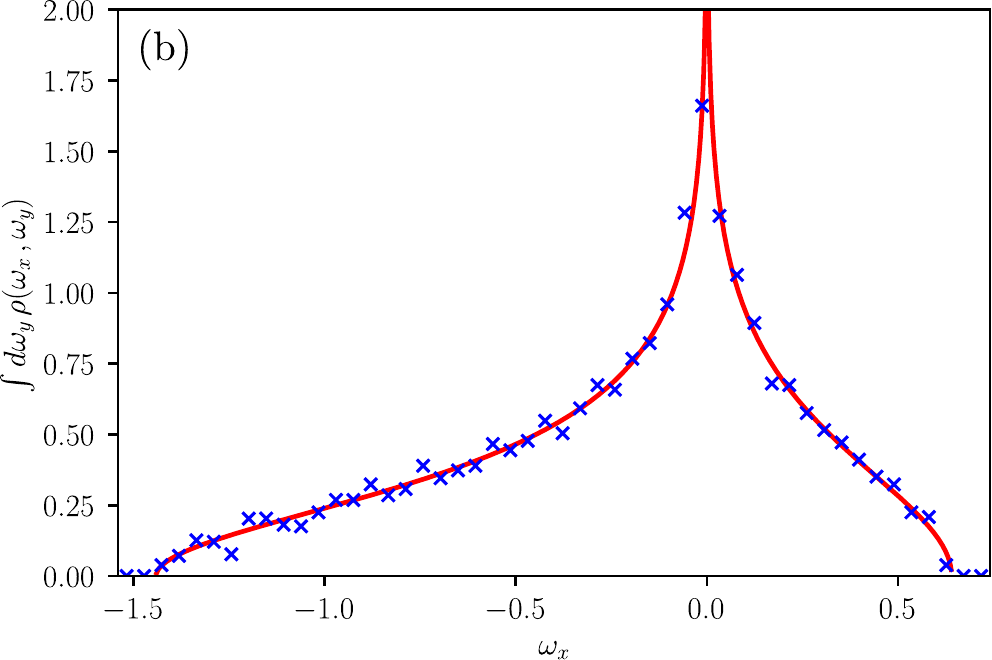}
	\caption{Eigenvalue spectra of a random matrix product as defined in Eq.~(\ref{productstats}). Parameters are $\sigma_a = 0.5$, $\sigma_b = 2$, $\Gamma_a = -0.3$, $\Gamma_b = -0.5$, $\Gamma_{ab^T} = 0.4$, $\Gamma_{ab} = -0.2$, $N = 4000$. Panel (a): Boundary of the eigenvalue spectrum. Blue crosses are the results of numerical diagonalisation and the red line is given by Eqs.~(\ref{boundaryproductsquare}). Panel (b): Density of the real parts of the eigenvalues. The red line comes from integrating Eq.~(\ref{densityproductsquare}) with respect to $y$ inside the boundary given by Eqs.~(\ref{boundaryproductsquare}).  }\label{fig:product}
\end{figure}

We first take the case where $M = N$ and we allow for the possibility for correlations between the matrix elements $a_{ij}$, $a_{ji}$, $b_{ij}$, $b_{ji}$. Specifically, we have
\begin{align}
	\overline{ a_{ij}^2 } = \frac{\sigma^2_a}{N}, \,\,\,\,\, \overline{ a_{ij}a_{ji} } = \frac{\Gamma_a \sigma_a^2}{N} , \nonumber \\
	\overline{ b_{ij}^2 } = \frac{\sigma^2_b}{N}, \,\,\,\,\, \overline{ b_{ij} b_{ji} } = \frac{\Gamma_b \sigma_b^2}{N}, \nonumber \\
	\overline{ a_{ij} b_{ij} } = \frac{\Gamma_{ab^T} \sigma_a \sigma_b}{N}, \,\,\,\,\, \overline{ a_{ij} b_{ji} } = \frac{\Gamma_{ab} \sigma_a \sigma_b}{N} . \label{productstats}
\end{align}
In this case, we solve the self-consistent expression for $\mathh$ in main text Eq.~(26) to find two possible solutions for $C_{xx}$ (we do not reproduce the corresponding expressions for the other traces of the blocks of $\mathh$ because they are lengthy and unenlightening)
\begin{align}
	C_{xx} = \frac{\omega - \sqrt{\omega^2 - 4 \omega \Gamma_{ab}\sigma_a \sigma_b}}{2 \Gamma_{ab} \sigma_a \sigma_b \omega}
\end{align}
or
\begin{align}
	C_{xx} = \frac{-\Gamma_{ab}\omega  + \sqrt{\omega \omega^\star}}{\sigma_a \sigma_b \omega (1-\Gamma_{ab}^2)}.
\end{align}
The latter solution for $C_{xx}$ is non-analytic, and so must correspond to the bulk region, whereas the former is analytic and corresponds to the outside region. Finally, by finding the values of $\omega$ for which these two solutions coincide, we see that the boundary of the bulk region is given by the following ellipse
\begin{align}
	\left[ \frac{y}{1-\Gamma_{ab}^2}\right]^2 + \left[\frac{x - 2 \Gamma_{ab} \sigma_a \sigma_b}{1 + \Gamma_{ab}^2} \right]^2 = \sigma_a^2 \sigma_b^2, \label{boundaryproductsquare}
\end{align}
(one notes that $\vert \Gamma_{ab} \vert<1$) and the density of eigenvalues inside the bulk region is given by [using Eq.~(3) of the main text]
\begin{align}
	\rho(\omega) = \frac{1}{2\pi \sigma_a \sigma_b (1-\Gamma_{ab}^2)\vert \omega \vert}. \label{densityproductsquare}
\end{align}
The expressions in Eqs.~(\ref{boundaryproductsquare}) and (\ref{densityproductsquare}) are verified in Figs. \ref{fig:product} (a) and (b) respectively.

One notes that the only statistics that matter in determining the eigenvalues in the limit $N\to \infty$ are the variances of the elements $b_{ij}$ and $a_{ij}$ and the correlations between $a_{ij}$ and $b_{ji}$. Further, unlike the spectrum of a single dense random matrix, the spectrum is shifted by an amount $2 \Gamma_{ab} \sigma_a \sigma_b$, which can be either positive or negative depending on the sign of $\Gamma_{ab}$. Also, the ellipse is always broader along the real axis than along the imaginary one. We see that the left-most and right-most eigenvalues are given by (respectively)
\begin{align}
	\lambda_l &= -\sigma_a \sigma_b \left(  1 -\Gamma_{ab}\right)^2, \nonumber \\
	\lambda_r &= \sigma_a \sigma_b \left(1 +\Gamma_{ab}\right)^2.
\end{align}
In the case $a_{ij} = b_{ij}$, these results must agree with those that we would obtain by simply squaring the eigenvalues of $\underline{\underline{a}}$. Indeed, we can see that this is the case as follows. 

When $a_{ij} = b_{ij}$, we have that $\Gamma_{ab} = \Gamma_a$. The eigenvalues of $\underline{\underline{a}}$ fall inside an ellipse $x_a^2(1-\Gamma_{ab})^2+ y_a^2(1+\Gamma_{ab})^2= (1-\Gamma_{ab}^2)^2 \sigma_a^2$, and have density $\rho_a = 1/[\sigma_a^2 \pi (1-\Gamma_{ab}^2)]$. Since the eigenvalues are squared, we make a change of coordinates $x = x_a^2-y_a^2$ and $y = 2 x_a y_a$. Substituting these expressions into the aforementioned equation for the ellipse for the eigenvalues of $\underline{\underline{a}}$, one recovers Eq.~(\ref{boundaryproductsquare}) with $\sigma_b = \sigma_a$. By noting that the Jacobian determinant of this transformation is $\vert J\vert = 1/[4 \sqrt{x^2 + y^2}]$ and also that both $(x_a, y_a)$ and $(-x_a, -y_a)$ are mapped onto the same $(x, y)$, giving an extra factor of $2$, we also recover the density in Eq.~(\ref{densityproductsquare}) with $\sigma_b = \sigma_a$.

\subsection{Products of matrices with arbitrary dimension}
We now examine the case where $M \neq N$ and recalling $\alpha = M/N$, which we suppose remains finite as $N \to \infty$, with $\alpha\leq 1$. We no longer allow for correlations between matrix elements $a_{ij}$ and $a_{ji}$ or $b_{ij}$ and $b_{ji}$ [which did not affect the eigenvalue spectrum in the $M = N$ case anyway], but we still allow for correlations between $a_{ij}$ and $b_{ji}$. In the limiting case $a_{ij} = c b_{ji}$ for constant $c$, we recover the Marchenko-Pastur law.   

Following a similar line of reasoning to the previous section, we find the following analytic expression
\begin{align}
	C_{xx} = \frac{(1-\alpha) \sigma_a \sigma_b \Gamma_{ab} + \omega + \sqrt{[(1-\alpha) \Gamma_{ab} \sigma_a \sigma_b + \omega]^2 - 4 \Gamma_{ab}\sigma_a \sigma_b \omega}}{2 \Gamma_{ab} \omega \sigma_a \sigma_b}, \label{analyticproductresolvent}
\end{align}
and the following non-analytic expression
\begin{align}
	C_{xx} = \frac{(1-\alpha)}{2 \omega} - \frac{\Gamma_{ab}}{\sigma_a \sigma_b (1-\Gamma_{ab}^2)} + \frac{\sqrt{\sigma_a^2 \sigma_b^2 (1-\alpha)^2 (1-\Gamma_{ab}^2)^2 + 4 \omega \omega^\star}}{2 \sigma_a \sigma_b (1-\Gamma_{ab}^2)\omega }. 
\end{align}
In this case, we obtain the following shifted ellipse for the boundary of the bulk region
\begin{align}
	\left[ \frac{y}{1-\Gamma_{ab}^2}\right]^2 + \left[\frac{x - (1+\alpha) \Gamma_{ab} \sigma_a \sigma_b}{1 + \Gamma_{ab}^2} \right]^2 = \alpha \sigma_a^2 \sigma_b^2,  \label{boundaryproductapp}
\end{align}
with the following eigenvalue density
\begin{align}
	\rho(\omega) = \frac{(1-\alpha)}{2}\delta(\omega)+ \frac{1}{\pi \sigma_a \sigma_b (1-\Gamma_{ab}^2)\sqrt{\sigma_a^2 \sigma_b^2 (1-\alpha)^2(1-\Gamma_{ab}^2)^2 + 4 \vert \omega \vert^2}}. \label{densityproductapp}
\end{align}
The left-most and right-most leading eigenvalues are (respectively)
\begin{align}
	\lambda_l &= \sigma_a \sigma_b \left[ (1+\alpha) \Gamma_{ab} - (1+\Gamma_{ab}^2)\sqrt{\alpha}\right], \nonumber \\
	\lambda_r &= \sigma_a \sigma_b \left[ (1+\alpha) \Gamma_{ab} + (1+\Gamma_{ab}^2)\sqrt{\alpha}\right].
\end{align}
Integrating Eq.~(\ref{densityproductapp}) with respect to $y$ between the limits imposed by Eq.~(\ref{boundaryproductapp}), one obtains for $\rho_x(x) = \int dy \rho(x,y)$
\begin{align}
	\rho_x(x) &= \frac{1}{\pi \sigma_a \sigma_b (1-\Gamma_{ab}^2)} \log\left[\frac{2 y_{+} + \sqrt{\sigma_a^2\sigma_b^2(1-\Gamma_{ab}^2)^2(1-\alpha)^2 + 4(x^2 + y_+^2)}}{\sqrt{\sigma_a^2\sigma_b^2(1-\Gamma_{ab}^2)^2(1-\alpha)^2 + 4x^2}} \right], \nonumber \\
	y_+(x) &= \frac{(1-\Gamma_{ab}^2)}{(1 + \Gamma_{ab}^2)} \sqrt{(x- \lambda_l)(\lambda_r- x)} . \label{xdensproductapp}
\end{align}
This is the generalised Marchenko-Pastur law to which we referred in the main text. 

That this indeed reduces to the usual Marchenko-Pastur law can be seen as follows. In the case where $\Gamma_{ab}= 1$, the product $\sum_k a_{ik}b_{kj}$ becomes symmetric and the eigenvalues concentrate on the real line. When this is the case, we expand the logarithm in Eq.~(\ref{xdensproductapp}), keeping only leading-order terms in $(1-\Gamma_{ab}^2)$, to obtain
\begin{align}
	\rho_x \approx  \frac{1}{\pi \sigma_a \sigma_b (1-\Gamma_{ab}^2)} \left[ 1 + \frac{y_+}{x}\right] .
\end{align}
We thus recover the usual Marchenko-Pastur law
\begin{align}
	\rho_x(x) = \frac{(1-\alpha)}{2} \delta(x) + \frac{\sqrt{\left[x - \sigma_a \sigma_b (1-\sqrt{\alpha})^2  \right] \left[ \sigma_a \sigma_b (1+\sqrt{\alpha})^2 - x \right]}}{2 \pi  x \sigma_a \sigma_b} .
\end{align}
One can also extract this expression for the real eigenvalue density directly from the analytic expression for the resolvent in Eq.~(\ref{analyticproductresolvent}) via Eq.~(4) of the main text in the case $\Gamma_{ab} \to 1$.

\begin{figure}[H]
	\centering 
	\includegraphics[scale = 0.5]{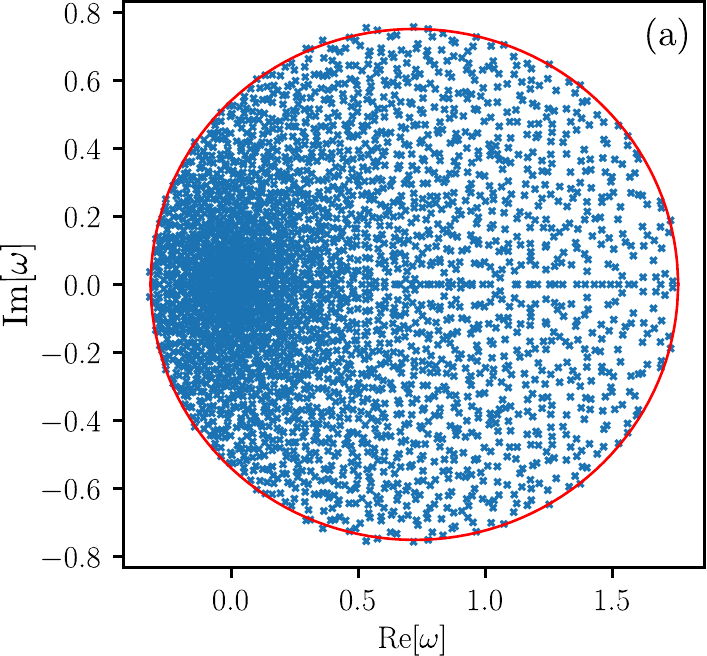}\hspace{5mm}
	\captionsetup{justification=raggedright,singlelinecheck=false}
	\includegraphics[scale = 0.5]{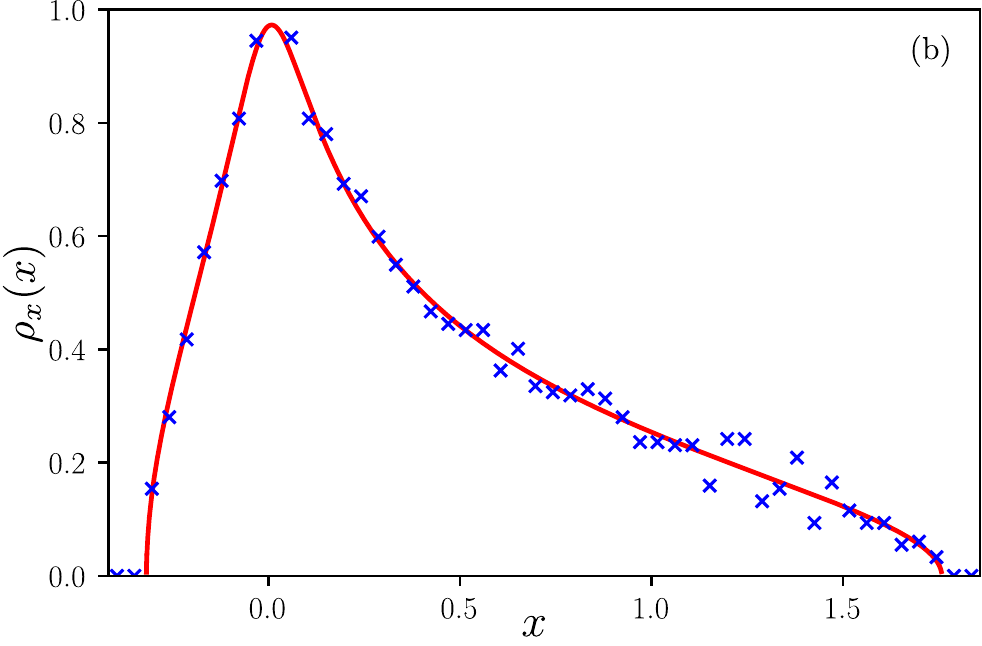}
	\caption{Eigenvalue spectrum of a (dense) random matrix product as defined in Eq.~(\ref{productstats}). Parameters are $\sigma_a = 0.5$, $\sigma_b = 2$, $\alpha = 0.8$, $\Gamma_{ab} = 0.4$, $N = 4000$. Panel (a): Boundary of the eigenvalue spectrum. Blue crosses are the results of numerical diagonalization and the red line is given by the shifted ellipse in Eq.~(\ref{boundaryproductapp}). Panel (b): Integrated eigenvalue density as a function of the real part, averaged over $y = \mathrm{Im} \,\omega$. The red line is given by the generalized Marchenko-Pastur law in Eq.~(\ref{xdensproductapp}).  }\label{fig:productnonsquare}
\end{figure}

\section{Block-structured random matrices}\label{appendix:blockstructure}
We now demonstrate how one can use the dynamic method to recover results that were previously obtained for block-structured random matrices in Ref. \cite{baron2020dispersal}. 

Consider a block-structured random matrix of dimension $N\times N$. The $N$ components are divided into groups of size $N_\alpha$ such that $\sum_\alpha N_\alpha = N$. We define $\gamma_\alpha = N_\alpha/N$. The matrix elements have statistics that are block dependent
\begin{align}
	\overline{ a_{ij}^{\alpha\beta}}  =0, \,\,\,\,\,\overline{( a_{ij}^{\alpha\beta})^2} = \frac{\sigma^2}{N}, \,\,\,\,\, \overline{ a^{\alpha\beta}_{ij}a^{\beta \alpha}_{ji}} = \frac{\Gamma_{\alpha\beta}\sigma^2}{N},
\end{align}
where $a_{ij}^{\alpha\beta}$ is the $(i,j)$ component of the $(\alpha, \beta)$ block. 

In this case, we define the following dynamical system
\begin{align}
	\dot x_i^{(1) \alpha} &= - \omega x_i^{(2)\alpha} + \sum_{j} a_{ij}^{\alpha\beta} x^{(2)\beta}_j + h_i^{(1)\alpha}, \nonumber \\
	\dot x_i^{(2) \alpha} &= - \omega^\star x_i^{(1)\alpha} + \sum_{j} a_{ij}^{\alpha\beta} x^{(1)\beta}_j + h_i^{(2)\alpha}.
\end{align}
The resolvent we desire is given by
\begin{align}
	C(\omega) = \lim_{\eta\to 0}\mathcal{L}_t\left[ \frac{1}{N}\sum_{i,\alpha} \frac{\delta x_i^{(1)\alpha}}{\delta h_i^{(2)\alpha}}\right](\eta)
\end{align}
In a similar fashion as the preceding section, we identify the following matrices in the $2$-block case, although the extension to more blocks is straightforward
\begin{align}
	\mathinv &\equiv \begin{bmatrix}
		0 & 0  & \omega \underline{\underline{\id}}_{N_1} & 0 \\
		0  & 0 &0 &\omega  \underline{\underline{\id}}_{N_2} \\
		\omega^\star  \underline{\underline{\id}}_{N_1} & 0 & 0 & 0 \\
		0 & \omega^\star  \underline{\underline{\id}}_{N_2} & 0 & 0
	\end{bmatrix} , \nonumber \\
	\mati &\equiv \begin{bmatrix}
		0& 0& \underline{\underline{a}}^{(1,1)}& \underline{\underline{a}}^{(1,2)} \\
		0 & 0 & \underline{\underline{a}}^{(2,1)} & \underline{\underline{a}}^{(2,2)}\\
		[\underline{\underline{a}}^{(1,1)}]^T & [\underline{\underline{a}}^{(2,1)}]^T  & 0 & 0 \\ 
		[\underline{\underline{a}}^{(1,2)}]^T  & [\underline{\underline{a}}^{(2,2)}]^T  & 0 & 0
	\end{bmatrix}, \nonumber \\
	\mathh &\equiv 	\left\langle \left[\eta \underline{\underline{\id}}_{N} - \underline{\underline{\mati}} + \underline{\underline{\mathcal{H}}}_0^{-1} \right]^{-1} \right\rangle \nonumber \\
	&\equiv
	\begin{bmatrix}
		\underline{\underline{A}}_{11} & \underline{\underline{A}}_{12} & \underline{\underline{B}}_{11} & \underline{\underline{B}}_{12} \\
		\underline{\underline{A}}_{21} & \underline{\underline{A}}_{22} & \underline{\underline{B}}_{21} & \underline{\underline{B}}_{22} \\
		\underline{\underline{C}}_{11} & \underline{\underline{C}}_{12} & \underline{\underline{D}}_{11} & \underline{\underline{D}}_{12} \\
		\underline{\underline{C}}_{21} & \underline{\underline{C}}_{22} & \underline{\underline{D}}_{21} & \underline{\underline{D}}_{22} \\
	\end{bmatrix}.\label{matrixdefsblock}
\end{align}
Again, we define a set of $4\times4$ matrices $\mathinv$, $\matj$, and $\mathh$ corresponding to those in Eq.~(\ref{matrixdefsblock}). For example
\begin{align}
	\matj &\equiv \begin{bmatrix}
		0& 0& a_{12}^{(1,1)}& a_{12}^{(1,2)} \\
		0 & 0 & a_{12}^{(2,1)} & a_{12}^{(2,2)}\\
		a_{21}^{(1,1)} & a_{21}^{(2,1)}  & 0 & 0 \\ 
		a_{21}^{(1,2)}  & a_{21}^{(2,2)}  & 0 & 0
	\end{bmatrix}.
\end{align}
As a result, we see that we can once again write the generating functional of the system in exactly the same form as in Eq.~(14) of the main text. This means that the hermitized resolvent is again given by main text Eq.~(26). Plugging in Eqs.~(\ref{matrixdefsblock}) into Eq.~(26) yields (defining $C_{\alpha \alpha} = N^{-1}_\alpha \mathrm{Tr}[ \underline{\underline{C}}_{\alpha\alpha}]$, etc.)
\begin{align}
	A_{\alpha\alpha} &= - \frac{\sigma^2}{q_\alpha}\sum_{\beta} \gamma_\beta A_{\beta\beta},\nonumber \\
	B_{\alpha\alpha} &= -\frac{1}{q_\alpha} \left( \omega - \sigma^2\sum_\beta \Gamma_{\alpha\beta} \gamma_\beta C_{\beta\beta}\right), \nonumber \\
	C_{\alpha\alpha} &= -\frac{1}{q_\alpha} \left( \omega^\star - \sigma^2\sum_\beta \Gamma_{\alpha\beta} \gamma_\beta B_{\beta\beta}\right), \nonumber \\
	D_{\alpha\alpha} &= - \frac{\sigma^2}{q_\alpha}\sum_{\beta} \gamma_\beta D_{\beta\beta}, \nonumber \\
	\frac{1}{q_\alpha} &= A_{\alpha\alpha} D_{\alpha \alpha} - B_{\alpha\alpha} C_{\alpha \alpha} .
\end{align}
with $\underline{\underline{A}}_{\alpha\beta} = \underline{\underline{B}}_{\alpha\beta} = \underline{\underline{C}}_{\alpha\beta} = \underline{\underline{D}}_{\alpha\beta} = 0$ for $\alpha \neq \beta$. One finds that the above expressions are fully general, regardless of the number of blocks.

There are two solutions to the above equations for the resolvent $C = \sum_\alpha \gamma_\alpha C_{\alpha\alpha}$. One for which $A_{\alpha\alpha} = 0$ and one for which $\sum_{\beta} \frac{\gamma_\beta}{q_\beta} = -\frac{1}{\sigma^2}$. In the first case, we have
\begin{align}
	C_{\alpha\alpha} &= \left( \omega - \sigma^2\sum_\beta \Gamma_{\alpha\beta}\gamma_\beta C_{\beta\beta} \right)^{-1}. \label{analyticblock}
\end{align}
The other solution yields instead (letting $A \equiv \sum_{\alpha} \gamma_\alpha A_{\alpha\alpha}$, etc.)
\begin{align}
	\sum_{\alpha} \frac{\gamma_\alpha}{q_\alpha} &= - \frac{1}{\sigma^2}, \nonumber \\
	C_{\alpha\alpha} B_{\alpha\alpha} &= \frac{AD}{q_\alpha^2 \left( \sum_\beta \frac{\gamma_\beta}{q_\beta}\right)}-\frac{1}{q_\alpha}, \nonumber \\
	C_{\alpha\alpha} &= -\frac{1}{q_\alpha} \left[\omega - \sigma^2\sum_\beta \Gamma_{\alpha\beta} \gamma_\beta B_{\beta\beta} \right], \nonumber \\
	B_{\alpha\alpha} &= C_{\alpha\alpha}^\star, \label{nonanlyticblock}
\end{align}
which constitute $3 N_B +1$ equations for the $3 N_B +1$ unknowns $C_\alpha$, $B_\alpha$, $q_\alpha$ and $AD$, where $N_B$ is the number of values that $\alpha$ can take. Eqs.~(\ref{analyticblock}) and (\ref{nonanlyticblock}) are equivalent to Eqs.~(S45) and (S47) respectively in the Supplement of Ref. \cite{baron2020dispersal}. In the aforementioned work, these expressions for the resolvent were solved numerically to yield the boundary of the bulk of the eigenvalue spectrum. We have thus succeeded in using the dynamic approach of the main text to recover previous results for the case of block structured matrices.